\documentclass[12pt,a4paper]{article}
\usepackage[left=2.3cm,top=2.4cm,right=2.3cm,bottom=2.4cm]{geometry}
\usepackage[utf8x]{inputenc}
\usepackage{ucs}
\usepackage{authblk}
\usepackage[authoryear]{natbib}
\usepackage{amsmath,amsfonts,bm,amssymb,latexsym,amsthm}
\usepackage{graphicx, graphics}
\usepackage{multirow}
\usepackage{microtype}
\usepackage[none]{hyphenat}
\usepackage{setspace} 
\usepackage{array}
\usepackage{tikz}
\usetikzlibrary{tikzmark}
\usetikzlibrary{decorations.pathreplacing,calc}
\usepackage{amsfonts}
\usepackage{makecell}
\usepackage[bottom]{footmisc}
\usepackage{color}
\usepackage{url}
\usepackage{float}
\usepackage{subfigure}
\usepackage{rotating}
\usepackage{mathtools}

\usepackage{enumerate}

\usepackage{morefloats}
\usepackage{booktabs}
\usepackage{multirow}
\usepackage[para,online,flushleft]{threeparttable}
\usepackage[justification=justified,font=small]{caption}

\usepackage{pifont}

\usepackage{subcaption} 

\usepackage{svg}

\usepackage{footnote}
\usepackage[flushleft]{threeparttable}

\newcolumntype{C}[1]{>{\centering\let\newline\\\arraybackslash\hspace{0pt}}m{#1}}

\usepackage{hyperref}
\hypersetup{colorlinks, breaklinks, linkcolor=[rgb]{0,0.3,1}, citecolor=[rgb]{0,0.3,1}, urlcolor=[rgb]{0,0.3,1} }
\usepackage{longtable}

\title{Assessing the Heterogeneous Impact of Economy-Wide Shocks: A Machine Learning Approach Applied to Colombian Firms\thanks{We would like to thank Nicola Lacetera, Florian Mayneris, Isabelle Mejean, Asier Minondo, Jes\'us Peir\'o-Palomino, Michele Ruta, Dario Sansone, and Hylke Vandenbussche for valuable conversations, and seminar participants at the ASSA meeting, ETSG, ISGEP, EEA Annual Congress, Causal Data Science Meeting, University of Valencia - INTECO, University of Massachusetts Lowell, UC Louvain-KU Leuven Trade Workshop, Nebrija University Madrid, Externado University of Colombia and Central Bank of Colombia for insightful comments.} 
}

\author[1,2]{Marco Due\~{n}as}
\author[2]{Federico Nutarelli}
\author[2]{V\'ictor Ortiz-Gim\'enez}
\author[2]{Massimo Riccaboni}
\author[2]{Francesco Serti \thanks{Corresponding author: \href{mailto:francesco.serti@imtlucca.it}{francesco.serti@imtlucca.it}. Laboratory for the Analysis of Complex Economic Systems, IMT School for Advanced Studies, Piazza San Francesco 19 - 55100 Lucca, Italy.}}

\affil[1]{\small CLIMAFIN -- Climate Finance Alpha, Paris, France}
\affil[2]{\small IMT School for Advanced Studies Lucca}

\date{}

\begin{document}
\maketitle
\vspace{-40pt}

\begin{abstract}
\noindent 
Our paper presents a methodology to study the heterogeneous effects of economy-wide shocks and applies it to the case of the impact of the COVID-19 crisis on exports. This methodology is applicable in scenarios where the pervasive nature of the shock hinders the identification of a control group unaffected by the shock, as well as the ex-ante definition of the intensity of the shock's exposure of each unit. In particular, our study investigates the effectiveness of various Machine Learning (ML) techniques in predicting firms' trade 
and, by building on recent developments in causal ML, uses these predictions to reconstruct the counterfactual distribution of firms' trade under different COVID-19 scenarios and to study treatment effect heterogeneity. Specifically, we focus on the probability of Colombian firms surviving in the export market under two different scenarios: a COVID-19 setting and a non-COVID-19 counterfactual situation. On average, we find that the COVID-19 shock decreased a firm's probability of surviving in the export market by about 20 percentage points in April 2020. We study the treatment effect heterogeneity by employing a classification analysis that compares the characteristics of the firms on the tails of the estimated distribution of the individual treatment effects.


\end{abstract}



\clearpage
\section{Introduction}

This paper presents a methodology to study the heterogeneous effects of economy-wide shocks, applicable in scenarios where neither a control group unaffected by the shock nor an ex-ante definition of each unit's exposure intensity is possible. We apply this methodology to the impact of the COVID-19 crisis on exports. In particular, we aim to estimate the causal effect of COVID-19 on a firm's probability of survival in the export markets and to study the heterogeneity of this effect. The main hurdles for this evaluation task are related to the pervasiveness of the COVID-19 shock. On the one hand, the fact that all firms are eventually directly or indirectly exposed to the effects of the COVID-19 crisis makes it hardly possible to find a control group of firms to be used to build a counterfactual non-COVID-19 scenario. On the other hand, adopting a continuous treatment variable would imply defining ex-ante the main patterns through which the COVID-19 shock has affected firm-level trade. This task is highly demanding, given that the economy-wide impact of the shock is coupled with complex interdependencies between firms and products across sectors and countries, as underlined below. The paper's main idea is to address these evaluation challenges, which are present when studying the heterogeneous impact of economy-wide shocks, by leveraging the predictive capabilities of Machine Learning (ML) techniques.

The COVID-19 outbreak has affected the world economy, generating unprecedented health, human, and economic crises. To face the health crisis, governments implemented social distancing and lockdown policies, exacerbating supply and demand shocks \citep{world_bank2020}. In a highly interconnected world, the impact of the pandemic on international trade has gained great attention \citep{felbermayr2020implications, 655456, bonadio2020global, evenett2020sicken}. 
 Global trade, which is typically more volatile than output and tends to fall sharply during a crisis, has shown the biggest fall since the 2009 global financial crisis. From the beginning of the COVID-19 epidemic, scholars underlined that, though its impact on international trade could have been comparable to the Great Trade Collapse of 2008-2009, this time, the demand side shock is accompanied by a supply-side shock \citep{baldwin2020thinking}. Moreover, this supply-side effect could be reinforced by a supply-side contagion via importing/supply chains, which have grown in relevance during the last decade \citep{antras2021}. In other words, supply disruptions in the countries providing intermediate inputs to a given country are likely to hurt also its export performance \citep{halpern2015imported, navas2020role}.

We focus on Colombian exporters because of the vulnerability of the Colombian economy to the COVID-19 shock and the availability of customs data. Similar to many other developing and developed countries, in 2020, Colombia has witnessed domestic supply and demand shocks related to factory closures, cessation of some public services, and disruptions in the supply chains. 
We find significant sectoral heterogeneity in the impact of COVID-19. We highlight that more affected firms tend to be small-sized (in terms of previous trade flows) and more exposed to export destinations and import source countries that are more severely hit by the containment policies related to the COVID-19 shock. Finally, we unveil that the degree of geographical diversification and  involvement in the import market are important determinants of resilience to the COVID-19 shock.

By interpreting exporters' dynamics as a complex learning process,\footnote{Firms have heterogeneous and incomplete information about the trade opportunities. This is true both on the exporting and the importing side of firm activities. For example, in \cite{albornoz2012sequential} and \cite{eslava2015search} exporting firms are uncertain and learn about the appeal of their products and, more in general, about the profitability of exporting their products on the international markets. By searching for clients and observing their realized profitability, firms update their beliefs about their capabilities in international markets.} 
this paper's first contribution is exploring and comparing the effectiveness of different ML techniques in predicting firms' trade status in two different scenarios, a COVID-19 and a non-COVID-19 setting. ML techniques have been successfully applied to predict firm performances in high-dimensional contexts \citep{bargagli2020supervised} in which the number of potentially relevant explanatory variables is very high. 
Our paper fits into a nascent literature that is applying ML techniques to study international trade patterns \citep{breinlich2022machine} and, up to what we know, in our study for the first time ML techniques are used to investigate firm-level international trade performance. Estimating more accurately the likelihood of a firm's success in exporting could be useful to increase the effectiveness of export promotion agencies \citep{van2015impact} by helping them target their activities. However, the effectiveness of ML in improving the prediction of a firm's success cannot be taken for granted, especially for developing countries, as shown by \cite{mckenzie2019predicting}.  

This paper's second and main contribution is to show how to use these predictions to estimate the causal effect of the COVID-19 shock at the firm level and to study its possible heterogeneity. We use the estimated ML model with the best performance in predicting the 2019 export status of firms exporting in 2018 to build a 2020 non-COVID-19 counterfactual outcome for firms exporting in 2019. Then, we compare these counterfactual non-COVID-19 firm-level export probabilities with the predicted probabilities of the best-performing ML model using the characteristics of 2019 exporters to predict their export status in 2020. The latter estimated probabilities summarize the information on the observed COVID-19 scenario and express it in a metric comparable with the estimated counterfactual non-COVID-19 outcomes. In the literature using ML counterfactuals  \citep{cerqua2020local, fabra2020degrowth}, it is instead common to estimate causal effects by comparing the counterfactual predictions with the observed outcome in case of treatment, following the so-called ``consistency assumption'': if the outcome in case of treatment is observed then it also represents the potential outcome under treatment. We follow \cite{chernozhukov2020generic} by using ML techniques to reconstruct firm potential outcomes in case of no treatment and to predict the outcomes in the treatment scenario. From a methodological standpoint, our study represents the pioneering application and adaptation of the generic machine learning tools proposed by \cite{chernozhukov2020generic} in a context where a control group is unavailable.\footnote{For an application of the generic machine learning methodology to the standard setting of a randomized controlled trial
 see \cite{magnan2021information}.}
Furthermore, we provide guidance on utilizing in-time placebo tests to assess the credibility of counterfactual estimates. Additionally, we compare the estimation results of the average treatment effect and treatment effect heterogeneity obtained by employing the predicted outcomes in the case of treatment, as proposed by \cite{chernozhukov2020generic}), with those obtained using the observed outcomes (i.e., following the \cite{cerqua2020local} and \cite{fabra2020degrowth} approach). Our findings suggest that while the estimates of the average treatment effect remain robust across methodologies, the former approach should be preferred when the objective is to identify the observations with the highest and lowest treatment effects, and subsequently determine the factors contributing to treatment effect heterogeneity.

Examining the heterogeneous effects of economy-wide shocks is a crucial undertaking as it represents the foundational stage in devising policy interventions intended to mitigate their deleterious outcomes and reactivate economic operations. 
However, from a methodological point of view, investigating the treatment effect heterogeneity  is not a straightforward task when its potential determinants are many. 
The traditional approach splits the sample into groups to assess the significance of the difference in the treatment effects of the groups. Unfortunately, this approach is prone to overfitting, and finding statistically significant differences out of all possible splits might be entirely due to random noise. 
Recently, new tools based on ML have been developed to identify subgroups that are particularly responsive to the treatment \citep{athey2019generalized, chernozhukov2020generic}. 
Building on the recent progress in causal ML application to the analysis of heterogeneous effects, in this paper we adopt an agnostic ML model to investigate treatment effect heterogeneity. In particular, we interpret the estimated effects stemming from our ML counterfactual empirical model by using the Sorted Effects method \citep{chernozhukov2018sorted, chernozhukov2020generic}. This method focuses on the tails of the estimated distribution of the firm-level treatment effect to identify the units that are most affected and those that are least affected by the treatment (whose characteristics are compared). 
We provide evidence that contrasting the estimated counterfactual outcomes with the outcomes predicted for the treatment scenario (and not directly with the observed outcomes under treatment) is crucial to correct the estimation error arising from the imperfect reconstruction of the unobservable counterfactual.   


Our paper is connected to the literature on the heterogeneous impact of the COVID-19 shock on trade. Using firm-level monthly data on Spanish trade in goods, \cite{DeLucio2020} find that  exports decreased more in countries that introduced strict policies to contain COVID-19 and for goods that are consumed
outside the household, particularly between March and May, showing how Spain's export performance during the pandemic depends on COVID-19-induced demand shocks in export markets and the characteristics of products. 
Using monthly bilateral product-level trade flows that cover three-quarters of world trade, \cite{berthou2022trade} also find that the impact of the COVID-19 shock on exports was particularly strong in the spring of 2020, and that demand shocks related to COVID-19 impacted exports directly (shocks in importing countries) but also indirectly (shocks in third countries). Using a sector-level gravity model, \cite{espitia2021pandemic} show that, during the COVID-19 crisis, sectors that tend to be relatively less internationally integrated suffered less from foreign shocks but were more vulnerable to domestic shocks. Using data on 
Chinese imports at the country-product level, also \cite{liu2021trade} show that the COVID-19 effects are heterogeneous, being weaker for medical goods and stronger for durable consumption goods.
All these papers base their identification strategy of the average COVID-19 effect on the cross-country differences in the implementation of lockdown measures over time and study treatment effect heterogeneity by focusing on subsamples or interacting the treatment variable with other possible determinants of heterogeneity. We share with these studies the ambition to estimate the causal impact of COVID-19 on trade and its possible heterogeneity. However, we use a different approach based on constructing a counterfactual using the predictive power of ML that, as explained above, recognises that all firms are directly or indirectly affected by this economy-wide shock and that it is very challenging to define ex-ante a variable summarising the (differential) exposure of firms to COVID-19. Moreover, we implement the heterogeneity analysis by using a classification analysis that safeguards against the risks of overfitting and multiple testing. Among the possible determinants of heterogeneity, we also consider a firm's diversification on the export and import side. Therefore, our study is also related to the international trade literature on the role of diversification in mediating the impact of adverse shocks \citep{kramarz2020volatility, grossman2021supply,lafrogne2022supply}.



The paper is structured as follows. Section \ref{methodology} details our empirical strategy. Section \ref{sec:data} presents the firm-level data,  variables employed in the analysis, and descriptive statistics. Section \ref{results} reports the estimation results, and Section \ref{conclusions} summarizes our findings and discusses the relevance and limitations of our analysis.

\section{Methodological framework}\label{methodology}

This section lays out our empirical approach to estimating the effect of an economy-wide shock on firms' survival probabilities in export markets and exploring its heterogeneity based on firms' observable attributes. 

As with any evaluation study, the primary identification task is to find a counterfactual outcome for the affected units, which is unobserved. In the presence of an economy-wide shock such as COVID-19, no subset of unaffected firms can likely be selected as a control group because the shock impacts, at least indirectly, all firms. Furthermore, even an identification strategy based on comparing individual firms subject to different treatment intensities appears infeasible due to the complex and ex-ante unknown paths through which firms are potentially exposed to the treatment. Though we study whether treatment effect heterogeneity depends, inter alia, on measures of the exposure to the COVID-19 shock,\footnote{These indexes are described in detail in section~\ref{sec:data}. They are based on firms' past export and import activities in different countries and on the time-varying strength of the virus and the stringency of the policies aimed at mitigating its spread.} 
the intensity of treatment might also depend on other firms' characteristics, such as the identity of suppliers and clients, the characteristics of the traded final product, among many others, that we cannot know in advance and whose interactions are \textit{a priori} unknown.


Therefore, as it is often practised when studying large-scale shocks' effects, we must employ the information on the pre-shock behavior to estimate the counterfactual behavior (in the hypothetical situation of the absence of the shock) during the actual shock. This process involves forecasting the future conduct of entities based on their historical behavior, an application perfectly suited to ML techniques, which are designed for such out-of-sample prediction tasks.

In line with the reasoning of \cite{varian2016causal}, and drawing parallels with the applications employed by \cite{cerqua2020local}, and \cite{fabra2020degrowth}, we harness the predictive strength of ML techniques. This allows us to construct a hypothetical scenario for firm-level outcomes during the shock period, using pre-shock data concerning firms' export behaviors and attributes. We aim to study whether a firm that was exporting in a given month during the pre-shock year $t_{s-1}$ (that is 2019 in our specific application) will export again during the same month of the year of the shock  $t_{s}$  (that is 2020 in our specific application). The empirical analysis we shall describe is carried out for each month separately to allow the effects of the explanatory variables (e.g., the hypothesized determinants of firm export status) to vary throughout the year.


Let us define the potential outcome under the scenario $d \in \{0,1\}$ for a firm $i$ at the time $t$ as $Y_{it}^d$. In this case, $d$ represents an indicator variable signifying the presence of the shock. More precisely, the outcome of our interest is the firm's export status, which is equal to one if a firm exporting at year $t-1$ in a given month is also exporting at year $t$ during the same month (otherwise it will be zero). Note that the regressors, $X_{it}^d$, may also be influenced by the presence or absence of the shock. 

The initial step in our analysis involves estimating the counterfactual outcome at $t_s$: $Y_{i,t_s}^0$. The model we utilize to derive this counterfactual (and the counterfactual itself) is referred to as the ``Shock Unaware Machine'' (SUM), a term acknowledging the ML techniques employed in constructing the counterfactual and the fact that no information about the shock is used in the analyses. In particular, we use the outcomes and covariates observed at $t_s-1$ and $t_s-2$ to reconstruct $Y_{t_s}^0$ based on the following assumptions (for ease of notation, we shall omit the subscript $i$ from this point forward):
\begin{enumerate}[(i)]
\item Neither covariates nor outcomes of $t_s-2$ and $t_s-1$ are affected by the shock:
\begin{equation}\label{assumption1}
Y_t=Y_t^0=Y_t^1, \quad X_t=X_t^0=X_t^1 \quad \text{for }\quad t=(t_s-1, t_s-2) 
\end{equation}
\item Define $Y_t^0=f_t^{0}(X_{t-1}^0)+u_t^0$, where $f_t^{0}(\cdot)$ is a generic model or function representing the relationship between explanatory variables and the outcome in the absence of the shock such that $\mathbf{E}[Y_t^{0}|X_{t-1}^{0}]=f_t^{0}(X_{t-1}^{0})$. Under (i), for $t=t_s-1$ we have that $Y_{t_s-1}=f_{t_s-1}^0(X_{t_s-2})+u_{t_s-1}^0$ such that $\mathbf{E}[Y_{t_s-1}|X_{t_s-2}]=f_{t_s-1}^0(X_{t_s-2})$. The second assumption states that the function $f_t^0$ does not depend on $t$, i.e., it is stable over the two considered years: 
\begin{equation}\label{assumption2}
f_{t_s-1}^0=f_{t_s}^0=f^0.
    \end{equation}
\end{enumerate}
  
Therefore, under the above assumptions, we can write $Y_{t_s}^0=f^0(X_{t_s-1})+u_{t_s}^0$, such that $\mathbf{E}[Y_{t_s}^0|X_{t_s-1}]=f^0(X_{t_s-1})$, and we can use data on $t_s-2$ and $t_s-1$ to estimate $Y_{t_s-1}^0=f^0(X_{t_s-2})+u_{t_s-1}^0$ and retrieve $\hat{f}^0$. By applying this invariant estimated function to the covariates of $t_s-1$, we can obtain the predictions for the counterfactual (without the shock) outcome in $t_s$:
\begin{equation}\label{counterfactual}
  \hat{Y}_{t_s}^0=\hat{f}^0(X_{t_s-1})=Y_{t_s}^0-
  \overbrace{
  \mathcal{E}^0_{t_s}(X_{t_s-1})
  }^{\textit{Prediction error}}-
  \overbrace{
  u^0_{t_s}
  }^{\textit{Orthogonal error}}.
\end{equation}

In general, the estimated counterfactual outcome at time $t_s$, denoted by $\hat{Y}_{t_s}^0$, is not a perfect estimate of $Y_{t_s}^0$. This imprecision arises due to two aspects: firstly, $\hat{f}^0$ is not an exact estimate of $f^0$ thereby generating a prediction error, which we represent as $\mathcal{E}^0_{t_s}(X_{t_s-1})={f}^0(X_{t_s-1})-\hat{f}^0(X_{t_s-1})$ in the above equation. Secondly, there exist other determinants of the outcome that are orthogonal to the covariates, represented as $u^0_{t_s}$ in the formula. The imprecision emanating from the estimation of $f^0$, which could differ depending on a firm's characteristics $X_{t_s-1}$, can be mitigated by exploring various ML techniques and employing the one that delivers the best out-of-sample performance.\footnote{In order to simplify the notation, from now on, we will denote $\mathcal{E}^d_{t_s}(X_{t_s-1})$ as $\mathcal{E}^d_{t_s}$.} 

In the application we present in this paper, we rely on the ``$K$-fold'' cross-validation method (with $K=5$) to discriminate between the considered ML techniques. We randomly divide the set of exporters observed in $t_s-2 = 2018$  (considering the exporting success during the same month in $t_s-1 = 2019$ as the outcome) into 5 equally sized groups and obtain the predictions for the firms belonging to a group by estimating $Y_{2019}=f^0(X_{2018})+u_{2019}^0$ with different ML models on the firms belonging to the other groups. Then we compute the accuracy of the different models for each month and choose the model with the best average performance across months. Notice that this comparison is entirely based on the pre-pandemic accuracy of the ML models by comparing the predictions $\hat{Y}_{2019}$ with the observed ${Y}_{2019}$, not on its merits in predicting the firms' outcomes in 2020. Finally, we obtain the $\hat{Y}_{2020}^0$ by estimating $Y_{2019}=f^0(X_{2018})+u_{2019}^0$ on the entire set of $2018$ units (also in this case month by month) and, as shown in~\eqref{counterfactual}, applying the estimated function $\hat{f}^0$ to the set of $2019$ units. 
Given that during the first three months of 2020 Colombia was in practice not exposed to COVID-19 (and therefore $Y_{2020}={Y}_{2020}^0$), if assumption~\eqref{assumption2} holds, we expect that in those months the accuracy of the predictions $\hat{Y}_{2019}$ obtained in the cross-validation step for 2019 will be very similar to the accuracy of $\hat{Y}_{2020}^0$ for 2020.

Following \cite{cerqua2020local} and \cite{fabra2020degrowth}, we define as an estimator of the individual-specific shock effect $\alpha$ the simple comparison of the observed outcome under the shock in $t_s$ with the estimated counterfactual outcome. This comparison is represented as:  
\begin{equation}\label{double_hat_ind}
\hat{\hat{\alpha}}= Y_{t_s}-\hat{Y}_{t_s}^0. 
\end{equation}
This provides the full distribution of treatment effects. 

Starting from Eq.~\eqref{double_hat_ind}, by taking the expected value of the individual treatment effect $\hat{\hat{\alpha}}$ for those units with $X_{t_s-1}=x_{t_s-1}$, we can define the following estimator of the conditional average treatment effect (CATE; the average effect for those units with $X_{2019}=x_{2019}$) as
\begin{equation}\small
\label{double_hat_cate}
  \begin{split}
    \mathbf{E}[\hat{\hat{\alpha}}|X_{t_s-1}=x_{t_s-1}] & =  \mathbf{E}[(Y_{t_s}-Y_{t_s}^0)-\mathcal{E}^0_{t_s}-u^0_{t_s}|X_{t_s-1}=x_{t_s-1}] = \\
    &= \underbrace{\Delta(X_{t_s-1}=x_{t_s-1})}_{\textit{CATE}}-\mathbf{E}[\mathcal{E}^0_{t_s}|X_{t_s-1}=x_{t_s-1}]-\underbrace{\mathbf{E}[u^0_{t_s}|X_{t_s-1}=x_{t_s-1}]}_{\textit{=0 by assumption}}, \\
    & \text{ where, } \\
    &\Delta(X_{t_s-1}=x_{t_s-1})=\mathbf{E}[Y_{t_s}-Y_{t_s}^0|X_{t_s-1}=x_{t_s-1}].
  \end{split}
\end{equation}
Therefore $\mathbf{E}[\hat{\hat{\alpha}}]$ identifies the unconditional average treatment effect, $\mathbf{E}[\Delta(X_{t_s-1})]=\Delta$, if on average the prediction error is zero: $\mathbf{E}[\mathcal{E}^0_{t_s}]=0$. The conditional average treatment effect, $\Delta(X_{t_s-1}=x_{t_s-1})$, is identified by $\mathbf{E}[\hat{\hat{\alpha}}|X_{t_s-1}=x_{t_s-1}]$ if on average the prediction error is zero in the relevant sub-sample: $\mathbf{E}[\mathcal{E}^0_{t_s}|X_{t_s-1}=x_{t_s-1}]=0.$ 

Now let us decompose the outcome observed in $t_s$ in the presence of the shock, $Y_{t_s}^1$, in a generic model or function $f^1(X_{t_s-1}^1)$, which represents the relationship between explanatory variables and the outcome during the shock, and other determinants of the outcome, $u^1_{t_s}$, that are orthogonal to the covariates
\begin{equation}\label{eqn:Y^1_t_s}
      Y_{t_s}^1=f^1(X_{t_s-1}^1)+u^1_{t_s}, \quad \textit{s.t. } \mathbf{E}[Y^1_{t_s}|X^1_{t_s-1}]=f^1(X^1_{t_s-1}).
\end{equation}
Given that $Y_{t_s}^1=Y_{t_s}$ and $X_{t_s-1}^1=X_{t_s-1}$, then
\begin{equation}\label{eqn:Y_t_s}
  Y_{t_s}=f^1(X_{t_s-1})+u^1_{t_s}, \quad \textit{s.t. } \mathbf{E}[Y_{t_s}|X_{t_s-1}]=f^1(X_{t_s-1}).
\end{equation} 

At this point, we can define an alternative estimator of the individual-specific shock effect $\alpha$ as the comparison of the predicted outcome under the shock in $t_s$ with the estimated counterfactual outcome for a given firm:
\begin{equation}\label{hat_ind}
 \hat{\alpha} = \hat{Y}_{t_s}-\hat{Y}^0_{t_s},
\end{equation}
where $\hat{Y}_{t_s}=\hat{f}^1(X_{t_s-1})=Y_{t_s}-\mathcal{E}^1_{t_s}-u^1_{t_s}$. We call ``Shock Aware Machine'' (SAM) the model that we use to predict ${Y}_{t_s}$ (and the predictions $\hat{Y}_{t_s}$ themselves). The term  ``Shock Aware'' is employed because this model leverages information from the observed shock scenario. Importantly, the predictions from SAM are generated in a metric that allows them to be directly compared with the estimated outcomes (that do not account for the shock), which are produced by the SUM.\footnote{Notice that with $\hat{\hat{\alpha}}$, we are comparing a probability (counterfactual) with a binary value (observed outcome), while with $\hat{\alpha}$, we are comparing two estimated probabilities.} 

In our application, the SAM expresses the outcome in 2020 of exporters operating the foreign market in $2019$ as a function of their characteristics in $2019$ and the information about governments' shock-related stringency measures all over the world coming from \cite{hale2020variation}.\footnote{
See subsection~\ref{sec:cdata}. We do not introduce these variables explicitly as an argument of $f^1()$ to simplify notation.}  
Similarly to the procedure followed to select the best-performing SUM, we rely on a 5-fold cross-validation strategy to obtain a $2020$ prediction for each firm that exported in $2019$. We randomly group the $2019$ exporters into five equally sized subsets and we predict the $2020$ outcomes of the firms contained in one subset by using the information of firms contained in the remaining four subsets. In other words, we train the models on a random 80\% of the data and test them on the remaining 20\% and we repeat the process five times for each different 20\% subset, thus obtaining a $2020$ prediction for each $2019$ exporter.

Starting from Eq.~\eqref{hat_ind}, by taking the expected value of the individual treatment effect $\hat{\alpha}$ for those units with $X_{t_s-1}=x_{t_s-1}$, we can define the following alternative estimator of the conditional average treatment effect (for those units with $X_{t_s-1}=x_{t_s-1}$)
\begin{equation}\small
\label{hat_cate}
 \begin{split}
  \mathbf{E}[\hat{\alpha}|X_{t_s-1}=x_{t_s-1}] 
  = & \mathbf{E}[(Y_{t_s}-Y^0_{t_s})-(\mathcal{E}^1_{t_s}-\mathcal{E}^0_{t_s})-(u^1_{t_s}-u^0_{t_s})|X_{t_s-1}=x_{t_s-1}] \\ 
  = & \underbrace{\Delta(X_{t_s-1}=x_{t_s-1})}_{\textit{CATE}} -\mathbf{E}[\underbrace{(\mathcal{E}_{t_s}^1-\mathcal{E}_{t_s}^0)}_{\Delta\mathcal{E}}|X_{t_s-1}=x_{t_s-1}] - \\
  & \mathbf{E}[u_{t_s}^1-u_{t_s}^0|X_{t_s-1}=x_{t_s-1}].
 \end{split}
\end{equation}
Therefore, $\mathbf{E}[\hat{\alpha}]$ identifies the unconditional average treatment effect, $\mathbf{E}[\Delta(X_{t_s-1})]=\Delta$, if, on average, the difference in prediction errors is zero: $\mathbf{E}[\Delta\mathcal{E}]=0$. 
The conditional average treatment effect, $\Delta(X_{t_s-1}=x_{t_s-1})$, is identified by $\mathbf{E}[\hat{\alpha}|X_{t_s-1}=x_{t_s-1}]$ if on average the difference in prediction errors is zero in the relevant sub-sample: 
$\mathbf{E}[\Delta\mathcal{E}|X_{t_s-1}=x_{t_s-1}]=0$.

Given the definitions of SUM and SAM, to simplify the reasoning in the following, we refer to Eqs.~\eqref{double_hat_ind} and~\eqref{hat_ind}, respectively as
\begin{align}
\label{Y_SUM_ind} 
\hat{\hat{\alpha}} = & Y -\hat{Y}_{SUM} = Y-SUM. \\
\label{SAM_SUM_ind}
\hat{\alpha} = & \hat{Y}_{SAM}-\hat{Y}_{SUM} = SAM-SUM.
\end{align}
The assumptions behind these identification results are not directly testable as they are expressed in terms of the expected values of the prediction error $\mathcal{E}^0_{t_s}$ that is a function of the unobservable counterfactual $Y^0_{t_s}$. Table~\ref{multiscen} distinguishes the five different scenarios concerning the values of $\mathcal{E}^0_{t_s}$ and $\mathcal{E}^1_{t_s}$ that are relevant in determining whether applying the statistic $\mathbf{T}$ to $Y-SUM$ and $SAM-SUM$ is able to recover the corresponding treatment effect estimand (e.g., whether averaging the estimated individual treatment effects would recover the average treatment effect).  
\begin{table}[H]
\begin{center}
\caption{Identification of generic functions of the individual treatment effects, $\mathbf{T}$, according to the corresponding value taken by the prediction errors}
\label{multiscen}
\begin{tabular}{l cc }
\toprule
  & \makecell{$\mathbf{T}(SAM-SUM)$} & \makecell{$\mathbf{T}(Y-SUM)$} \\
  \midrule
    \multirow{1}{*}{$\mathbf{T}[\mathcal{E}^1_{t_s}]\neq 0$ and $\mathbf{T}[\mathcal{E}^0_{t_s}]=0$}& $\times$ &$\checkmark$\\ 
    \multirow{1}{*}{$\mathbf{T}[\mathcal{E}^1_{t_s}]=\mathbf{T}[\mathcal{E}^0_{t_s}]=0$}&$\checkmark$&$\checkmark$\\ 
    \multirow{1}{*}{$\mathbf{T}[\mathcal{E}^1_{t_s}]=0$ and $\mathbf{T}[\mathcal{E}^0_{t_s}]\neq 0$}& $\times$ & $\times$\\ 
    \multirow{1}{*}{$\mathbf{T}[\mathcal{E}^1_{t_s}]=\mathbf{T}[\mathcal{E}^0_{t_s}]\neq 0$}&$\checkmark$ & $\times$\\ 
    \multirow{1}{*}{$\mathbf{T}[\mathcal{E}^1_{t_s}] \neq \mathbf{T}[\mathcal{E}^0_{t_s}]\neq 0$}& $\times$ & $\times$ \\
\bottomrule
\end{tabular}
\end{center}
\end{table}

The estimators based on $Y$-$SUM$ identify the population parameters when $\mathbf{T}[\mathcal{E}^0_{2020}]=0$. The estimators based on $SAM$-$SUM$ are unbiased whenever $\mathbf{T}[\mathcal{E}^1_{2020}]=\mathbf{T}[\mathcal{E}^0_{2020}]$. Under the assumption that the strength of the COVID-19 effect on export propensity was at most very limited during the first quarter of 2020, we will use the out-of-sample prediction errors for the first quarter of 2020 as a proxy for the unobservable behavior of $\mathcal{E}^0_{2020}$ in the following months. Moreover, as explained in detail in section \ref{results2}, the distribution of the estimated treatment effects during the first quarter will be used to check the credibility of the above assumptions for the set of all 2019 exporters and for different subsets of 2019 exporters defined according to their characteristics $X_{2019}$ or to their position in the distribution of such effects.

As a final step, we perform the heterogeneity analysis by adapting the Sorted Partial Effect (SPE) method introduced in \cite{chernozhukov2018sorted}. Formally, the SPEs are defined as percentiles of the Treatment Effects (TE) and can supply a more detailed summary of the distribution of TE than the Average Treatment Effects (ATE), commonly employed in econometric analysis. The SPEs are defined as
\begin{equation}\label{SPE}
\alpha^{*}(u)=u^{t h}-\text { percentile of } \alpha. 
\end{equation}
In our setting, $\alpha^{*}(u)$ is a function of $X_{t_s-1}$ defined over its distribution in the population of $t_s-1$ exporters. 

The SPEs are used to do a classification analysis (CA) that allocates the $t_s-1$ exporters into two groups, the most and the least affected by the shock, according to whether their $\alpha$ are lower than $\alpha^{*}(25)$ or greater than $\alpha^{*}(75)$, respectively. Notice that, since the shock effect is negative, we have defined as the most (negative) affected units those whose $\alpha$ lie in the left tail of the sorted distribution of treatment effects. Finally, to study the determinants of treatment effect heterogeneity, we focus on the difference in means ($CADiff$) of the $X_{t_s-1}$ across the most and least affected groups. In the estimation, we use sample analogues of $\alpha^{*}(u)$ and $CADiff$. We calculate standard errors of $\alpha^{*}(u)$ and $CADiff$ by bootstrapping the entire estimation process, starting from the initial $\alpha$ estimation step. 

The application of the SPE technique presents several advantages in our setting. First, the estimated $\alpha^{*}(u)$\textit{s} provide a summary of the distribution of the estimated treatment effects and, therefore, of treatment effect heterogeneity. Second, the CA identifies the subgroup of the population that is more affected by the treatment and the $CADiff$ studies how the heterogeneity of the treatment effect depends on observables without imposing (additional) functional form assumptions. Third, the $CADiff$ step provides p-value adjustments to account for the joint testing of all the covariates that are considered to detect if observables are associated with treatment effect heterogeneity. In other words, the main idea is to test the null hypothesis of no difference between the value of the covariates in the most and the least affected groups by also taking into account that we conduct simultaneous inference on multiple variables.\footnote{See Online Appendix \ref{app:jpvals}.} 

Finally, notice that our approach in estimating the individual treatment effect and in performing the heterogeneity analysis is similar to the generic ML technique presented in \cite{chernozhukov2020generic}, which is adapted to a situation in which there is no available (contemporaneous) control group (i.e., it is difficult to identify ex-ante firms that are not affected by the shock).\footnote{See Online Appendix \ref{app:comparison}.}

\section{Data}\label{sec:data}

This study focuses on the social and economic disruption caused by the COVID-19 pandemic and its effect on Colombian exporters. This global health crisis served as a notable example of a large-scale economic shock that profoundly impacted global trade, with the dynamics of exporters in Colombia being significantly affected. Applying our machine learning strategy to data collected from Colombian exporting firms during this period can yield substantial insights into how such entities adapt and persist in the face of such extensive disruption. This provides an understanding of market resilience and firm survival dynamics in the context of global trade shocks. By grounding our research in a tangible case study, we maintain its relevance to the specific scenario while preserving its potential for broader applications.

We use monthly export transaction data reported at the Colombian Customs Office (Direcci\'on de Impuestos y Aduanas Nacionales, DIAN) for 2018, 2019, and 2020. For each transaction, we consider the exporter ID as the firm identifier; the date; a 10-digit Harmonized System code (HS) characterizing the product; the product origin within Colombia (department level); the means of transportation of the shipment; the country of destination; and, the free on board value of the export transaction in US dollars. This data set also contains information about the value and origin country from which a given exporter imports. We remove all transactions related to re-exports of products elaborated in other countries. As a result, we ended up with 386,132 customs reports in 2018 (7741 firms), 402,140 in 2019 (7831 firms), and 365,626 in 2020 (7518 firms).

\subsection{Control Variables}\label{sec:cdata}

We classify products at the six-digit level of the HS code. We consider different features of exporters according to their monthly exports: the total export (and import) value, the number of products ($NP$), the number of export destinations ($ND$), the number of import origin countries ($NO$), the Herfindahl-Hirschman indexes at the product level ($HH_p$) and the destination level ($HH_d$), and a set of dummies for the destinations and origin countries and continents. We create a set of dummies according to the Colombian department from which the product comes, a set of dummies for the means of transportation used, and a set of dummies classifying the product HS-chapter and HS-section. Moreover, we build two sets of dummy variables indicating whether a firm has experience exporting in specific destinations and product sectors. We also account for the accumulated exporting (importing) experience by summing up the total value exported (imported) during the last twelve months. Furthermore, we create four \textit{size} dummies classifying firms according to the quartiles of the firm-level distribution of the total monthly log-value of exports.

To measure the COVID-19 demand and supply shock, we use the information on government contention measures coming from \cite{hale2020variation}, which consists of four indexes (ranging from 0 to 100) representing the strength of the measures taken by countries to contain the COVID-19 outbreak. The authors provide an economic index summarizing economic policies ($E$), a health index summarizing health policies ($H$), a government index describing the strictness of ‘lockdown style’ policies ($G$), and an overall government response index called stringency index ($S$). The value of these indexes ranges from 0 to 100.\footnote{These indexes are released daily. We average this information at the monthly level.} We build two variables at the firm level for each of the four indexes, one at the export and one at the import side, by taking a weighted average of the country-level scores according to the proportion of the total monthly value of exports (imports) that a firm ships (source) in each country in 2019. We call these firm-level indexes for a firm $i$  ``$\text{Containment Index}_{i,j,z}$'', with $j=\{E,H,G,S\}$ and $z=\{\text{Imp},\text{Exp}\}$.\footnote{The value of the Containment Stringency Index Import for firms that are not importing corresponds to the value of the Containment Stringency Index for Colombia (as firms are sourcing all their inputs domestically).} 

Our final data set is composed of 1,975 covariates. They are presented in detail in Table \ref{tb:variables} of Online Appendix \ref{app:data}.

\section{Results}\label{results}

\subsection{Selection of the machine learning algorithm} \label{results1}

We evaluate and contrast the outcomes of several ML techniques against a benchmark logistic regression, aiming to identify the model with superior prediction performance. The out-of-sample predictive efficacy of our empirical models is crucial, given our goal to reconstruct an unobserved counterfactual. The complexity of this task arises from its high dimensionality and complex interdependencies between firms and products from various sectors and export destinations. While an approach focusing on in-sample prediction accuracy might overfit, ML techniques optimally balance the bias-variance trade-off for out-of-sample predictions.\footnote{Hyperparameter tuning through cross-validation or other theory-driven methods is often critical in order to avoid overfitting.}


We examine four distinct models: Logit, Logit-LASSO, Logit-Ridge, and Random Forest (RF). The traditional choice for binary dependent variables, Logit, serves as our baseline. Even though literature often shows ML techniques outperforming traditional models with numerous predictors, we have included Logit results for comparison. The main idea of Logit-LASSO is to mitigate overfitting by introducing a penalty
term in the Logit log-likelihood function that forces the parameters associated with the less
relevant predictors to be exactly zero. 
On the other hand, Logit-Ridge reduces the coefficients of less significant predictors without eliminating any of them, proving especially useful when many variables play an important role. The main idea behind Random Forest is the wisdom of crowds because it combines the predictions of many uncorrelated models (the trees) obtained by randomly re-sampling observations and explanatory variables.\footnote{Note that it is important to optimize (tune) the hyper-parameters of Logit-Ridge and Random Forest for an accurate predicting exercise. The hyperparameter to optimize in Logit-Ridge and Logit-LASSO is $\lambda$, which controls the impact of the penalty or shrinkage on parameters (when $\lambda=0$ we are in a Logit scenario when $lambda$ increases the penalty impact grows). We find the optimal hyper-parameter for Logit-Ridge and Logit-LASSO by choosing the $\lambda$ that minimizes the mean cross-validated error. One of the main RF parameters is the number of random trees used. Because of the RF design, it is very difficult to have over-fitted predictions when using this model. Therefore, we set the number of trees to a large enough number (500). We find this number is large enough after repeating the same Random Forest model with a different number of random trees. Even with less than 500 random trees, the error rate of the model remains unchanged.} For Logit, Logit-Ridge, and Logit-LASSO models we include interactions between the \textit{size} of the company and some of the main product characteristics, \textit{industry}, \textit{sector}, \textit{means of transportation} as well as with \textit{destination country} dummies. Notice that Random Forest uses the variables sequentially and, therefore, with a large enough number of trees, it is not necessary to explicitly introduce interactions as explanatory variables, i.e., the model automatically takes into account the interactions that are useful to accurately predict the outcome.\footnote{For more information about all the features included to build the SUM and SAM see Table \ref{tb:variables} in Online Appendix \ref{app:data}.} The prediction analysis is repeated for all months between January-December 2020.



Table~\ref{tb:GoF} shows the accuracy of the model's predictions through two widely-used classification performance metrics: Area Under the Receiver Operating Curve (AUC) and Root-Mean-Square Error (RMSE). The AUC achieves a value of 0.5 for random predictions and 1 when outcomes are classified without error. Meanwhile, RMSE's best score is 0, indicating optimal accuracy with no fixed upper bound. 

The table's upper part displays the accuracy of predictions for the probability of exporting in 2019 based on 2018 exporter data, serving as an out-of-sample performance benchmark in a pre-COVID-19 context using cross-validation. Here, the Logit-LASSO and RF models arise as top performers. The table's middle section
also shows the accuracy of models estimated using the exporters’ characteristics in 2018 to explain their observed outcomes in 2019; however, these models are now tested using the set of exporters of 2019 and their observed outcomes in 2020. If the functions $f_t^0$ representing the relationship between explanatory variables and the outcome in the absence of the pandemic are sufficiently similar for the pre-pandemic year and 2020 ($f_{2019}^0$ and $f_{2020}^0$, respectively), we expect that the accuracy of $\hat{f}_{2019}^0$ in the first three months of 2020 (when arguably no relevant COVID-19 effect is in place in Colombia) to be similar to that one which is observed during the same months of 2019. Indeed, during January, February, and March, the accuracy of Logit-LASSO and RF remains unchanged, as expected, compared to the accuracy obtained in the upper part of the table. However, a decline in accuracy appears post-April in the middle part of the tables, demonstrating the challenges of a model not trained on COVID-19 data, predicting in a COVID-19-impacted environment.


Models in the bottom part of Table~\ref{tb:GoF} are trained and tested with the universe of exporters in 2019 and their observed outcomes in 2020. Using these models, we construct the SAM predictions. The accuracy of the predictions is very similar to the one obtained with the SUM for 2019 and for the first three months of 2020. Our analysis is crucial to reach accurate predictions because the unbiasedness of our treatment effects estimators depends on the quality of the (counterfactual) prediction accuracy. Both the SUM and the SAM show acceptable levels of accuracy when predictions are made with Logit-LASSO and Random Forest.
\begin{table}[H]
\begin{center}
\caption{Goodness of Fit for SUM and SAM in 2018/19 and 2019/20}
\label{tb:GoF}
\resizebox{0.8\textwidth}{!}{\renewcommand{\arraystretch}{0.9}
\begin{tabular}{l cccc c cccc} 
\toprule
& \multicolumn{4}{c}{AUC} & \multicolumn{5}{c}{RMSE} \\ 
\cline{2-5} \cline{7-10}
 & Logit-LASSO & Logit-Ridge & Random Forest & Logit && Logit-LASSO & Logit-Ridge & Random Forest & Logit \\ 
 \midrule
& \multicolumn{9}{c}{Goodness of Fit for SUM in 2018/19} \\ 
\midrule
  Jan & 0.73 & 0.53 & 0.73 & 0.59 && 0.40 & 0.45 & 0.41 & 0.64 \\ 
  Feb & 0.70 & 0.50 & 0.71 & 0.58 && 0.41 & 0.45 & 0.41 & 0.64 \\ 
  Mar & 0.70 & 0.56 & 0.71 & 0.57 && 0.41 & 0.44 & 0.41 & 0.65 \\ 
  Apr & 0.73 & 0.59 & 0.73 & 0.60 && 0.40 & 0.43 & 0.40 & 0.63 \\ 
  May & 0.72 & 0.52 & 0.71 & 0.59 && 0.40 & 0.44 & 0.41 & 0.64 \\ 
  Jun & 0.71 & 0.50 & 0.72 & 0.59 && 0.40 & 0.45 & 0.41 & 0.64 \\ 
  Jul & 0.73 & 0.50 & 0.73 & 0.55 && 0.40 & 0.45 & 0.40 & 0.66 \\ 
  Aug & 0.70 & 0.51 & 0.72 & 0.58 && 0.41 & 0.45 & 0.40 & 0.64 \\ 
  Sep & 0.72 & 0.50 & 0.71 & 0.58 && 0.41 & 0.45 & 0.40 & 0.64 \\ 
  Oct & 0.73 & 0.58 & 0.74 & 0.58 && 0.40 & 0.44 & 0.41 & 0.64 \\ 
  Nov & 0.71 & 0.51 & 0.72 & 0.57 && 0.41 & 0.45 & 0.41 & 0.64 \\ 
  Dec & 0.70 & 0.50 & 0.71 & 0.58 && 0.41 & 0.45 & 0.41 & 0.64 \\
\midrule
& \multicolumn{9}{c}{Goodness of Fit for SUM in 2019/20} \\ 
\midrule
  Jan & 0.72 & 0.53 & 0.72 & 0.49 && 0.41 & 0.45 & 0.41 & 0.75 \\
  Feb & 0.69 & 0.50 & 0.69 & 0.56 && 0.41 & 0.45 & 0.42 & 0.64 \\
  Mar & 0.72 & 0.54 & 0.73 & 0.59 && 0.40 & 0.44 & 0.41 & 0.63 \\
  Apr & 0.67 & 0.56 & 0.66 & 0.51 && 0.48 & 0.50 & 0.49 & 0.70 \\
  May & 0.69 & 0.51 & 0.69 & 0.60 && 0.46 & 0.48 & 0.46 & 0.63 \\
  Jun & 0.68 & 0.50 & 0.68 & 0.59 && 0.43 & 0.47 & 0.44 & 0.63 \\
  Jul & 0.70 & 0.50 & 0.69 & 0.59 && 0.42 & 0.46 & 0.43 & 0.63 \\
  Aug & 0.68 & 0.51 & 0.69 & 0.58 && 0.42 & 0.45 & 0.43 & 0.63 \\
  Sep & 0.69 & 0.50 & 0.70 & 0.59 && 0.42 & 0.45 & 0.42 & 0.63 \\ 
  Oct & 0.71 & 0.59 & 0.70 & 0.60 && 0.42 & 0.45 & 0.43 & 0.63 \\ 
  Nov & 0.71 & 0.51 & 0.71 & 0.59 && 0.41 & 0.45 & 0.41 & 0.63 \\
  Dec & 0.69 & 0.50 & 0.69 & 0.58 && 0.42 & 0.46 & 0.42 & 0.63 \\  
\midrule
& \multicolumn{9}{c}{Goodness of Fit for SAM in 2019/20} \\ 
\midrule
   Jan & 0.73 & 0.58 & 0.74 & 0.50 && 0.41 & 0.45 & 0.41 & 0.71 \\
   Feb & 0.70 & 0.50 & 0.70 & 0.49 && 0.41 & 0.46 & 0.42 & 0.70 \\
   Mar & 0.73 & 0.50 & 0.73 & 0.50 && 0.40 & 0.46 & 0.40 & 0.71 \\
   Apr & 0.74 & 0.66 & 0.73 & 0.52 && 0.42 & 0.47 & 0.42 & 0.69 \\
   May & 0.76 & 0.74 & 0.77 & 0.50 && 0.41 & 0.46 & 0.41 & 0.71 \\
   Jun & 0.73 & 0.69 & 0.73 & 0.48 && 0.42 & 0.46 & 0.42 & 0.72 \\
   Jul & 0.73 & 0.63 & 0.72 & 0.51 && 0.41 & 0.45 & 0.42 & 0.69 \\
   Aug & 0.72 & 0.50 & 0.72 & 0.53 && 0.41 & 0.46 & 0.42 & 0.69 \\
   Sep & 0.71 & 0.50 & 0.70 & 0.55 && 0.42 & 0.47 & 0.42 & 0.67 \\
   Oct & 0.72 & 0.50 & 0.71 & 0.52 && 0.42 & 0.46 & 0.42 & 0.70 \\
   Nov & 0.72 & 0.52 & 0.72 & 0.49 && 0.41 & 0.45 & 0.41 & 0.71 \\
   Dec & 0.71 & 0.51 & 0.70 & 0.51 && 0.41 & 0.45 & 0.42 & 0.70 \\
\bottomrule
\end{tabular}%
}
\end{center}
\end{table}

\subsection{Evaluation of the COVID-19 effect} 
\label{results2}
\begin{figure}[h]
\centering
\includegraphics[width=0.7\linewidth]{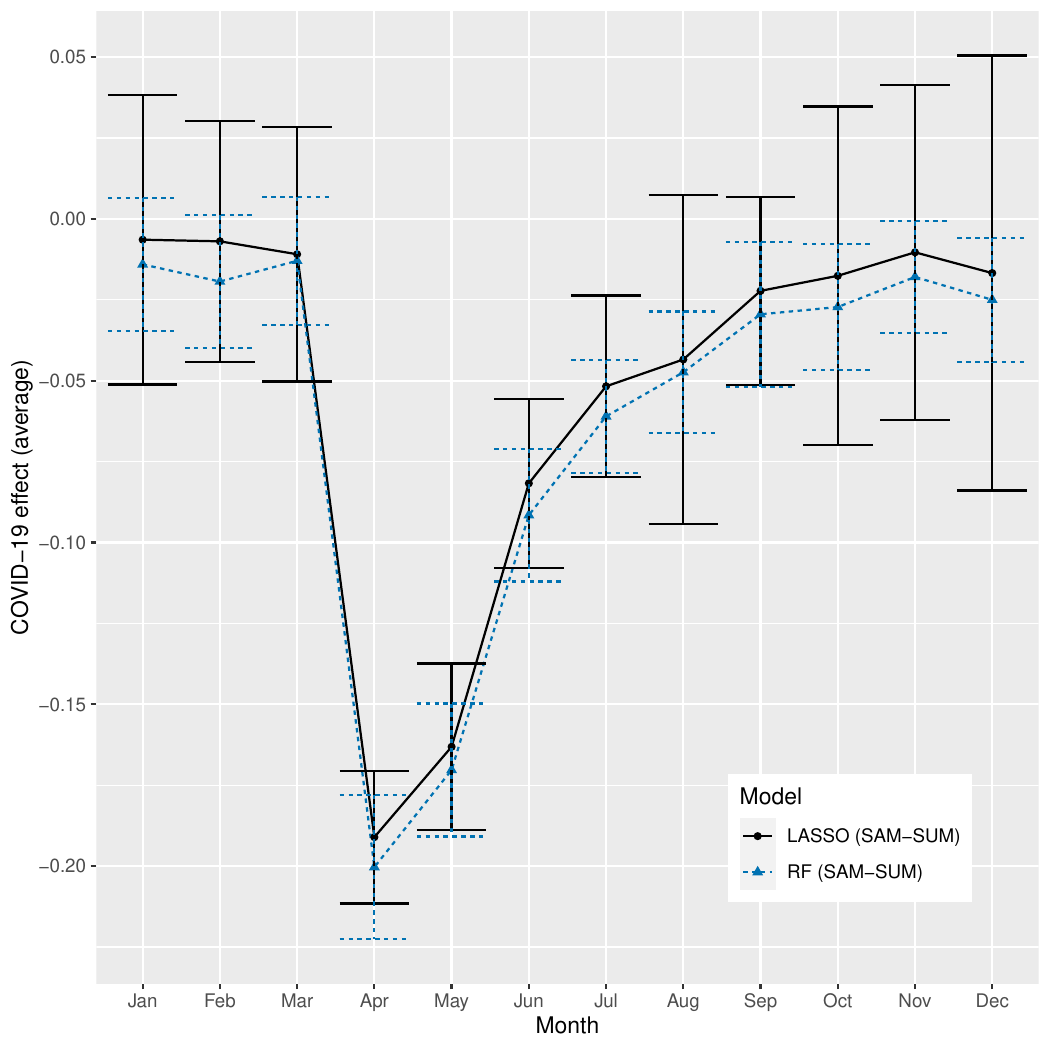}
\caption{Average Individual Treatment Effect, by months, comparing Logit-LASSO and RF. Standard errors were obtained with 100 bootstrap replications. Confidence intervals for a $5\%$ significance level.}
\label{fig:TE_months}
\end{figure}

Both Logit-LASSO and Random Forest reach high accuracy levels in the export status prediction. As explained in section \ref{methodology}, the predicted probabilities are used to estimate the average monthly effect of the COVID-19 shock as the monthly average of $\hat{\alpha}$ (the difference between the firm-level probabilities of success predicted by the SUM and the SAM.). They are presented in Figure~\ref{fig:TE_months}. 

Given the presumption that firms suffered a negligible COVID-19 shock impact during the initial three months of 2020, the treatment effect estimates for this period can be viewed as a placebo test, reminiscent of the in-time placebo test routinely employed in Synthetic Control Methods \citep{abadie2015comparative}. Detecting a significant COVID-19 effect in the months preceding the actual economic shock would suggest that our model mechanically estimates a COVID-19 effect even in the absence of the stated shock. We conduct these placebo studies also conditioning on exogenous firms' characteristics observed in 2019 by estimating COVID-19 effects for selected subsamples of firms according to such characteristics. We interpret these additional placebo studies as a robustness check on our results on treatment effect heterogeneity.

As shown in Figure~\ref{fig:TE_months}, the probabilities obtained from the SUM and the SAM are almost identical on average for January, February, and March. This result is reassuring since only from March 25, 2020, the Colombian government implemented a complete and mandatory lockdown. More in general, we can conclude that our identification strategy is not mechanically recovering COVID-19 effects for a period with low incidence in Colombia and in the rest of the world. We find that the peak of the COVID-19 effect is in April 2020, when we estimate an average difference between the predicted probabilities of exporting of nearly 20 percentage points. In the following months, the estimated average effect declines with time.

The results indicate that both Logit-LASSO and RF models yield comparable performances. Given their good performance and considering that Logit models are frequently used in similar contexts, we opt for Logit-LASSO. It aligns with the conventional approaches and offers greater interpretability as an extension of the traditional model.\footnote{Non-reported results using RF are equivalent and available upon request.} 

\begin{figure}[h]
\centering
\includegraphics[width=0.6\linewidth]{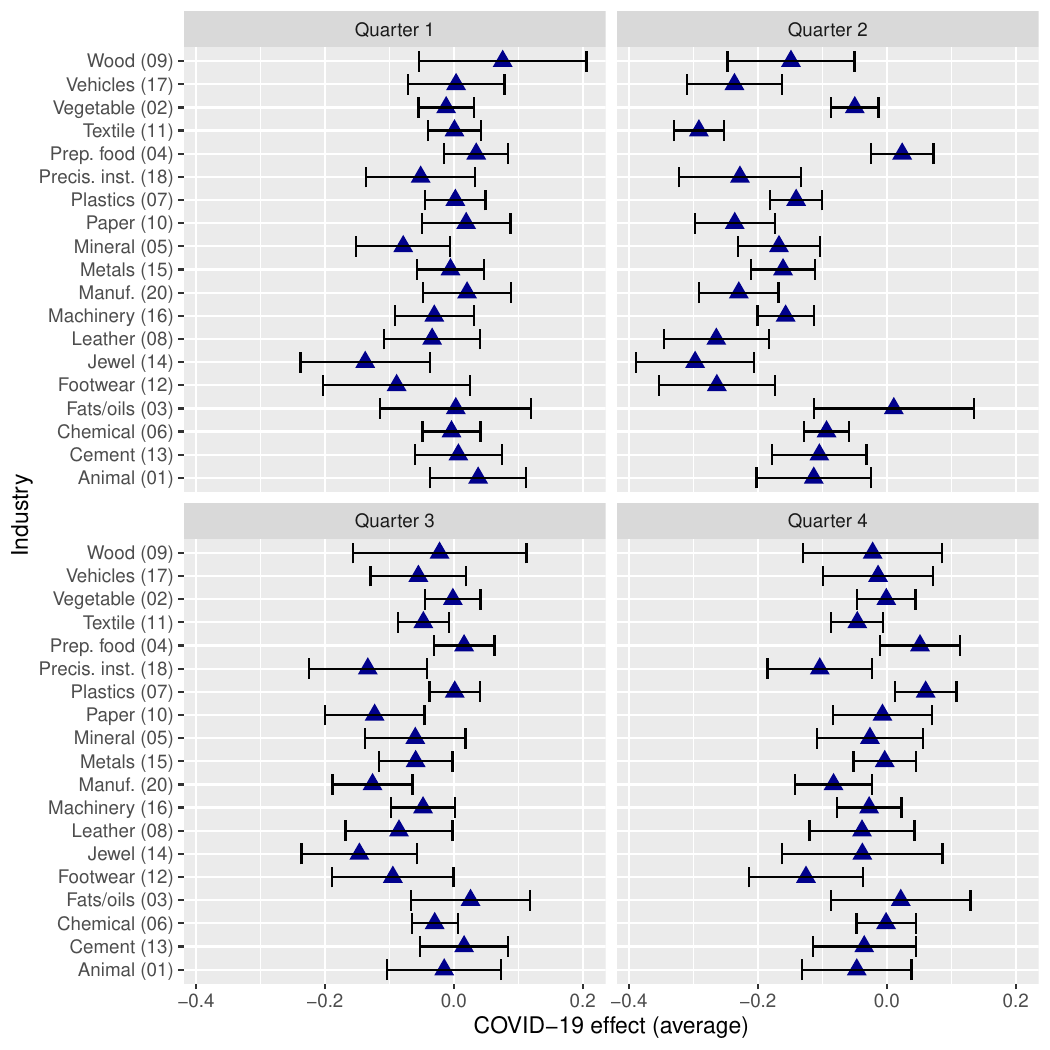}
\caption{The quarterly mean difference in the predicted probability of success (SAM vs. SUM) by industry, using the Logit-LASSO predictions. Standard errors were obtained with 100 bootstrap replications. Confidence intervals for a $5\%$ significance level.}
\label{fig:diff_predictions_industry}
\end{figure}

Figure \ref{fig:diff_predictions_industry} shows evidence of substantial variations in the quarterly estimated average individual treatment effect by industry. On the one hand, during the first, third, and fourth quarters of 2020, there is no evidence supporting the existence of sectoral heterogeneity in the COVID-19 effect, and the COVID-19 shock is economically and statistically insignificant. Therefore, concentrating on the results for the first quarter, we are able to reject the existence of an effect even within sectors.\footnote{We have conducted other similar placebo studies conditioning on other variables (e.g., the main destination of exports, the main origin of imports,...) and in all the considered subsamples we do not estimate any significant effect of COVID-19.} On the other hand, during the second quarter of 2020, Colombian exporters belonging to almost every industry are found to significantly reduce their probability of surviving in the international markets. This decline is particularly pronounced in industries such as Textiles, Footwear, and Jewelry. However, industries like Food Preparations and Vegetables saw minimal changes in their survival probabilities due to the COVID-19 shock.


\subsection{Heterogeneity of the COVID-19 effect on Colombian exporters} \label{results3}

\begin{figure}[h]
\centering
\includegraphics[width=0.5\linewidth]{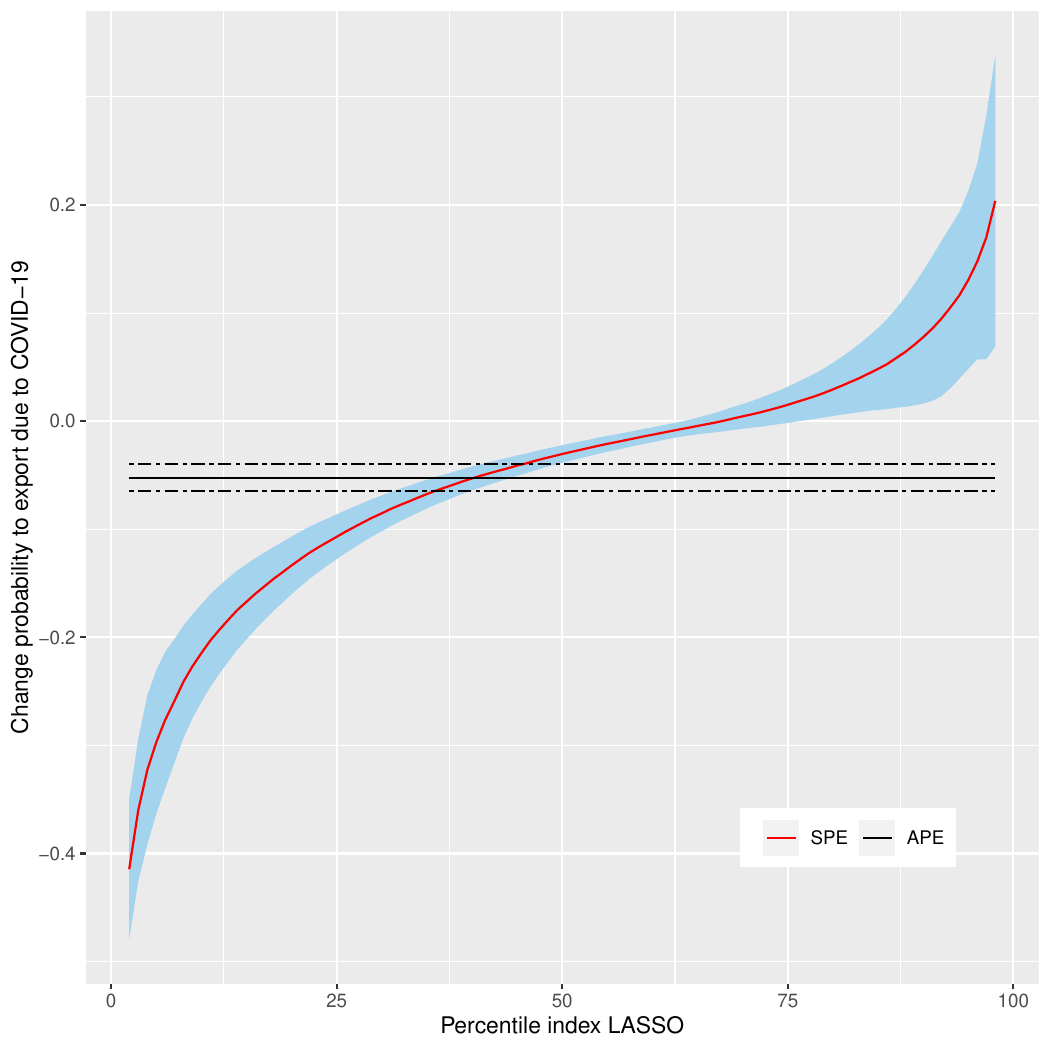}
\caption{Annual Sorted Partial Effects (SPE) and Average Partial Effects (APE) of COVID-19 on Colombian firm export's status. The treatment effect is calculated as a difference between SAM and SUM predictions. Standard errors were obtained with 100 bootstrap replications. Confidence intervals for a $5\%$ significance level.}
\label{fig:bootstrap_ann}
\end{figure}

\begin{figure}[t]
\centering
\includegraphics[width=0.9\linewidth]{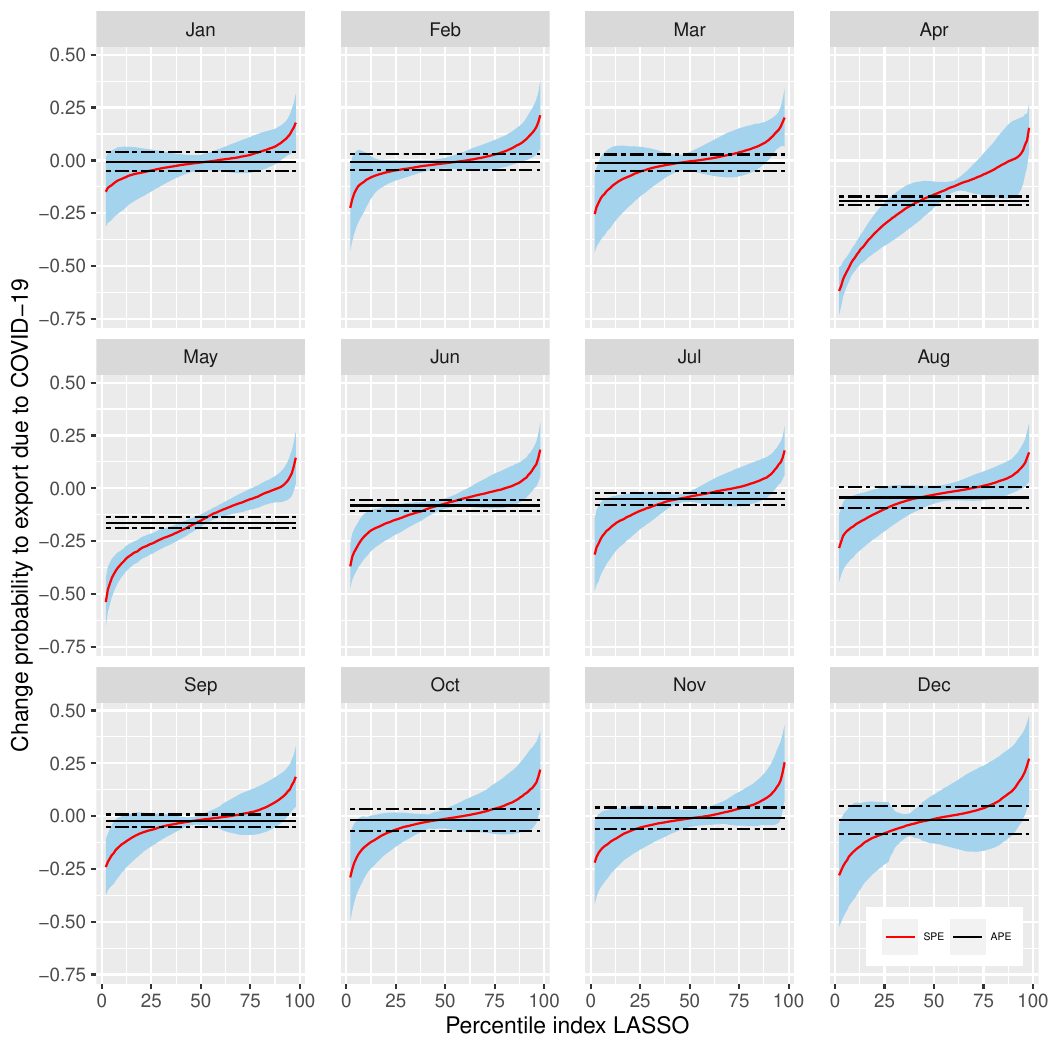}
\caption{Monthly Sorted Partial Effects (SPE) and Average Partial Effects (APE) of COVID-19 on Colombian firm export status. The treatment effect is calculated as a difference between SAM and SUM predictions. Standard errors were obtained with 100 bootstrap replications. Confidence intervals for a $5\%$ significance level.}
\label{fig:Sorted_LASSO_MONTHS}
\end{figure}

In this section, we investigate the determinant of possible treatment effect heterogeneity.  Figures \ref{fig:bootstrap_ann} 
 and \ref{fig:Sorted_LASSO_MONTHS} show the estimated Sorted Partial Effects (SPE) and Average Partial Effects (APE), which are obtained as explained in section \ref{methodology} by month and aggregating all the months, respectively. The two figures also report the $95\%$ confidence intervals with blue bands for SPE and black dashed lines for APE. 

Significant treatment effect heterogeneity is observed for April and May, with June showing a milder effect. The statistically significant (negative) estimated values of $\alpha^{*}(u)$ are primarily confined to the distribution's left tail. However, from July onwards, the confidence intervals of the SPEs overlap with those of the APEs, indicating an absence of treatment effect heterogeneity. Interestingly, in the pre-pandemic months, the SPEs closely aligned with the APEs (estimated to be zero). This demonstrates that individual placebo treatment effects are not statistically significant throughout the distribution, not just on average, reinforcing the robustness of our methodology across the entire distribution of treatment effects.

To identify the determinants of treatment effect heterogeneity, we examine the difference in means ($CADiff$) of firm characteristics between the most and least affected groups in Table \ref{tb:CA}. These groups are defined by whether their estimated $\alpha$ is lower than $\alpha^{*}(25th)$ or greater than $\alpha^{*}(75th)$, respectively. Therefore, we compute the raw difference in the means of the covariates between the most and the least affected firms by regressing the variables of interest on a constant and a dummy $q={\bf 1}_{ \{\alpha \leq \alpha^{*}(25th) \}}$ for the observations for which  $\alpha \leq \alpha^{*}(25th)$ or $\alpha \geq \alpha^{*}(75th)$. Then, we also provide the difference in adjusted means once we have controlled for the firm sector and month of the year. Controlling for sector and month allows us to perform a \textit{ceteris paribus} analysis, i.e., to dig into the effects of COVID-19 within specific sectors and specific months.
\begin{table}[ht!]
\centering
\caption{Estimated differences in means of the estimated treatment effect and other covariates between the group of more affected and the group of less affected firms ($CADiff$) applying the classification analysis to the $SAM - SUM$ estimates} \label{tb:CA}
\scalebox{0.8}{
\begin{tabular}{l ccc} 
\toprule  
Outcome variable & (1) & (2) & (3) \\ 
\midrule
TE  & $-0.3130^{***}$ & $-0.3060^{***}$  & $-0.2790^{**}$ \\
Agriculture  & -0.1940 &    &\\
Chemicals  & -0.0057 &   & \\
Manufacturing  & -0.0092 &   & \\
Metals  & 0.0134 &    &\\
Special  & $0.0056^{***}$ &   & \\
Textile  & $0.1600^{***}$ &    & \\ 
Wood  & $0.0292^{***}$ &   &\\
Air  & $0.2030^{*}$ & $0.1680^{***}$  & $0.2040^{***}$\\
Land  & 0.0340 & 0.0249 &  0.0170\\
Sea  & $-0.2360^{***}$ & $-0.1920^{***}$ & $-0.2200^{***}$\\
Jan  & -0.0738 & $-0.0766^{***}$ &  \\
Feb  & -0.0710 & $-0.0768^{***}$ &  \\
Mar  & -0.0751 & $-0.0773^{***}$ &  \\
Apr  & $0.1860^{***}$ & $0.1950^{***}$ & \\
May  & $0.1770^{***}$ & $0.1820^{***}$ &  \\
Jun  & 0.0754 & $0.0784^{***}$ &  \\
Jul  & 0.0132 & 0.0159 &  \\
Aug  & 0.0021 & 0.0008 & \\
Sep  & $-0.0412^{***}$ & $-0.0406^{**}$ &  \\
Oct  & $-0.0604^{***}$ & $-0.0609^{**}$ &  \\
Nov  & $-0.0723^{***}$ & $-0.0763^{**}$ &  \\
Dec  & -0.0557 & $-0.0621^{**}$ & \\
Number of export destinations (ND)  & -0.1990 & -0.1640 &  -0.2480\\
Number of import origins (NO)  & -1.7470 & $-1.9820^{***}$ & $-2.4440^{**}$ \\
Number of exported products (NP)  & 0.2400 & -0.2570 & -0.3440\\
Containment Index Stringency Export & $19.3600^{***}$ & $19.5100^{***}$ & $7.1800^{*}$ \\
Containment Index Stringency Import  & $19.1100^{***}$ & $20.8000^{***}$ & $7.2490^{***}$ \\
Value Exported (log)  & $-0.5110^{***}$ & -0.4490 &  $-0.5700^{*}$\\
Value Imported (log) & $-1.8160^{***}$ & $-2.2020^{***}$ & $-2.6860^{***}$ \\ 
\hline
Deviation from sectoral mean & & $\checkmark$ & $\checkmark$ \\
Deviation from monthly mean & & & $\checkmark$ \\ 
\bottomrule
\multicolumn{4}{p{16cm}}{\textit{Notes:} column (1) does not include sector or month variables in the regression; column (2) includes sectors in the regression; and, column (3) includes both the sector and month variables. $^{***}$ means significant at $1\%$, $^{**}$ at $5\%$, and $^{*}$ at $10\%$. Standard errors are obtained by bootstrapping the whole estimation process, and joint p-values are adjusted to consider the simultaneous testing of all variables. 
}
\end{tabular}}
\end{table}

Table \ref{tb:CA} is divided into 3 columns according to the control variables included in the regressions: in the first column, we show the unconditional average difference in the firms' characteristics between the most and least affected firms; in the second column, we control for the firm sector; and, in the third column, we control for firm sector and month of observation. The firm characteristics that we consider to explore the sources of COVID-19 treatment effect heterogeneity among Colombian exporters are observed in 2019 (the year before receiving the treatment). First, we check whether the estimated individual treatment effect (TE) differs between the firms contained in the two groups by using the TE as the dependent variable. We then move to firm-sector specific characteristics. In particular, the first set of firm characteristics that we use as dependent variables are dummies indicating the industry where the exporters operate.\footnote{We aggregate the 22 industries defined in the main analysis as follows. ``Agriculture'' contains Animals (01), Vegetables (02), Fats/oils (03), and Prepared Foodstuffs (04). ``Chemicals'' includes Chemical (06), and Plastics (07). ``Manufacturing'' contains Machinery (16), Vehicles (17), and Manufactured (20). ``Metals'' aggregates Mineral (05), Cement (13), Jewelries (14), and Metals (15). ``Special'' includes Precision Instruments (18), Arms (19), Art (21), and Special (22). ``Textile'' contains Leather (08), Textile (11), and Footwear (12). Finally, ``Wood'' aggregates Wood (09), and Paper (10). See Table \ref{tb:HS_section_chapter} in the Online Appendix for the complete industry names.} We also investigate the \textit{CADiff} for the means of transportation and the months when firms operate. Moreover, to account for the role of diversification patterns, we also consider as dependent variables the number of export destinations ($ND$), import origins ($NO$), and products ($NP$) exported. The weighted Containment Stringency Index that exporters face when exporting (importing) allows us to study to what extent treatment effect heterogeneity depends on these measures of direct exposure (to COVID-19 through their activities on international markets). A traditional continuous-DID strategy would have used these exposure variables as treatment variables, assuming that any COVID effect would emanate through them. Finally, including the total value exported (imported) by firms --expressed in logarithm-- among the variables for which the \textit{CADiff} is computed highlights the difference in the quantities sold (purchased) by most and least affected companies. A discussion of the main findings follows.

Considering the estimated individual treatment effects (TE) as a dependent variable, we find a negative and significant difference between most and least affected firms independently of the set of controls employed. These results show that the most affected exporters--i.e., those located in the first SPE quartile distribution--experienced a decrease in the probabilities of exporting between 27.9 and 31.3 percentage points lower than the one experienced by the least affected firms--i.e., those located in the last SPE quartile. 

We found significant differences among firms when examining how different aggregate sectors are affected. For instance, we detect that the share of textile firms among the most affected 2019 exporters is 16 percentage points higher with respect to the one estimated for the group of the least affected firms. Likewise, there is a difference of 2.9 percentage points in the presence of wood exporters between the most and least affected groups.

We also detect the existence of treatment effect heterogeneity associated with the means of transportation used by exporters in 2019. On the one hand, there are 16.8 to 20.4 percentage points more exporters using air transportation among the most affected than among the least affected firms. However, there are 19.2  to 23.6 percentage points fewer Colombian exporters using the sea for shipping among the most impacted firms compared to the least affected ones \citep{Nitsch22}. 

Looking at the treatment effect heterogeneity associated with months, the first pattern we notice is that only the months from April to August have a positive estimated parameter. However, only April and May estimated differences are statistically significant. There are 18.6 to 19.5 percentage points (17.7 to 18.2) more firms in April (May) among the most affected than among the least affected firms. From September to November, the coefficients become negative and significant, indicating the beginning stages of recovery. 

To evaluate how ex-ante exporter diversification affects the COVID-19 effect, we explore the estimated parameters associated with $ND$, $NO$, and $NP$. We want to investigate whether Colombian exporters' supply chain diversification and export destination diversification help mitigate the COVID-shock. We do not find compelling evidence that ex-ante diversification helps to face a shock of this kind, as we can evince from the estimated parameters associated with $ND$, $NP$, and, in the first column, to $NO$. Following the reasoning of \cite{lafrogne2022supply}, which exploits the COVID-19 crisis to study the export consequences of a country-specific supply-side shock by concentrating on the differential import exposure of French firms to the Chinese early lockdown, one possible explanation is that firms cannot substitute away the partner (or the product) under COVID-19. Another possible explanation, which they offer, is that exporters that do not diversify ex-ante can benefit from some form of ex-post diversification. However, when they restrict the analysis to homogeneous inputs, \cite{lafrogne2022supply} find weak evidence of a larger COVID-19 effect for firms with non-diversified inputs. They restrict the sample to homogeneous inputs because they want to analyze the COVID-19 effect among inputs expected to be substituted. Similarly, once we control for the sector and, therefore, inter alia, for the fact that some sector has relatively more diversification potential, the negative estimated difference turns statistically significant. Indeed, within sectors, the most affected Colombian exporters tend to import from $1.98$ fewer countries in 2019 than the least affected firms. The economic size of this estimate is large as approximately 60 per cent of Colombian exporters are not integrated into global value chains (they do not import), and the mean of $NO$ is approximately $4.16$ origins.

The $CADiff$ estimated when using the Export (Import) Containment Stringency Index as dependent variables provides insightful hints on the difficulties of Colombian firms in exporting (importing) to (from) countries adopting severe stringency measures. In particular, the most affected Colombian exporters face, on average, a higher Export (Import) Containment Stringency Index than those faced by least affected firms by 7.18 to 19.51 (7.25 to 20.80) points, depending on the column in the table.\footnote{Remember that the Index ranges from 0 to 100.}

Finally, the least affected firms exported (imported) 156.7\% to 176.83\% (614.7\% to 1467.3\%) more value in 2019 than the most affected firms. As expected, Colombian exporters trading in larger volumes (in value) are more resilient under a COVID-19 scenario. As with diversification, the comparison of the export and import side reinforces the idea that having more experience in sourcing inputs from abroad decreases the strength of the shock.

\subsection{Estimations based on $Y-SUM$} 
\begin{figure}[H]
\centering
\includegraphics[width=0.6\linewidth]{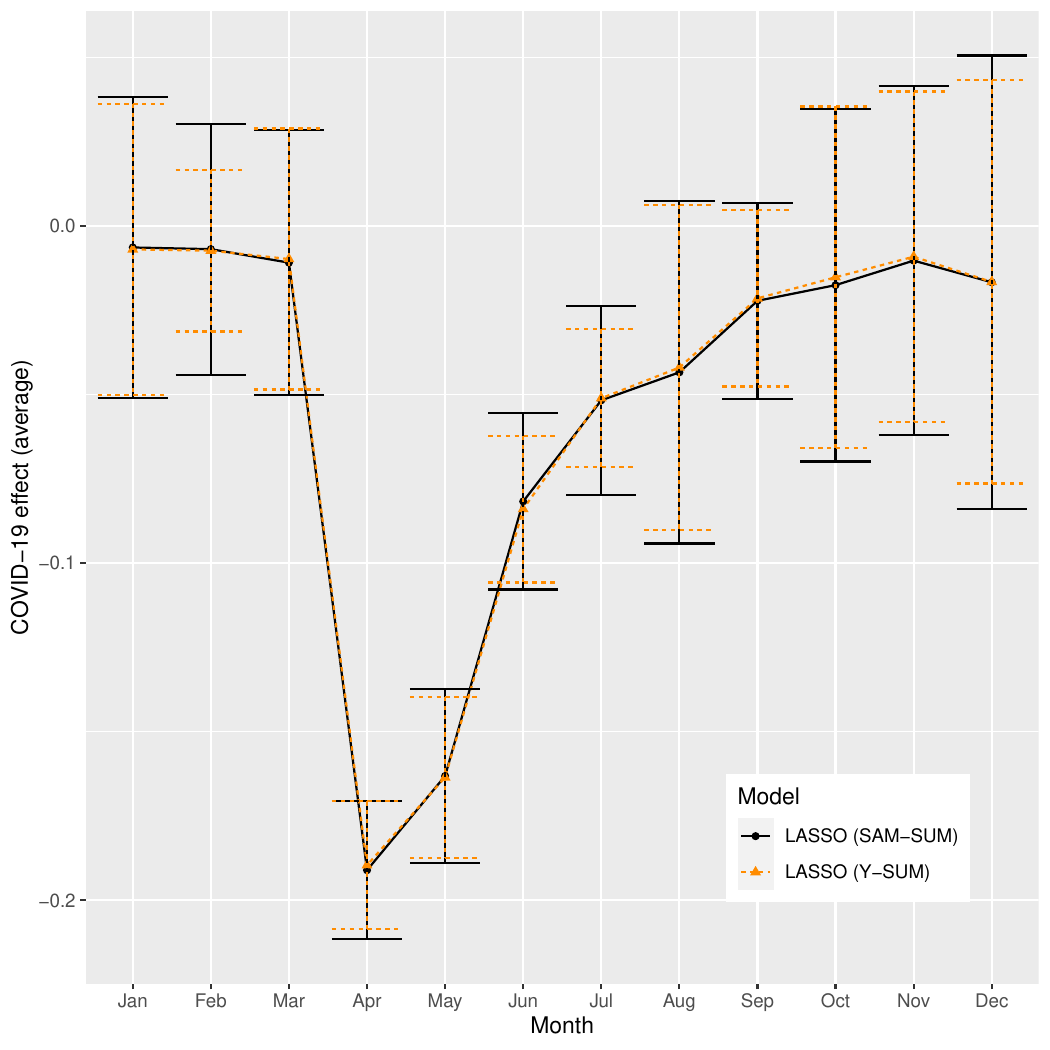}
\caption{Mean difference in the predicted probability of success (SAM vs. SUM / $Y$ vs. SUM) by month, using Logit-LASSO predictions and (SAM vs. SUM). Standard errors were obtained with 100 bootstrap replications. Confidence intervals for a $5\%$ significance level.}
\label{fig:diff_predictions_Y_SUM}
\end{figure}

In this paragraph, following \cite{fabra2020degrowth} and \cite{cerqua2020local}, we use the estimators based on Eq.~\eqref{Y_SUM_ind}. These estimators capture the differences between the observed outcome, $Y$ (binary variable accounting for the success of a Colombian exporter in 2020), and its counterfactual predictions (SUM). Figure \ref{fig:diff_predictions_Y_SUM} shows that the average individual treatment effect for COVID-19 is very similar when the individual treatment effect is estimated as the average of $\hat{\alpha}$ (black line, $SAM-SUM$) or as the average of $\hat{\hat{\alpha}}$ (yellow line, $Y-SUM$). As shown in Figure~\ref{fig:diff_predictions_Y_SUM}, when the interest lies in estimating the average treatment effects (by months in this case), the results based on $Y-SUM$ do not differ from those obtained by using $SUM-SAM$. We obtain similar results for the two methodologies also in terms of conditional treatment effects based on subgroups defined on firm characteristics (e.g., by industry or main export destination country).

The fact that the two estimators consistently find zero estimated effects for all 2019 exporters (and for subgroups based on the values of individual observables) during the first quarter suggests that the estimation error of both SUM and SAM, $\mathcal{E}_0$ and $\mathcal{E}_1$ respectively, goes to zero when we average the individual treatment effects across the whole distribution of 2019 exporters (or in subgroups defined by one of the possible dimension of treatment effect heterogeneity defined by observables; e.g., by industry or main export destination country).

\begin{figure}[h]
\centering
 \includegraphics[width=0.49\linewidth]{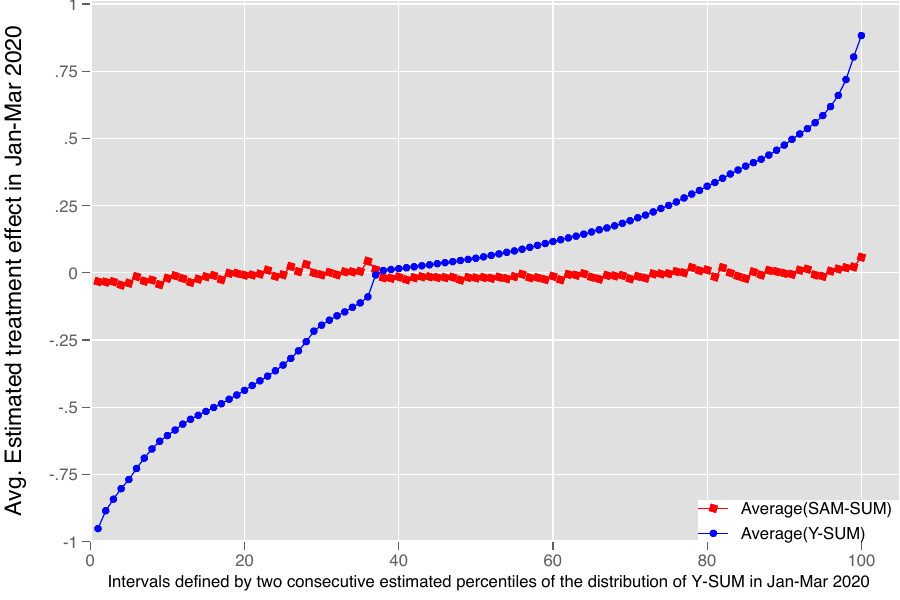}
\includegraphics[width=0.49\linewidth]{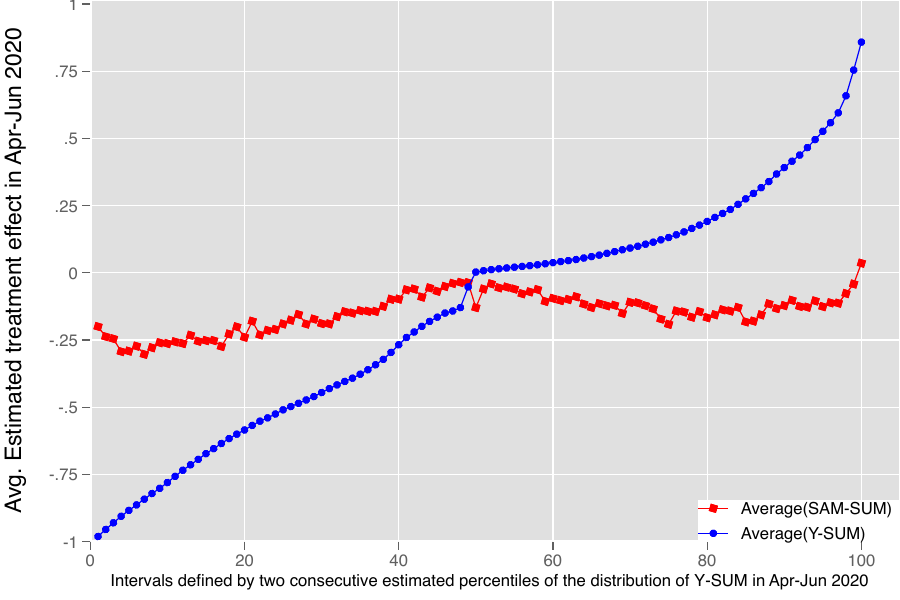}\\
\includegraphics[width=0.49\linewidth]{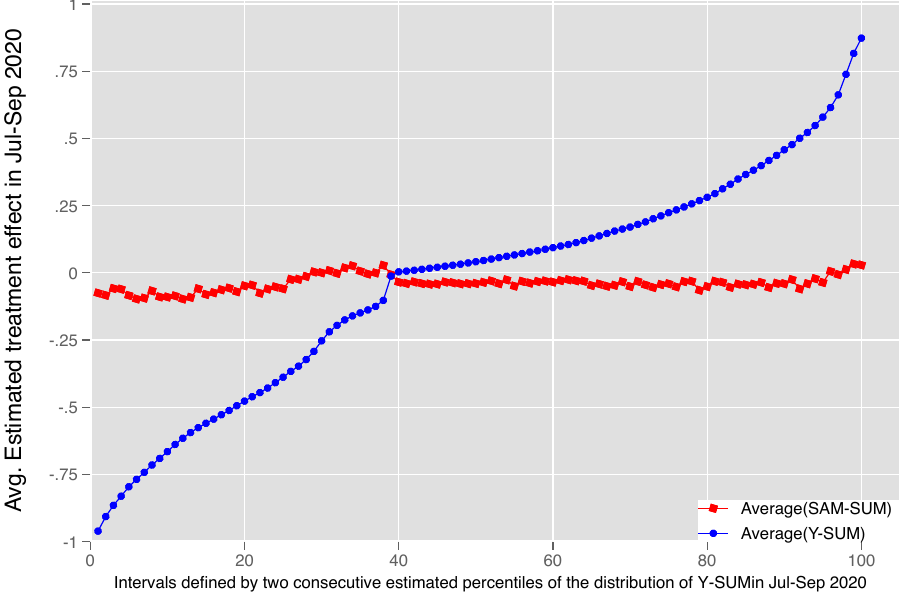}
\includegraphics[width=0.49\linewidth]{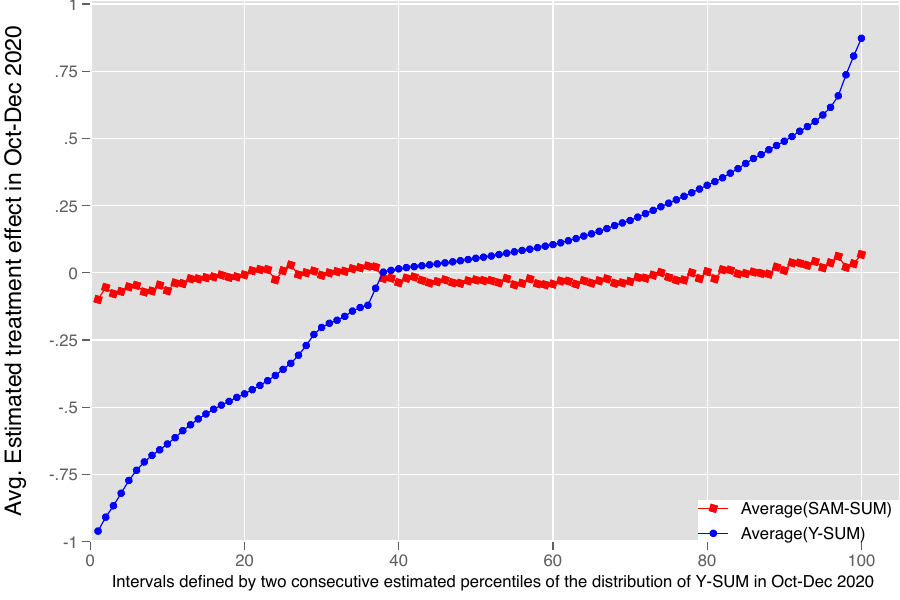}
\caption{Estimated average treatment effects ($SAM-SUM$, red line, and $Y-SUM$, blue line) by quarter for observations contained in intervals defined by the estimated percentiles of $Y-SUM$.}
\label{fig:SAM-SUM_vs_Y-SAM}
\end{figure}

However, since our goal is to identify the main dimensions of treatment effect heterogeneity by classifying units with the highest and lowest estimated treatment effects, we need also to evaluate how well these alternative estimation strategies perform in identifying treatment effects at the extremes of the distribution of treatment effects. 
Figure \ref{fig:SAM-SUM_vs_Y-SAM} shows the average of the estimated treatment effects obtained with the two estimators for the observations whose estimated treatment effects (by using $Y-SUM$) are contained in intervals defined by two consecutive values of the estimated percentiles of $Y-SUM$.
On the one hand, the estimator based on $Y-SUM$ is also identifying significant treatment effect heterogeneity in the first quarter, suggesting that the distribution's estimation error, $\mathcal{E}_0$, is not zero on average in the tails. Moreover, the shape of the $Y-SUM$ curve is similar across quarters, suggesting that this estimation method will be prone to misclassify units when using the Sorted Effects strategy suggested above. On the other hand, in the first quarter, the shape of the $SAM-SUM$ curve is flat, showing a constant average estimated effect that is zero along the whole distribution of the $Y-SUM$ estimated effects, suggesting that by using the $SAM$ we are able to wash out the estimation error of the $SUM$ because $\mathcal{E}_1$ = $\mathcal{E}_0$.

\begin{figure}[h]
\centering
\includegraphics[width=0.85\linewidth]{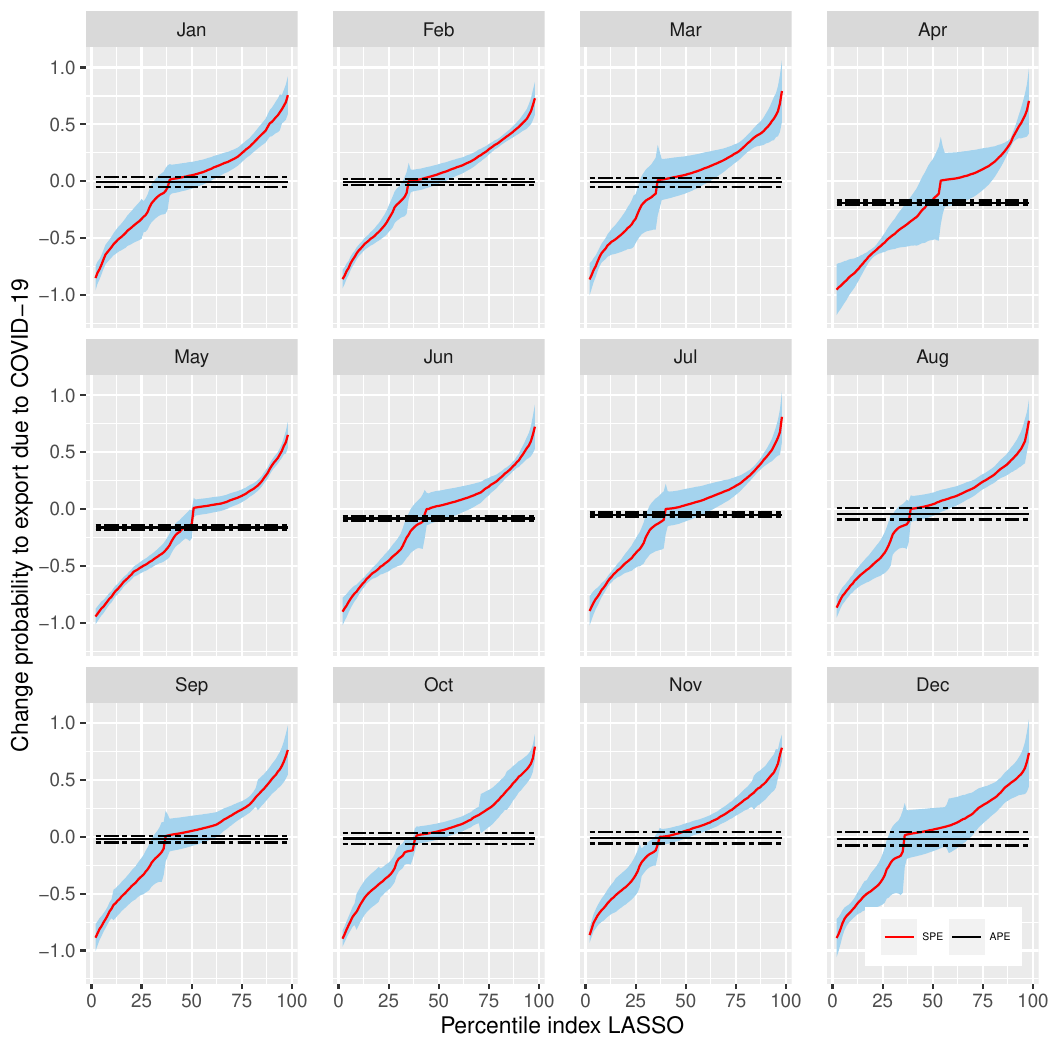}
\caption{Monthly Sorted Partial Effects (SPE) and Average Partial Effects (APE) of COVID-19 on Colombian firm export's status. TE is calculated as a difference between the observed outcome ($Y$) and SUM predictions. Standard errors were obtained with 100 bootstrap replications. Confidence intervals for a $5\%$ significance level.}
\label{fig:sorted_Y_SUM}
\end{figure}

This behavior of the estimators based on $SAM-SUM$ is consistent with the results shown in Figure \ref{fig:Sorted_LASSO_MONTHS} for the Sorted Effects analysis. Figure \ref{fig:sorted_Y_SUM} shows that the intuition on the inadequacy of the $Y-SUM$-based estimators to identify treatment effects on the tail of the distribution is also confirmed by the Sorted Effects analysis based on this estimation strategy. When using the $Y-SUM$ individual level estimates to feed the SPE methodology, we find economically and statistically significant effects of the COVID-19 shock all along the percentile distribution in the first quarter. While it is true that, on average, $\mathcal{E}_0$ tends to be zero across all observations, these findings suggest that this is not true when we focus on specific segments of the treatment effect distribution, particularly in the tails.

Table \ref{tb:CA_Y_SUM} presents the classification analysis results on the sources of treatment effect heterogeneity when the $CADiff$ is estimated using the ($Y-SUM$) approach. For all the firm characteristics we examined, we found no statistically significant difference between the most and least affected groups. This is consistent with the inability of the $Y-SUM$ approach to consistently estimate treatment effects in the tails of the $\alpha$'s distribution and, consequently, to identify the groups of the most affected and the least affected firms. In other words, such groups will be contaminated by the inclusion of firms wrongly classified due to the estimation error $\mathcal{E}_0$.

\begin{table}[h]
    \centering
\caption{Estimated differences in means of the estimated treatment effect and other covariates between the group of more affected and the group of less affected firms ($CADiff$) applying the classification analysis to the $Y - SUM$ estimates}
\label{tb:CA_Y_SUM}
\scalebox{0.8}{
\begin{tabular}{lccc} 
\toprule
\textbf{Outcome variable} & $\beta^{(1)}_{1,f}$ & $\beta^{(2)}_{1,f}$ & $\beta^{(3)}_{1,f}$   \\
\midrule
TE & -1.0910 & -1.0930*** & -1.0710  \\
Agriculture  & -0.0616 &    &\\
Chemicals  & -0.0192 &   & \\
Manufacturing  & 0.0112 &   & \\
Metals  & 0.0109 &    &\\
Special  & 0.0059 &   & \\
Textile  &0.0486 &    &\\
Wood  & 0.0041 &   &\\
Air & 0.0411 & 0.0271 &  0.0289 \\
Land & 0.0086 & 0.0062 &  0.0068\\
Sea & -0.0482 & -0.0321 & -0.0343  \\
Jan & -0.0190 & -0.0189 &   \\
Feb & -0.0242 & -0.0237 &   \\
Mar & -0.0181 & -0.0181 &   \\
Apr & 0.0631 & 0.0630 &   \\
May & 0.0620 & 0.0612 &   \\
Jun & 0.0166 & 0.0167 &   \\
Jul & 0.0033 & 0.0028 &   \\
Aug & -0.0050 & -0.0053 &   \\
Sep & -0.0169 & -0.0167 &   \\
Oct & -0.0216 & -0.0208 &   \\
Nov & -0.0218 & -0.0222 &   \\
Dec & -0.0183 & -0.0181 &   \\
Number of export destinations (ND) & 0.3310 & 0.3470 &  0.3306 \\
Number of import origins (NO) & 0.0350 & -0.0595 &  -0.1077 \\
Number of exported products (NP) & 0.6050 & 0.4670 &  0.4275 \\
Containment Index Stringency Export & -0.2280 & -0.0264 &  0.9690 \\
Containment Index Stringency Import & -4.2180 & -4.4910 & -0.0520 \\
Value Exported (log) & -0.2700 & -0.2760 & -0.1800 \\
Value Imported (log) & -0.0910 & 0.0296 & 0.0040  \\ 
\hline
Deviation from sectoral mean & & $\checkmark$ & $\checkmark$ \\
Deviation from monthly mean & & & $\checkmark$ \\ 
\bottomrule
\multicolumn{4}{p{12cm}}{\textit{Notes:} column 1 does not include sector or month variables in the regression; column 2 includes sector in the regression, and, column 3 includes both the sector and month variables. $^{***}$ means significant at $1\%$, $^{**}$ at $5\%$, $^{*}$ at $10\%$. Standard errors are obtained by bootstrapping the whole estimation process and joint p-values are adjusted to take into account the simultaneous testing of all the variables.}
\end{tabular}
}
\end{table}

\clearpage
\newpage
\section{Concluding discussion}\label{conclusions}

In this paper, we show the potential of ML techniques for building counterfactuals, identifying the most affected subpopulations and the sources of treatment effect heterogeneity in scenarios where a credible control group is unavailable and it is difficult to define ex-ante the varying degrees of exposure to a shock for each economic agent. 

In the application we consider, we concentrate on the effects of an economy-wide shock such as COVID-19 on a firm's export behaviour. Using data from the Colombian customs office, we estimate that, during 2020, on average, the COVID-19 shock decreased a firm's probability of surviving in the export market by about 15 to 20 percentage points in April and May and by approximately 5 to 8 percentage points in June and July. By analysing the estimated treatment effect distribution, we reveal that these average results hide considerable heterogeneity.
For example, in April 2020, we find that for some exporters COVID-19 decreased their survival probability by 55 percentage points.
%
We identify the firms most and least affected by COVID-19 and we compare their characteristics by integrating the Sorted Partial Effects methodology with our causal ML approach. We emphasize how the integration into global value chains on the import side, both in terms of the number of countries from which a firm sources and the value of imports, is an important factor of resilience for exporters facing the COVID-19 shock.




From a methodological point of view, we show practitioners how to apply the generic ML tools proposed by \cite{chernozhukov2020generic} to a context in which there is no control group available; we suggest how to use in-time placebo tests to check the credibility of counterfactual estimates; finally, we provide evidence indicating that in the Sorted Partial Effects analysis, in which the focus lies on the tails of the distribution of the treatment effects, it is critical to correct the estimation error arising from the necessarily imperfect reconstruction of the unobservable counterfactual.

While this method is specifically designed for analyzing the heterogeneous impacts of economy-wide shocks, there exists potential utility in employing this approach in less extreme situations where policies or shocks may exhibit unobservable spillovers that are challenging to model in advance. In such contexts, our empirical framework proves advantageous in detecting these potential heterogeneous indirect effects, as it circumvents the need for a priori identification of a control group of untreated units.

Finally, in this paper we also demonstrate that ML methods can be applied successfully to predict firms' trade potential. We consider ML methods a promising tool to assist firms and public agencies in their export decision-making processes. The bulk of countries possesses export promotion agencies whose objective is to sustain firms' internationalization activities by lowering the costs of information acquisition \citep{broocks2017impact, munch2018effect}. 
Given that exporters' dynamics can be understood as a complex learning process dense of interdependencies (complementarity or substitutability) between products and destination markets (from the perspective of technology/knowledge, local tastes, legal requirements, and marketing and distribution costs) 
and that ML techniques have been successfully applied to predict firm performances in such settings, we believe that an important avenue for future research is to test the effectiveness of using these techniques and firm-level data to build recommendation systems. These systems could help firms identify their latent comparative advantages and provide export diversification and differentiation recommendations.

\clearpage
\small
\newpage
\bibliographystyle{chicago}
\bibliography{biblio}


\clearpage
\newpage

\setcounter{page}{1}
\appendix

\section*{\textbf{ONLINE APPENDIX}}

\section{Appendix - The Colombian economy amidst the COVID-19 crisis} 
\label{app:colombia}
\vspace{0.5cm}

Colombia exports little compared to other countries in Latin America with similar development levels. In recent years, the share of total exports of Colombian GDP has oscillated around 15\%, well below other countries in the region such as Chile and Mexico \citep{cepeda2019colombian}. Colombia started to open its economy in the 1990s with several market-oriented reforms to liberalize financial and capital markets. Although the Colombian economy was still relatively closed during most of the twentieth century \citep{ocampo2000colombia}, it has been strongly affected by the global financial crisis in 2008-2009 \citep{zuluaga2009}. Nowadays, despite the growing number of trade agreements, partners, and volume of trade, the integration of Colombia into world trade markets is still modest. 

The main reason behind Colombia's poor performance is the low diversification of trade, with a prevalence of primary products, because of the relative abundance of natural resources and low-skilled labor. Besides, the emergence of raw products derived from mining has gained a larger share in total exports, reducing the importance of other products such as coffee, bananas, flowers, some labor-intensive manufactures, and petrochemicals \citep{garavito2020}. 

Since the outbreak of the COVID-19 pandemic, Colombia implemented early measures to contain the spread of COVID-19 and prepare the health system and mitigate the economic and social impact. The Colombian government issued non-compulsory requests for remote working to private companies on February 24, 2020; schools and universities were closed on March 16. On March 25, when there were less than a dozen deaths, the government implemented a complete and mandatory lockdown until April 13. During this period, only a few essential activities -- such as health services, public services, communications, banking and financial services, food production, pharmaceuticals, and cleaning and disinfection products -- were excluded. The partial lockdown implementation--between April 27 and May 11--allowed a gradual restoration of mobility, enabling a set of non-essential activities under security guidelines and protocols to guarantee social distancing. Most manufacturing activities were gradually allowed at this stage, while non-authorized activities were restricted to market their products through electronic commerce platforms. Finally, from May 28, restrictions to the services sector have been lifted, and on September 1, the government announced the confinement end, and airports were opened.

To better cope with the emergency, Colombian authorities have introduced transitory provisions to secure international trade of essential products. Along with the lockdown measures, medicines, supplies, and equipment in the health sector had zero-tariff for six months. Besides, the export and re-export of these products were forbidden. There was a zero-tariff from April 7 to June 30 for raw materials such as maize, sorghum, soybeans, and soybean cake.


\clearpage
\newpage

\section{Appendix - Data} 
\label{app:data}
\vspace{0.5cm}
\setcounter{table}{0}\renewcommand{\thetable}{Appx.\arabic{table}}
\setcounter{figure}{0}\renewcommand{\thefigure}{Appx.\arabic{figure}}
\begin{table}[H]
\begin{center}
\tiny
   \caption{Predictors for exporters success}
\label{tb:variables}
\resizebox{0.8\textwidth}{!}{\renewcommand{\arraystretch}{1.1}
\begin{threeparttable}
\begin{tabular}{p{1.5cm} p{11cm} p{3cm}} 
 \toprule
Variable &   Description &    Source \\ 
\midrule
\multicolumn{3}{c}{\textit{Models: SUM and SAM}}  \\
\midrule
$NP, ND, NO$ &   Number of products exported by, number of destinations where a company exports, and number of import origin countries for an exporter in a given month, respectively. &   Authors' own elaboration from Colombian Customs Office (DIAN).\\
$HH_p, HH_d$ &  Product-Herfindahl Index, and Destination-Herfindahl Index. Measure the concentration of products at 6-digits HS, and the concentration at destination by company-month, respectively. &   Authors' own elaboration from Colombian Customs Office (DIAN). \\
Total value (exports) &   Free on board value of the export transaction in US dollars for each company-month. &   Colombian Customs Office (DIAN)  \\
Total value (imports) &   Free on board value of the import transaction in US dollars for each company-month. &   Colombian Customs Office (DIAN)  \\
Size & 4 class dummies classifying firms according to the quartiles of the firm-level (Q1, Q2, Q3 and Q4) distribution of the total monthly value of exports (in $\ln$).  &  Authors' own elaboration \\
Destination & Factor variable with one level (dummy variable) for each destination country where Colombian exporters operate by month. &   Colombian Customs Office (DIAN) \\
Origin & Factor variable with one level (dummy variable) for each import origin country where Colombian exporters operate by month. &   Colombian Customs Office (DIAN) \\
Continent & Factor variable with one level (dummy variable) for each continent where Colombian exporters operate. &   Authors' own elaboration \\
Department & Factor variable with one level (dummy variable) for each department (region) in Colombia from which companies operate. &    Colombian Customs Office (DIAN) \\
Means of Transportation & 4 class dummies indicating the means of transportation a company use to perform a transaction (land, sea, air, others). &   Colombian Customs Office (DIAN) \\
Sector & 99 class dummies classifying company products at 2-digit HS code (corresponding to a HS-chapter). &   Authors' own elaboration \\
Industry &  22 class dummies indicating the industries (HS-sections) where companies operate. &   Authors' own elaboration from Colombian Customs Office (DIAN). \\
Sector Experience & Factor variable with one level (dummy variable) for each sector. Takes value 1 in all periods after a company exports for first time in a given sector (reflecting past experience in a sector).  &  Authors' own elaboration from Colombian Customs Office (DIAN). \\
Destination Experience & Factor variable with one level (dummy variable) for each destination. Takes value 1 in all periods after a company exports for first time in a given destination (reflecting past experience in a destination). &   Authors' own elaboration from Colombian Customs Office (DIAN). \\
Exporter (importer) Experience & Variable counting the accumulated value exported (imported) in the last twelve months. &   Authors' own elaboration from Colombian Customs Office (DIAN). \\
\midrule
\multicolumn{3}{c}{\textit{Models: SAM (COVID-19 variables)}}\\
\midrule
 Containment Economic Index & We consider the Economic Index from \cite{hale2020variation} that provides a measure of the strength of the economic policies set in place to deal with the pandemic (such as income support and debt relief) for each country in the world. It ranges from 0 to 100. At the firm level we define two variables, one at the export and one at the import side, by taking a weighted average of these country level scores according to the monthly value of exports(imports) that a company ships(source) in every country.\tnote{1} &    \cite{hale2020variation} and Colombian Customs Office (DIAN). \\
Containment Government Index  & We consider the Government Index from \cite{hale2020variation} that measures the strictness of 'lockdown' style policies that primarily restrict people's behavior. It ranges from 0 to 100. At the firm level we define two variables, one at the export and one at the import side, by taking a weighted average of these country level scores according to the monthly value of exports(imports) that a company ships(source) in every country. &    \cite{hale2020variation} and Colombian Customs Office (DIAN).\\
Containment Health Index & We consider the Health Index from \cite{hale2020variation} that combines 'lockdown' restrictions and closures with measures such as testing policy and contact tracing, short term investment in healthcare, as well investments in vaccine). Ranges from 0 to 100. At the firm level we define two variables, one at the export and one at the import side, by taking a weighted average of these country level scores according to the monthly value of exports(imports) that a company ships(source) in every country. &    \cite{hale2020variation} and Colombian Customs Office (DIAN).\\
Containment Stringency Index & We consider the Stringency Index from \cite{hale2020variation} that records how the response of governments has varied over all indicators, becoming stronger or weaker over the course of the outbreak. Ranges from 0 to 100. At the firm level we define two variables, one at the export and one at the import side, by taking a weighted average of these country level scores according to the monthly value of exports(imports) that a company ships(source) in every country. &    \cite{hale2020variation} and Colombian Customs Office (DIAN).\\
\midrule
\multicolumn{3}{c}{\textit{Models: SUM and SAM (variables only for Logit, Logit-LASSO, and Logit-Ridge)}}\\
\midrule
Size*Industry & Factor variables with 5 levels for each industry. Takes value 1 when the company size is Q1, value 2 when company size is Q2, value 3 when the size is Q3 and 4 when the size is Q4 while operating in a given industry. However, it takes value 0 if a company is not operating in this industry (for any size level). &   Authors' own elaboration from Colombian Customs Office (DIAN). \\
Size*Sector & Factor variables with 5 levels for each sector. Takes value 1 when the company size is Q1, value 2 when company size is Q2, value 3 when the size is Q3 and value 4 when the size is Q4 while operating in a given sector. However, it takes value 0 if a company is not operating in this sector (for any size level). &   Authors' own elaboration from Colombian Customs Office (DIAN). \\
Size*Means of Transportation & Factor variables with 5 levels for each sector. Takes value 1 when the company size is Q1, value 2 when company size is Q2, value 3 when the size is Q3 and value 4 when the size is Q4 while operating using a given means of transportation. However, it takes value 0 if a company is not operating using this means of transportation (for any size level). &   Authors' own elaboration from Colombian Customs Office (DIAN). \\
Size*Destination & Factor variables with 5 levels for each sector. Takes value 1 when the company size is Q1, value 2 when company size is Q2, value 3 when the size is Q3 and value 4 when the size is Q4 while operating in a given destination. However, it takes value 0 if a company is not operating in this destination (for any size level). &   Authors' own elaboration from Colombian Customs Office (DIAN). \\
\bottomrule
\multicolumn{3}{l}{* 
\url{https://data.europa.eu/euodp/en/data/dataset/covid-19-coronavirus-data}}
\end{tabular}
\begin{tablenotes}
    \item[1] When an exporter does not import we impute the corresponding internal Index (Economic, Government, Health, and Stringency) of Colombia to create the corresponding import side Index.
  \end{tablenotes}
\end{threeparttable}}
\end{center}
\end{table}

\newpage

\begin{table}[h!]
\caption{Sector-Industry mapping}
\centering
\label{tb:HS_section_chapter}
\resizebox{\textwidth}{!}{%
\begin{tabular}{@{}llc@{}}
\toprule
Section (Industry)             & Industry Name        & HS-Chapter (Sector)         \\ \midrule
1                 & Live Animals/ Animal Products &  1-5\\
2               & Vegetable Products &  6-14\\
3           & Animal or Vegetable Fats/Oils & 15 \\
4       & Prepared Foodstuffs&  16-24\\
5   & Mineral Products &  25-27\\
6         & Products of Chemical Industries &  28-38\\
7       & Plastics, Rubber & 39-40 \\
8        & Raw Hides, Skins and Leather & 41-43 \\
9          & Wood &  44-46\\
10          & Paper &  47-49\\
11            & Textile &  50-63\\
12        & Footwear &  64-67\\
13             & Art. of Stone, Cement & 68-70 \\
14   & Jewelries &  71\\
15   & Base Metals &  72-83\\
16   & Machinery Equipment &  84-85\\
17                  & Vehicles &  86-89\\
18                  & Precision Instruments &  90-92\\
19                  & Arms &  93\\
20     & Miscellaneous Manufactured Articles &  94-96\\
21             & Works of Art &  97\\
22               & Special Classification Provisions &  98-99\\ \bottomrule
\multicolumn{2}{l}{Source: Author's elaboration using \cite{pierce2011concordance} tables.}
\end{tabular}
}
\end{table}

\clearpage
\newpage

\section{Appendix - Descriptive Statistics} 
\label{app:des}

\begin{figure}[H]
\centering
\includegraphics[height=0.35\linewidth]{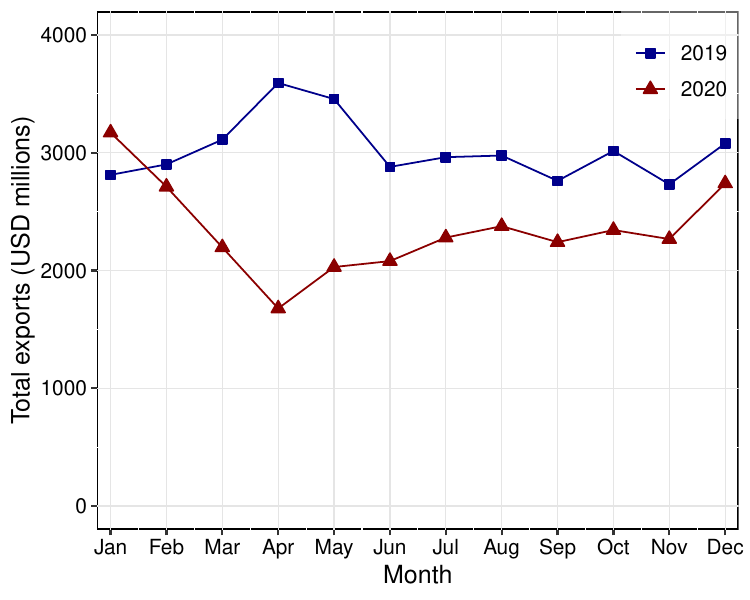}
\includegraphics[height=0.35\linewidth]{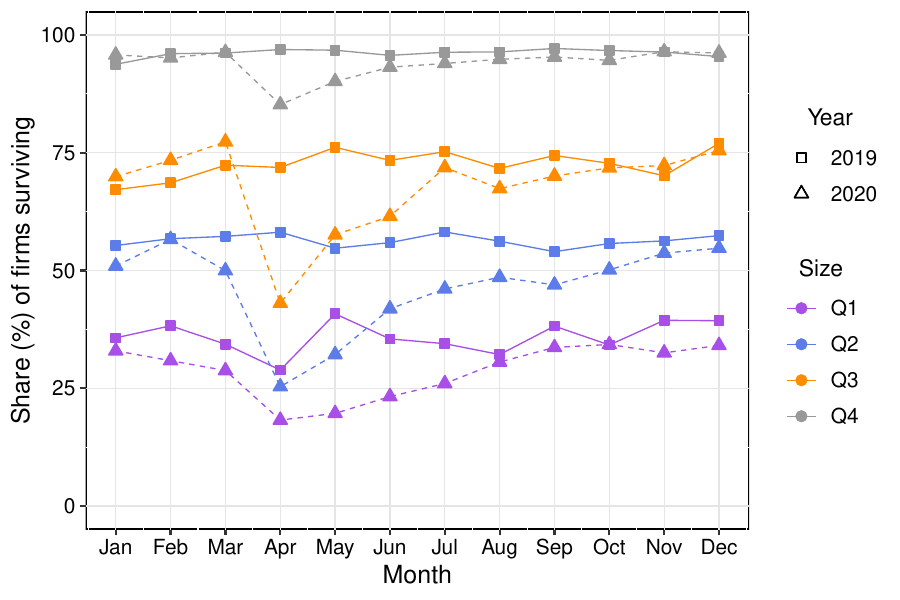}
\caption{The evolution of total exports (left) and the proportion of surviving exporting firms in year $t$ among those exporting in year $t-1$ within size classes defined at $t-1$ (right). Firm size classes are based on the quartiles of the firms' exports (in $\ln$) distribution in a given month.}
\label{fig:evolution}
\end{figure}

The left panel in Figure~\ref{fig:evolution} shows the evolution of total monthly exports during 2019 and 2020. The total monthly value of exports in 2020 is significantly lower than the one observed for the corresponding month in 2019, except for January and February. The lockdown measures implemented to contain the COVID-19 outbreak in Colombia and abroad had a severe impact between April and June--the value in April 2020 is half of the one observed in April 2019 (47\%). 

In a typical month, large firms get a lion's share of the total exports. A regular pattern in looking at customs data is that more prominent exporters trade for many months and ship more frequently than smaller firms, which make only a few shipments. The right panel in Figure~\ref{fig:evolution} shows the proportion of surviving exporting firms in year $t$ among those exporting in year $t-1$, by size classes defined at $t-1$. Comparing the figures for 2020 with those for 2019, it seems that the COVID-19 outbreak affected all firms regardless of their size. However, the effect looks proportionally stronger for small firms (Q1 and Q2 of the distribution). In contrast, larger firms are less affected and recover faster than the survival rates observed in 2019. 

In the following of this Appendix~\ref{app:des}, we show the growth patterns of the number of exporters and export volumes between 2019 and 2020 (and, as a comparison, between 2018 and 2019) segmented by country of destination and product sector, offering further insights into the heterogeneous impacts of the COVID-19 pandemic on Colombian exports.

\begin{figure}[h]
\centering
\includegraphics[width=0.20\linewidth]{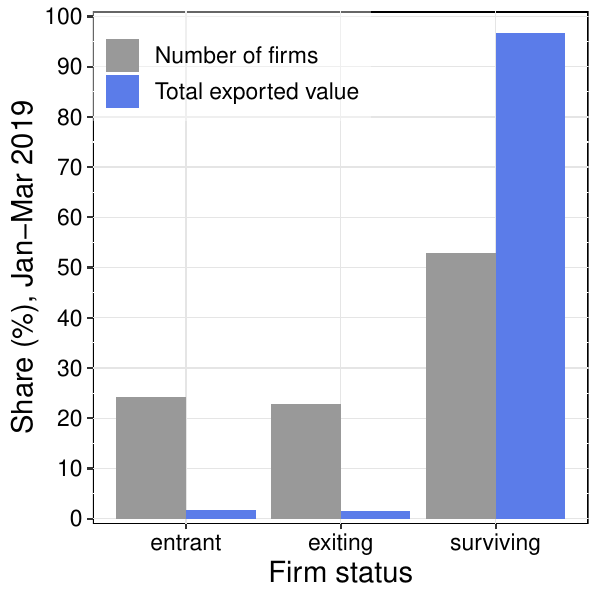}
\includegraphics[width=0.20\linewidth]{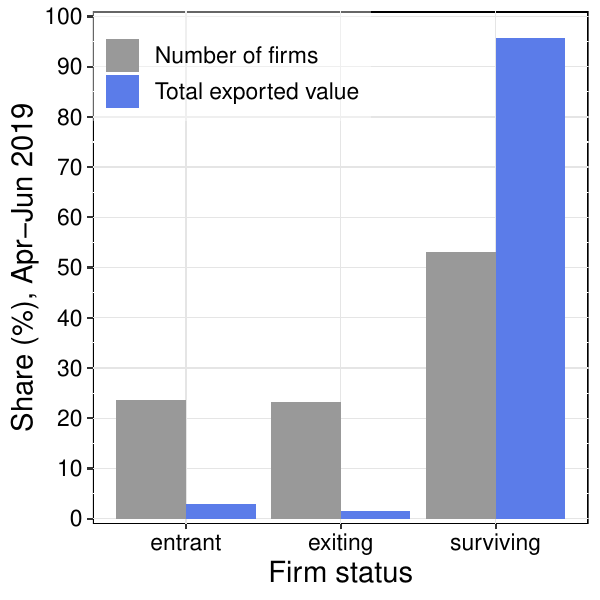} 
\includegraphics[width=0.20\linewidth]{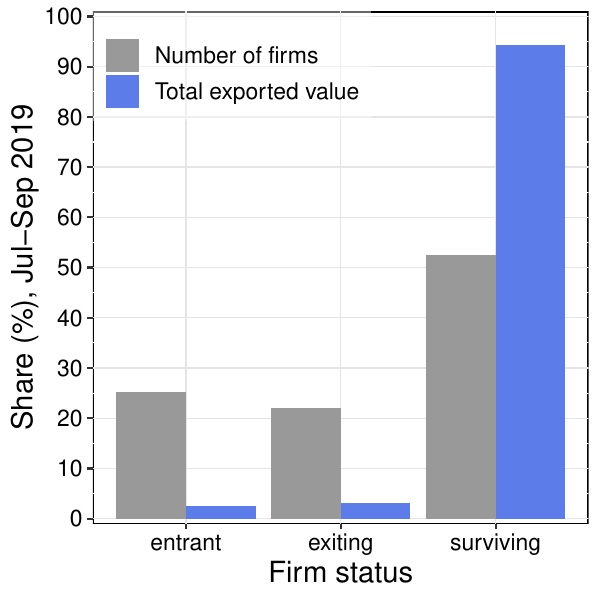}
\includegraphics[width=0.20\linewidth]{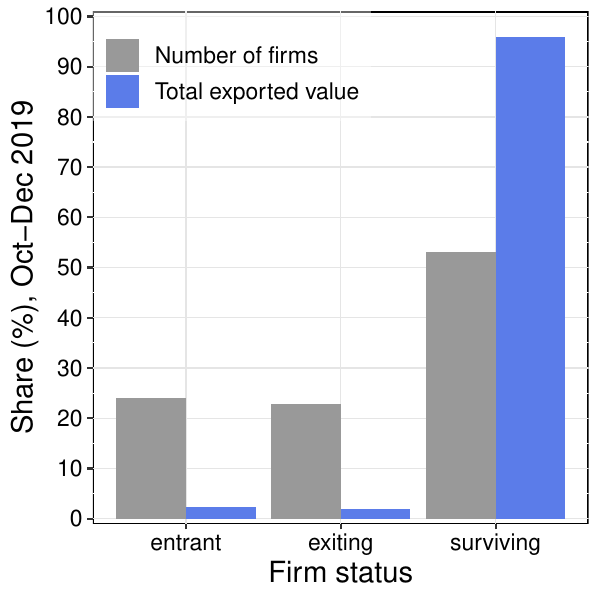}\\ \includegraphics[width=0.20\linewidth]{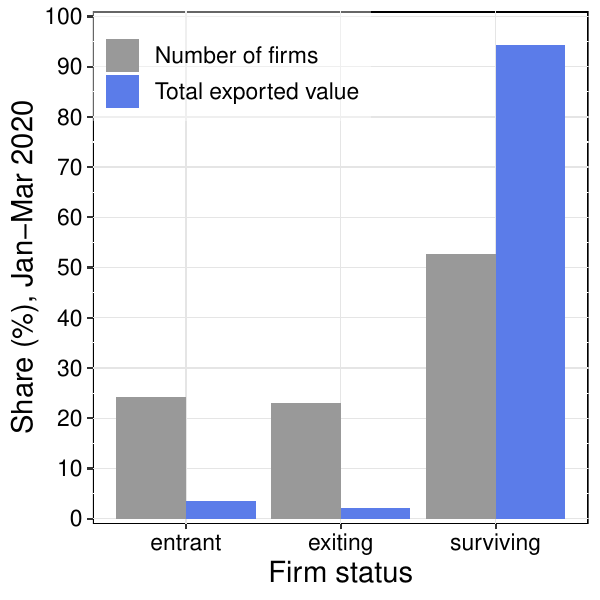}
\includegraphics[width=0.20\linewidth]{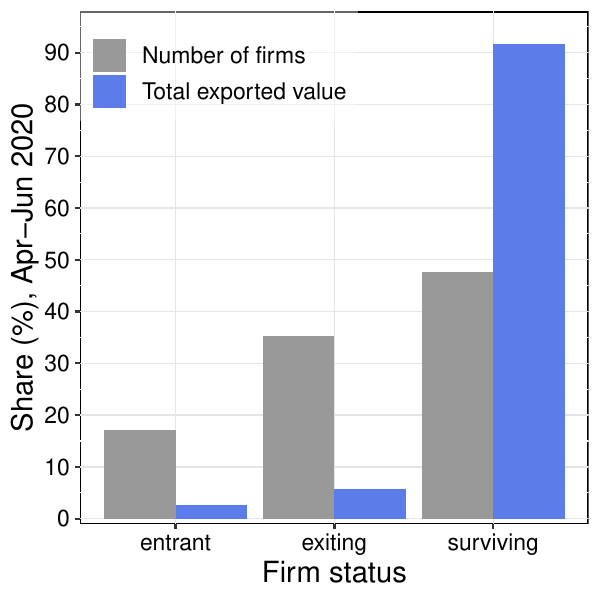} 
\includegraphics[width=0.20\linewidth]{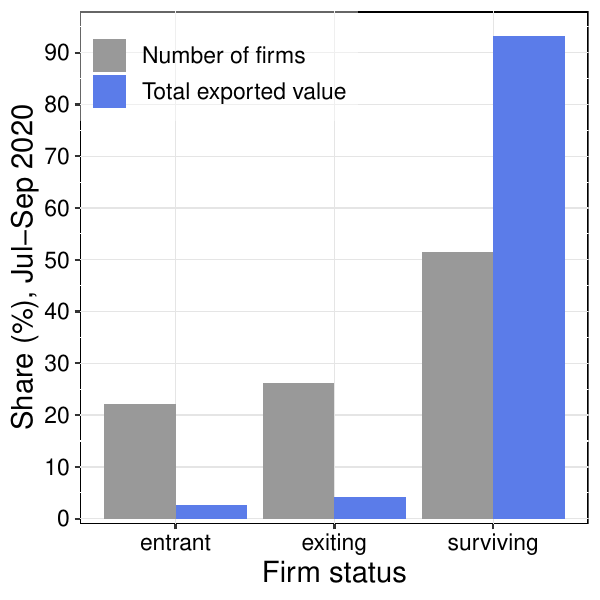}
\includegraphics[width=0.20\linewidth]{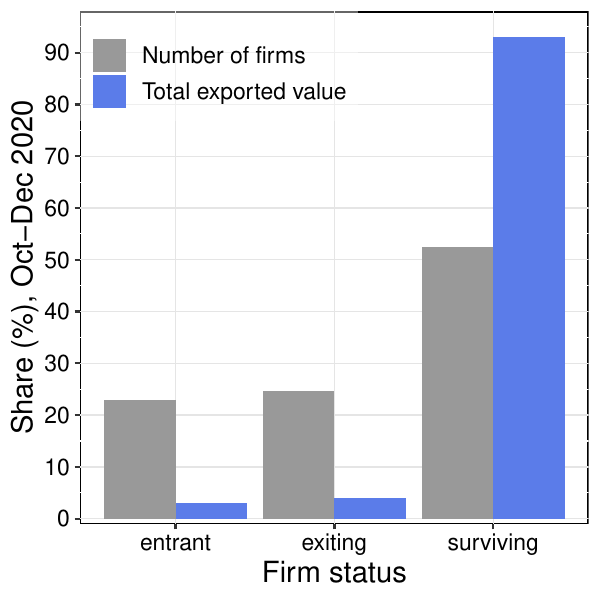}\\
\caption{Entry-exit dynamics of firms and total export value by firms that drop, enter or stay active, in 2019 (upper part of the figure) and in 2020 (bottom part of the figure) by quarters. Firm status is defined by looking at the previous year.}
\label{fig:firm_status}
\end{figure}

Figure~\ref{fig:firm_status} shows, separately for the first and second quarter of a year, the percentage of firms that survive, enter or exit the export market and their corresponding shares of total exports. Thus, for a given quarter in 2019 and the corresponding quarter in 2020, we label each firm as \textit{exiting} when it is present in 2019 and absent in 2020, \textit{entrant} when it is absent in 2019 and present in 2020, and \textit{surviving} when it is present in both years. We average the total value exported by each firm during the same quarter of two different years. Then, we sum the individual average value exported according to the firms' status. It turns out that surviving firms play an essential role in explaining total exports: they are around half of the total number of firms in both quarters and account for about 90\% of the total export value. The volume lost, during the second quarter of 2020, due to exiting firms is around 5\% (assuming they would have exported in 2020 similar export volumes as observed in 2019). Entrant firms almost made up this 5\% loss. Despite this, the firms' composition that participates in exports is very different. The number of existing firms in the second quarter of 2020 is much higher than the share of the first quarter of 2020 and the share of 2019 in the same period of the year. 

\begin{figure}[h]
\centering
\includegraphics[width=0.24\linewidth]{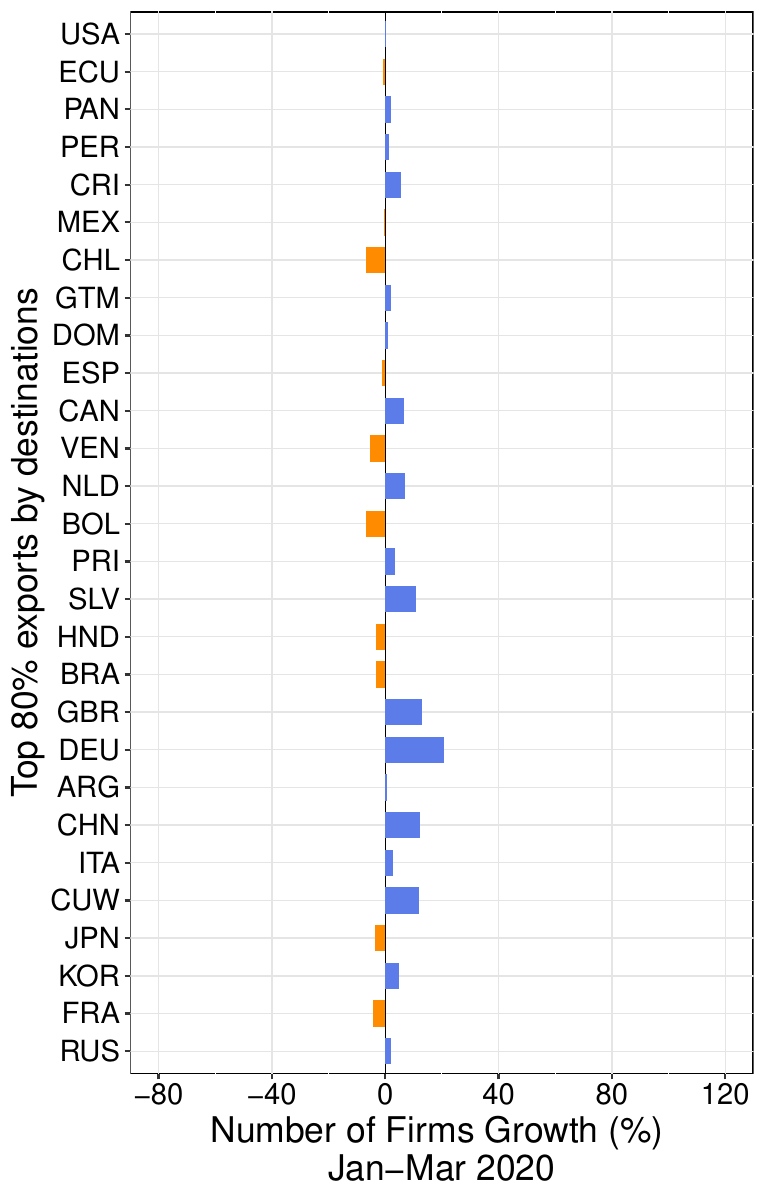}
\includegraphics[width=0.24\linewidth]{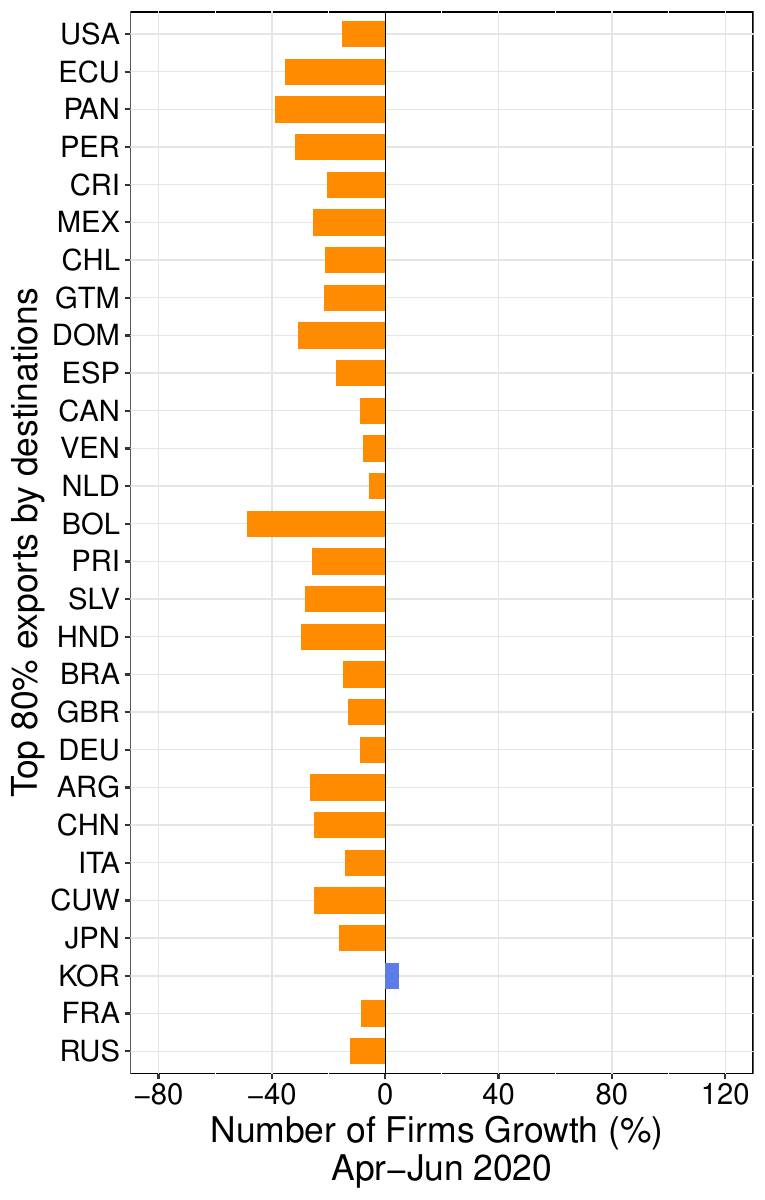}
\includegraphics[width=0.24\linewidth]{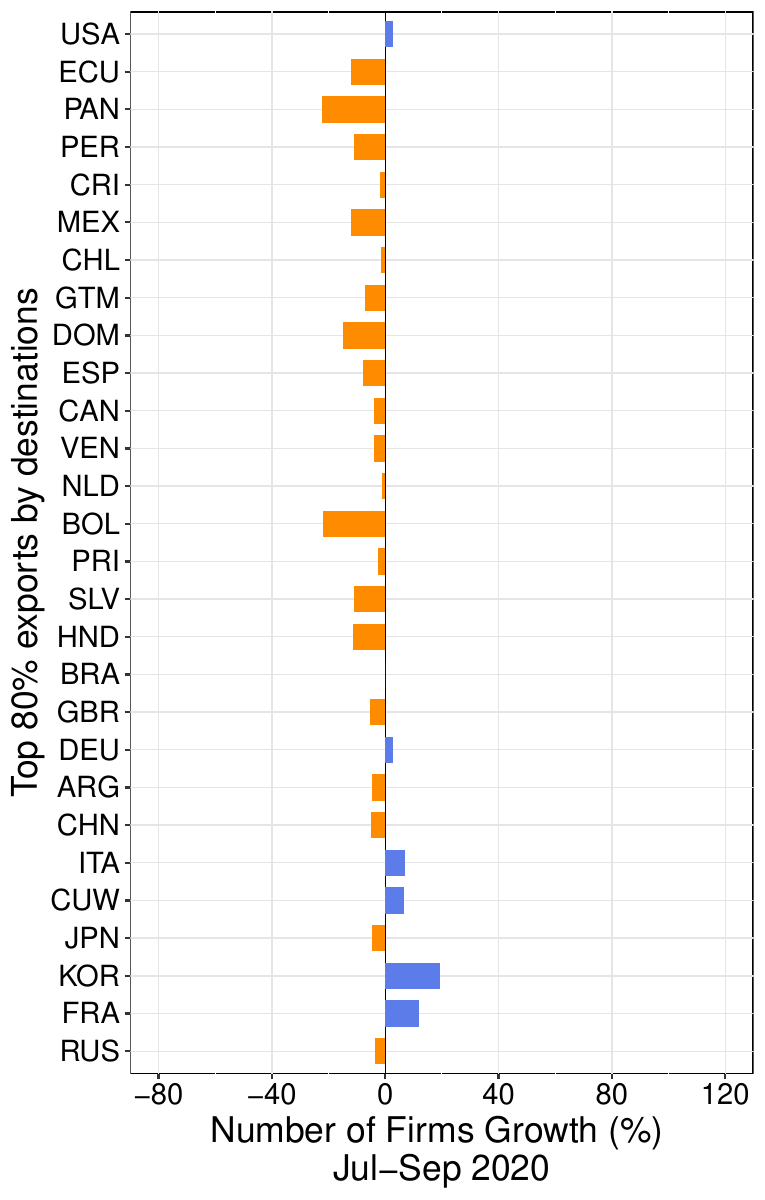}
\includegraphics[width=0.24\linewidth]{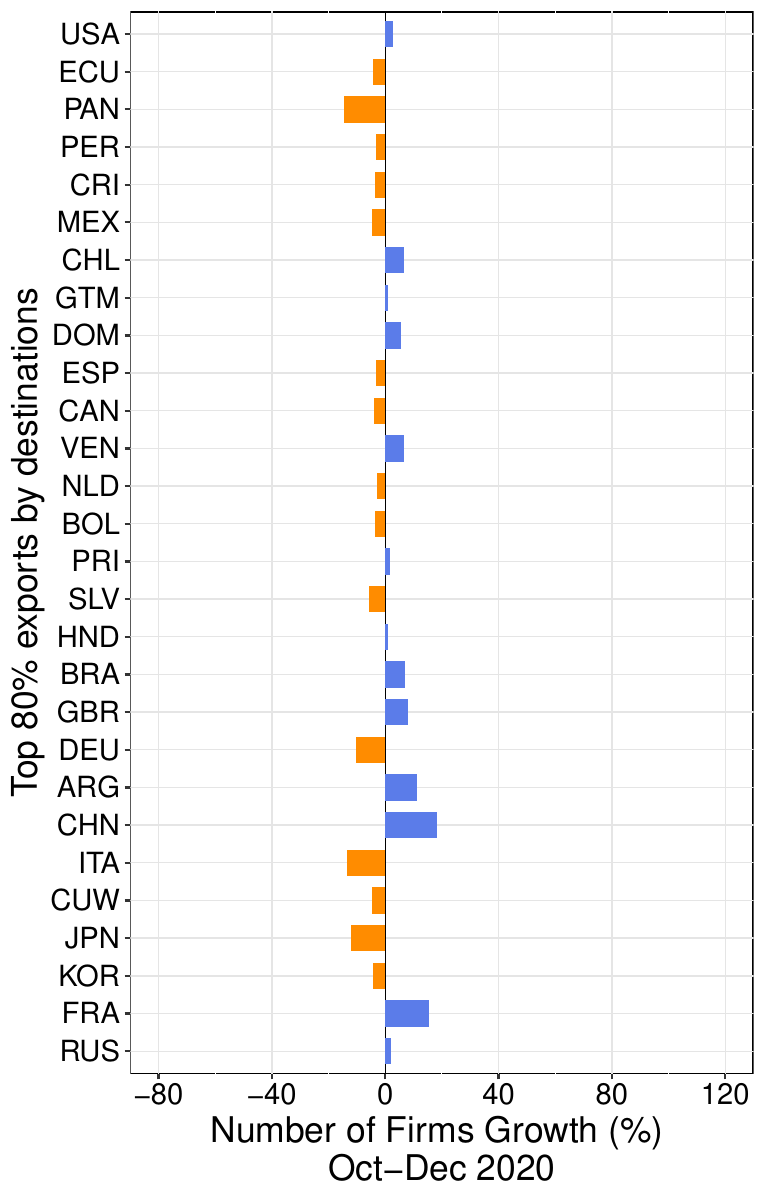} \\
\includegraphics[width=0.24\linewidth]{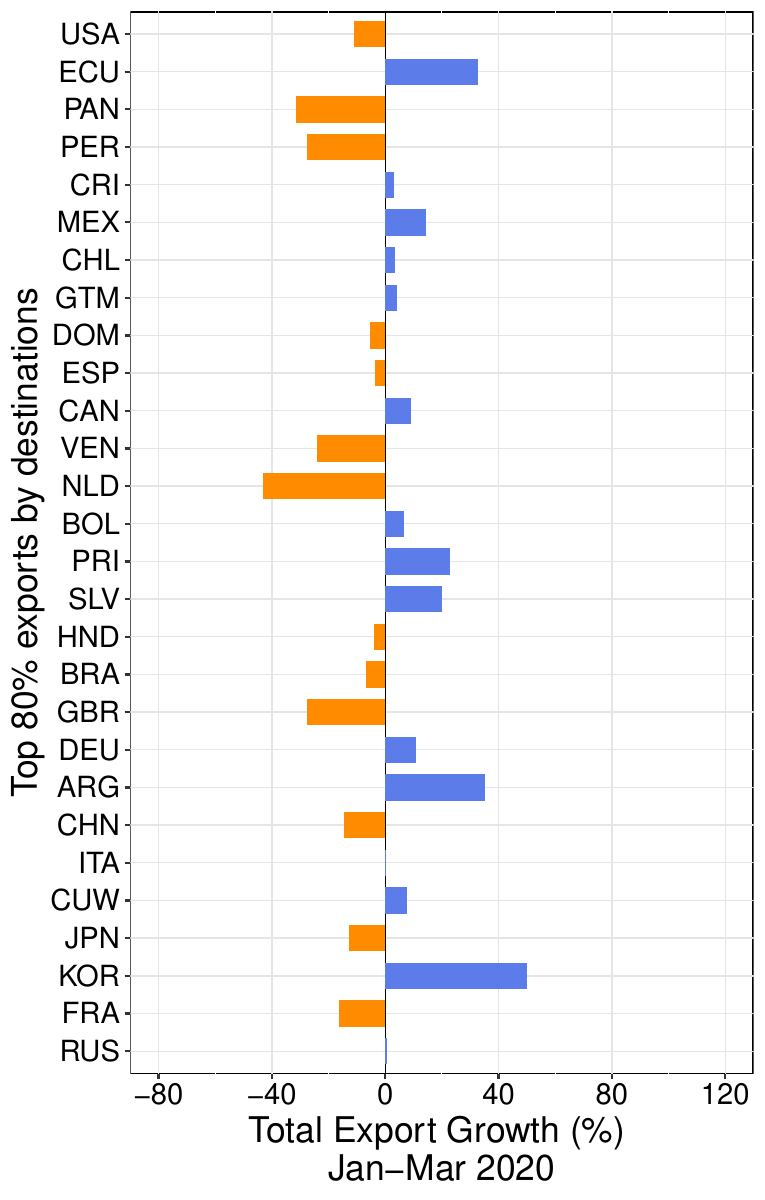}
\includegraphics[width=0.24\linewidth]{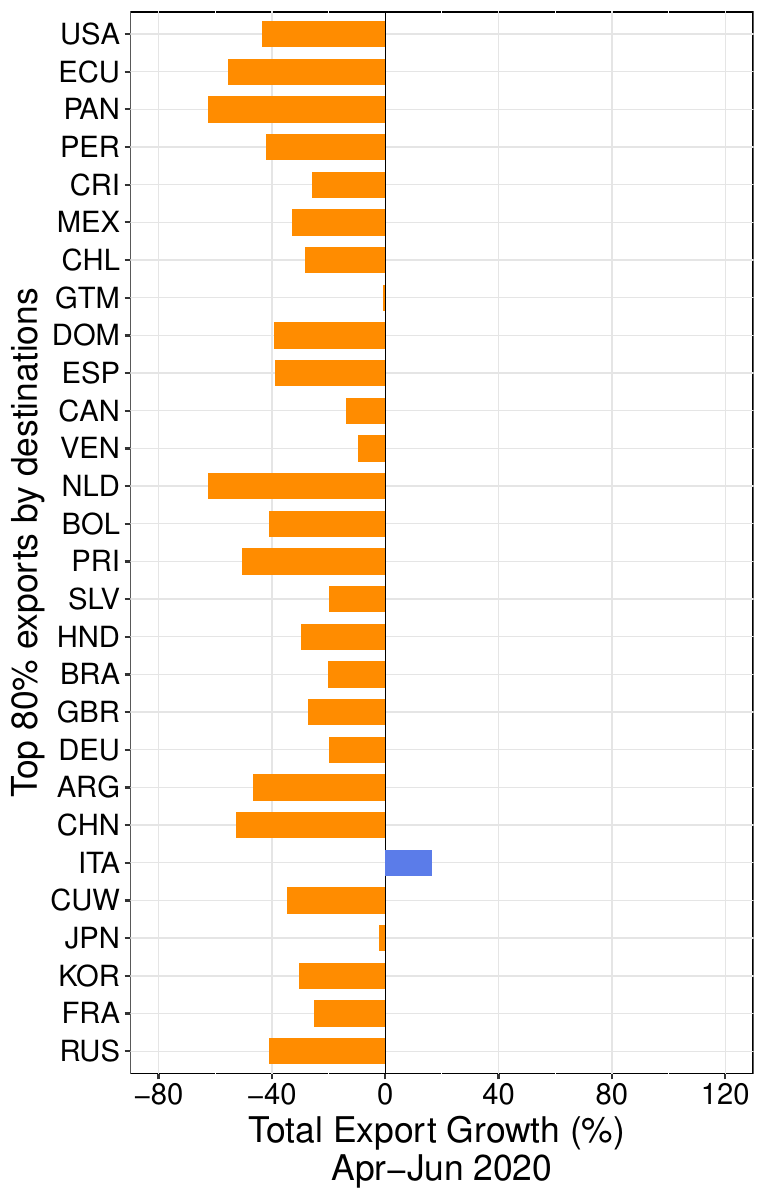}
\includegraphics[width=0.24\linewidth]{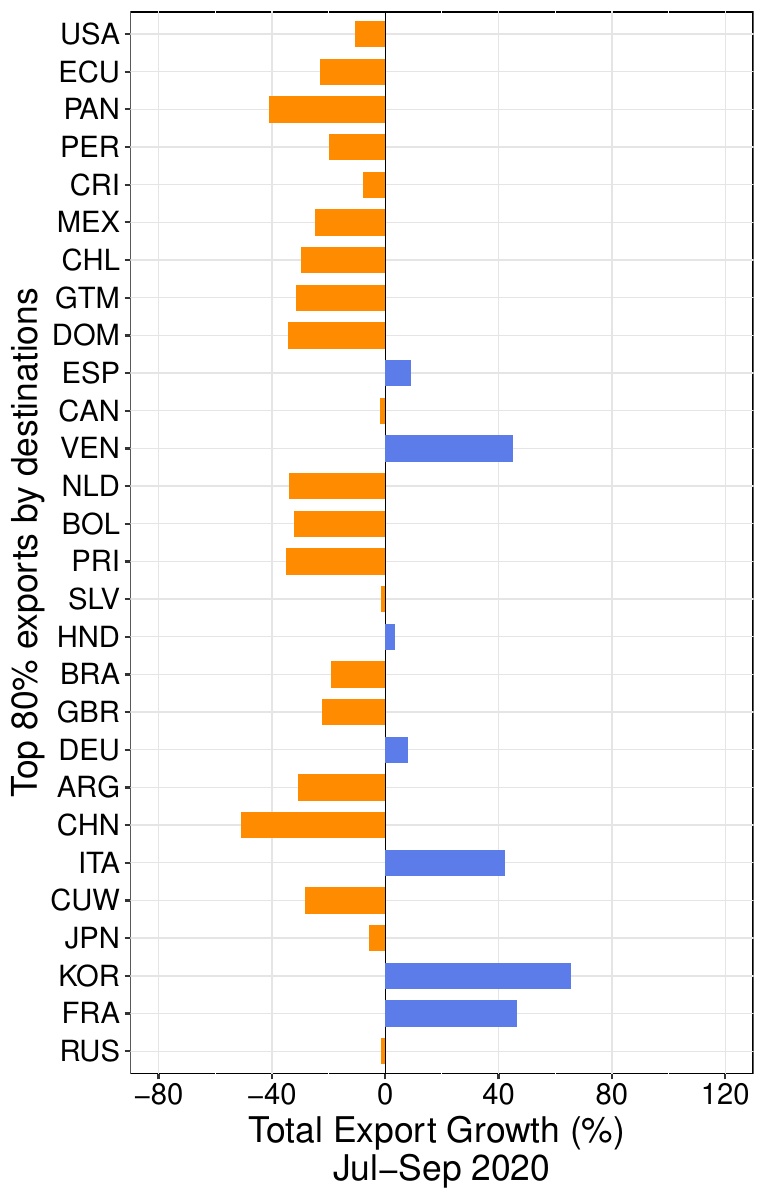}
\includegraphics[width=0.24\linewidth]{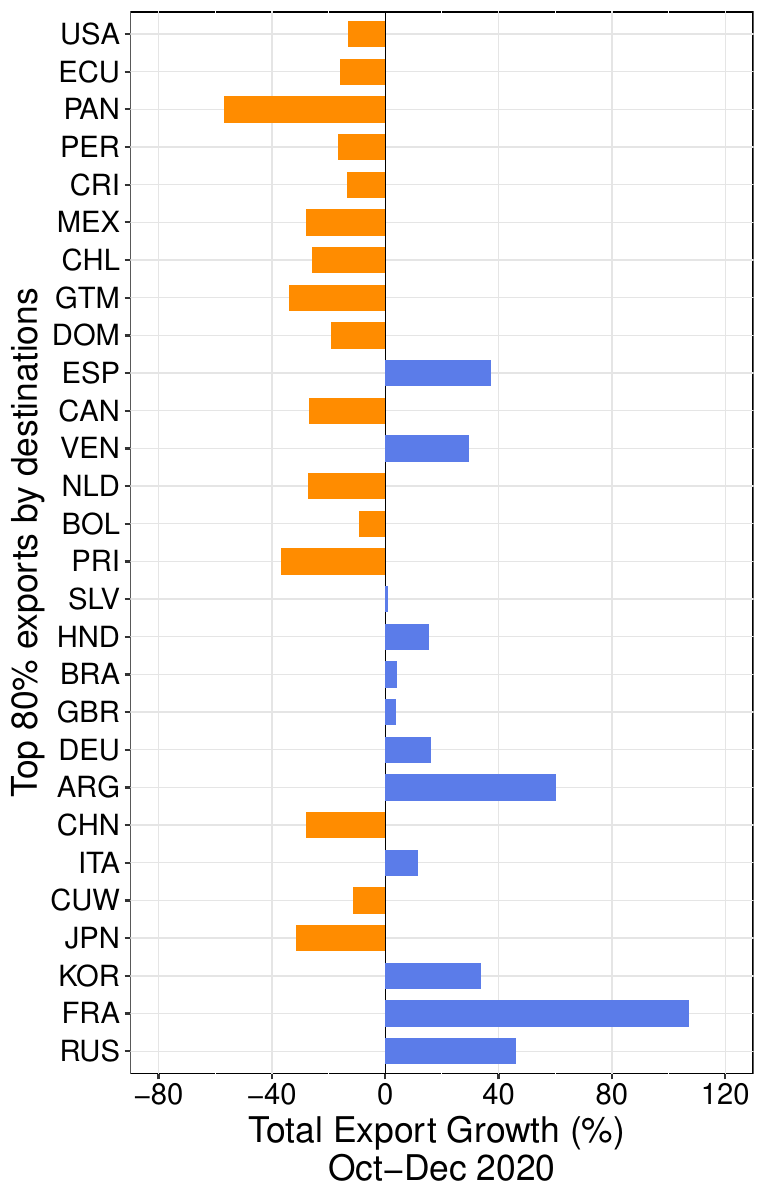} \\
\caption{The growth of the total number of exporters and the total value of exports by the destination country for the four quarters of 2020. Orange bars represent negative growth and blue bars positive growth. Destination countries are sorted from top to bottom accordingly to their importance in the share of the number of exporters in 2019.}
\label{fig:growth_iso}
\end{figure}
\begin{figure}[h]
\centering
\includegraphics[width=0.22\linewidth]{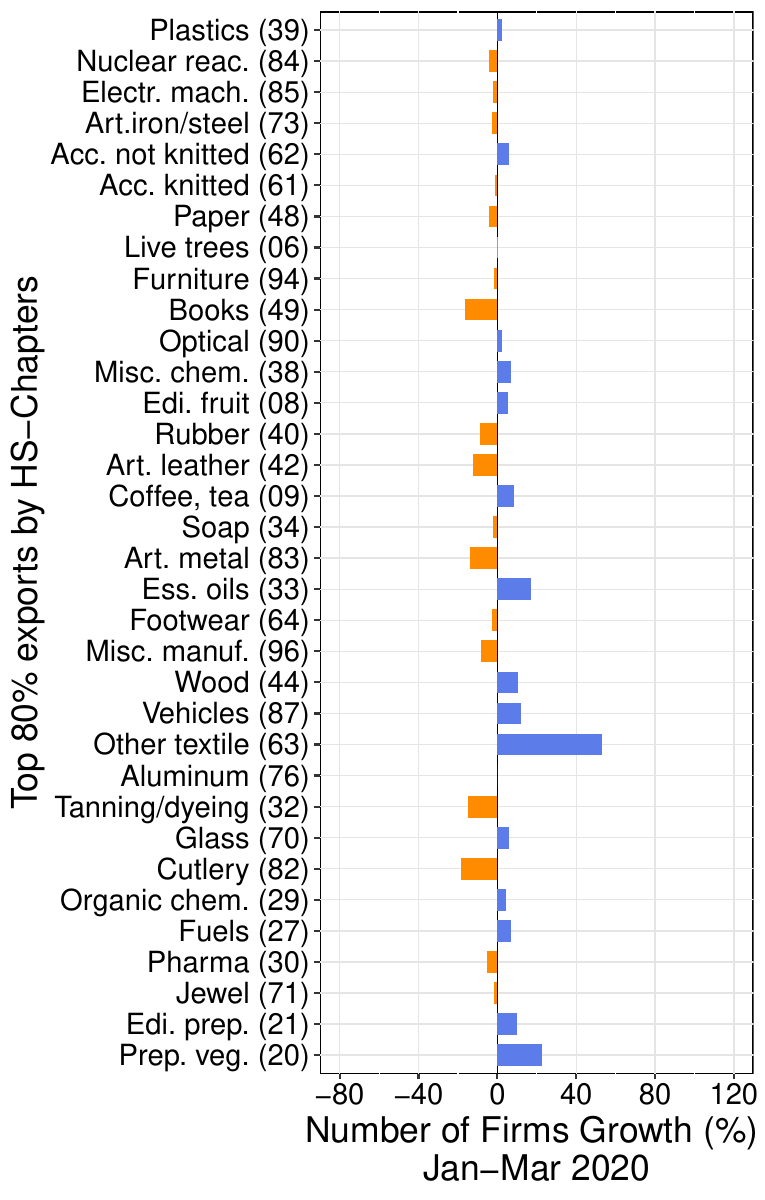}
\includegraphics[width=0.22\linewidth]{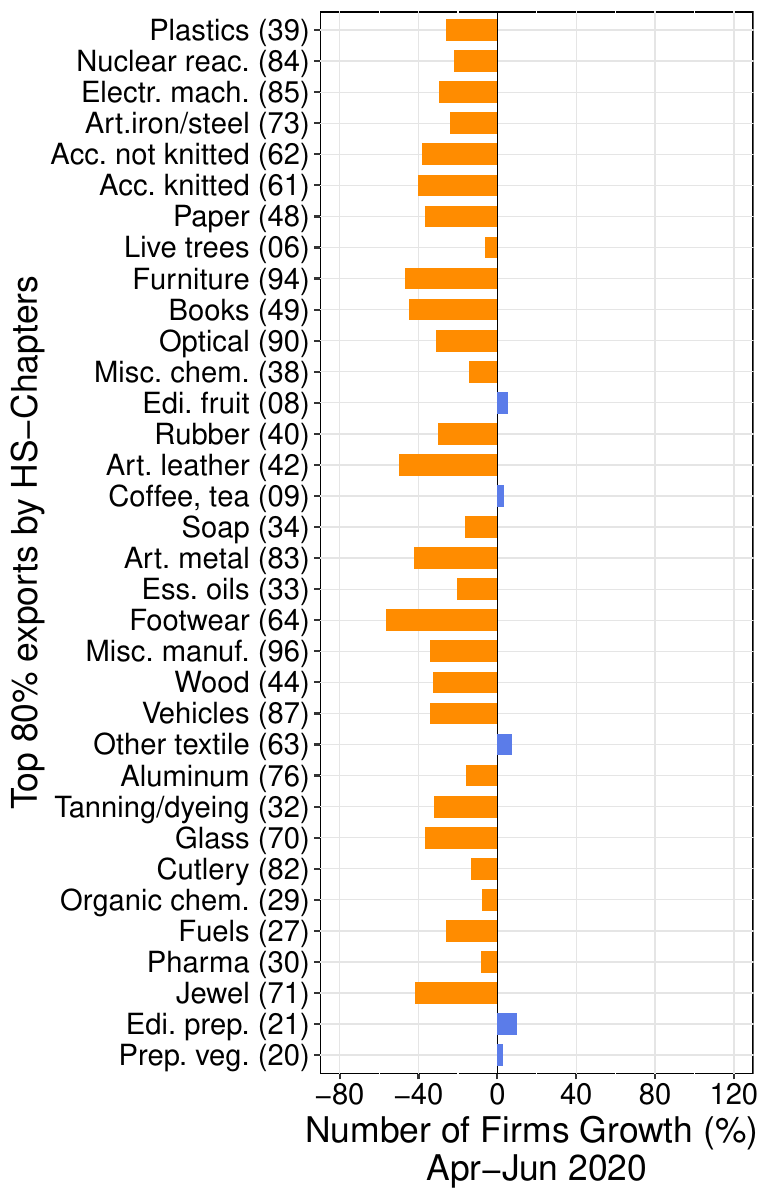} 
\includegraphics[width=0.22\linewidth]{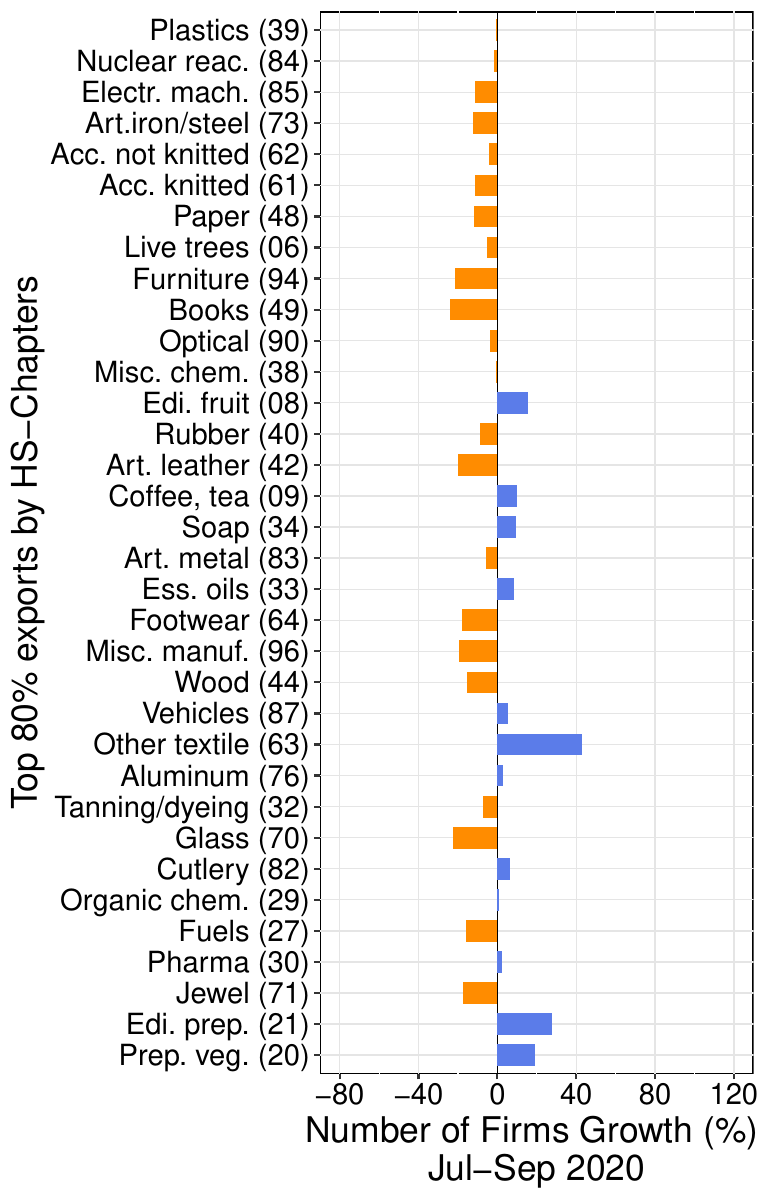}
\includegraphics[width=0.22\linewidth]{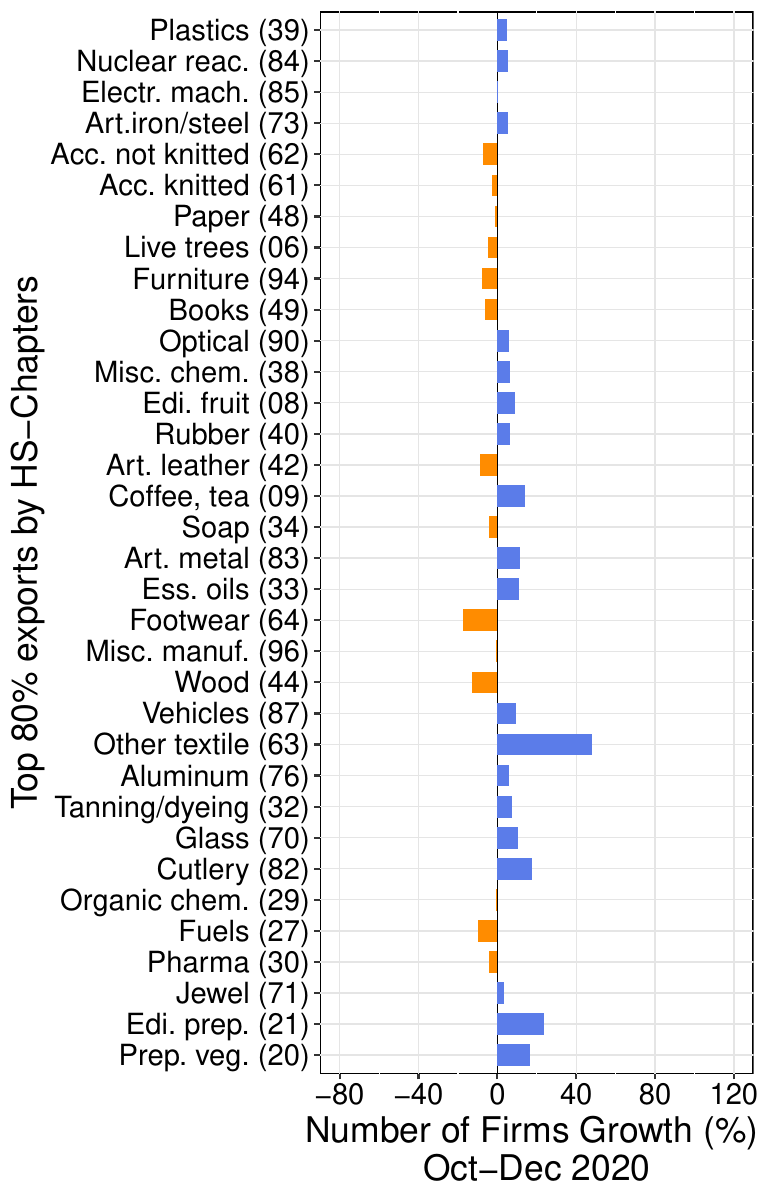}\\ \includegraphics[width=0.22\linewidth]{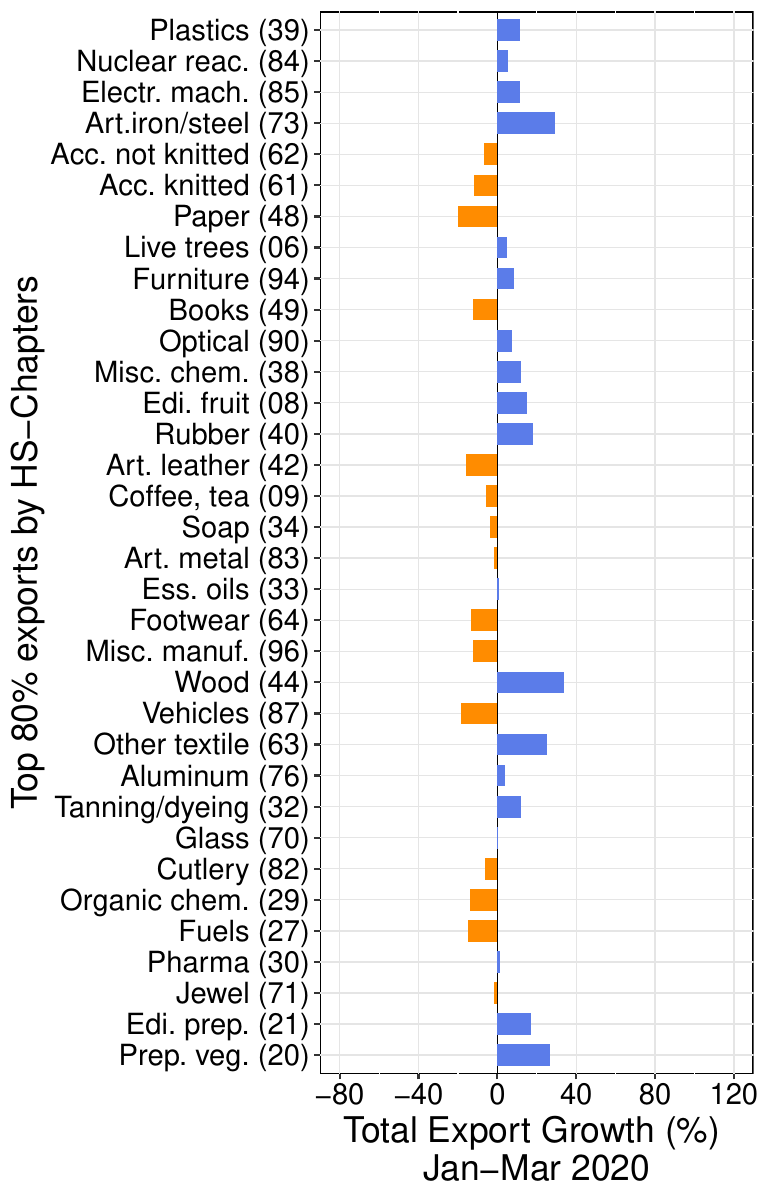}
\includegraphics[width=0.22\linewidth]{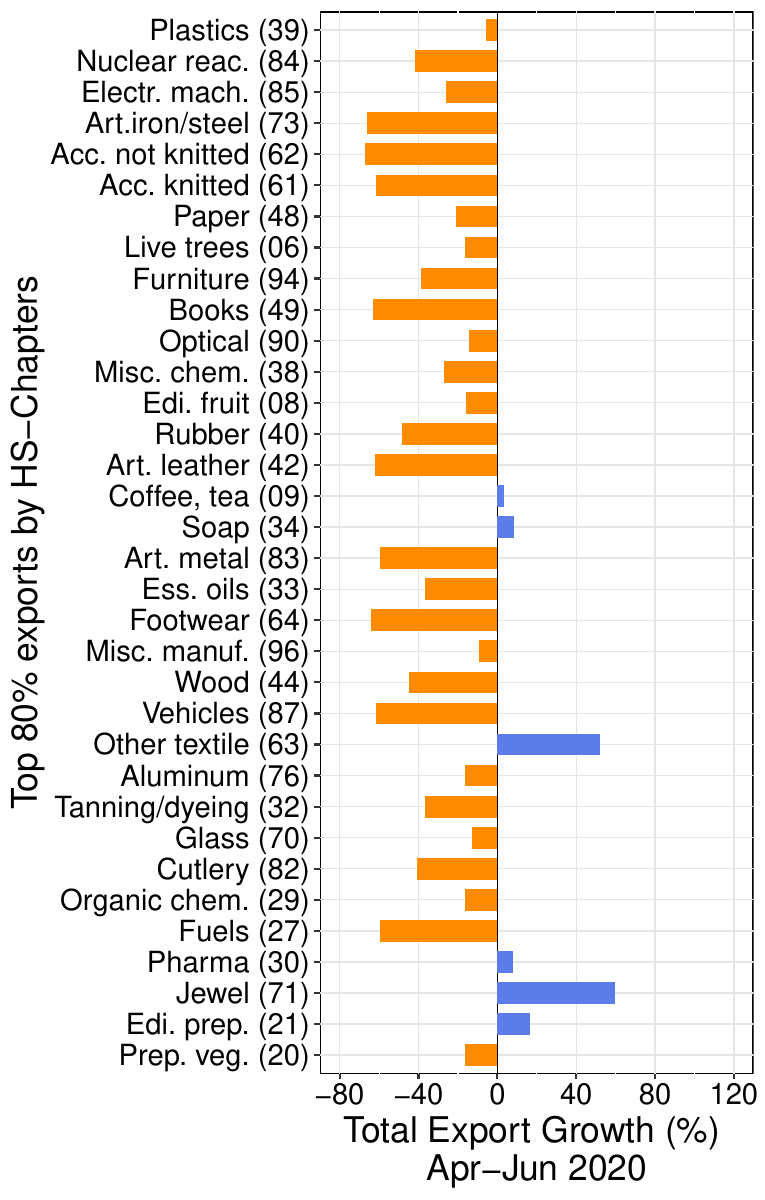} 
\includegraphics[width=0.22\linewidth]{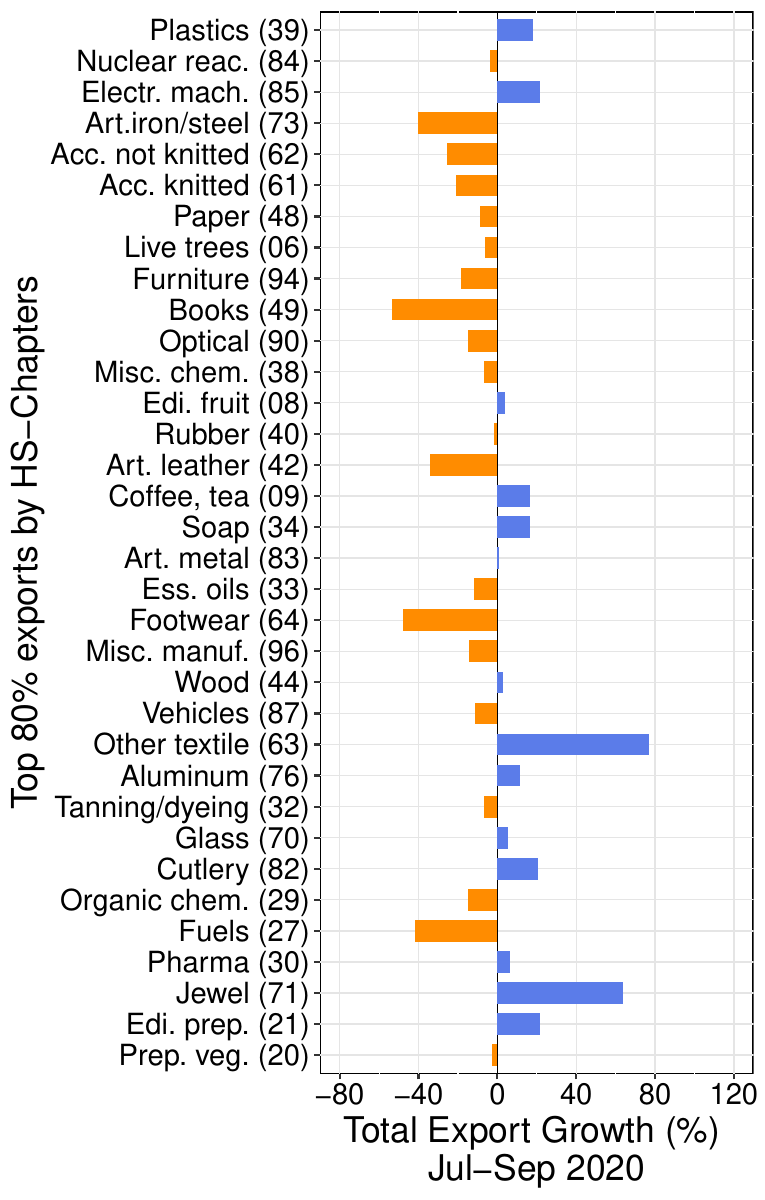}
\includegraphics[width=0.22\linewidth]{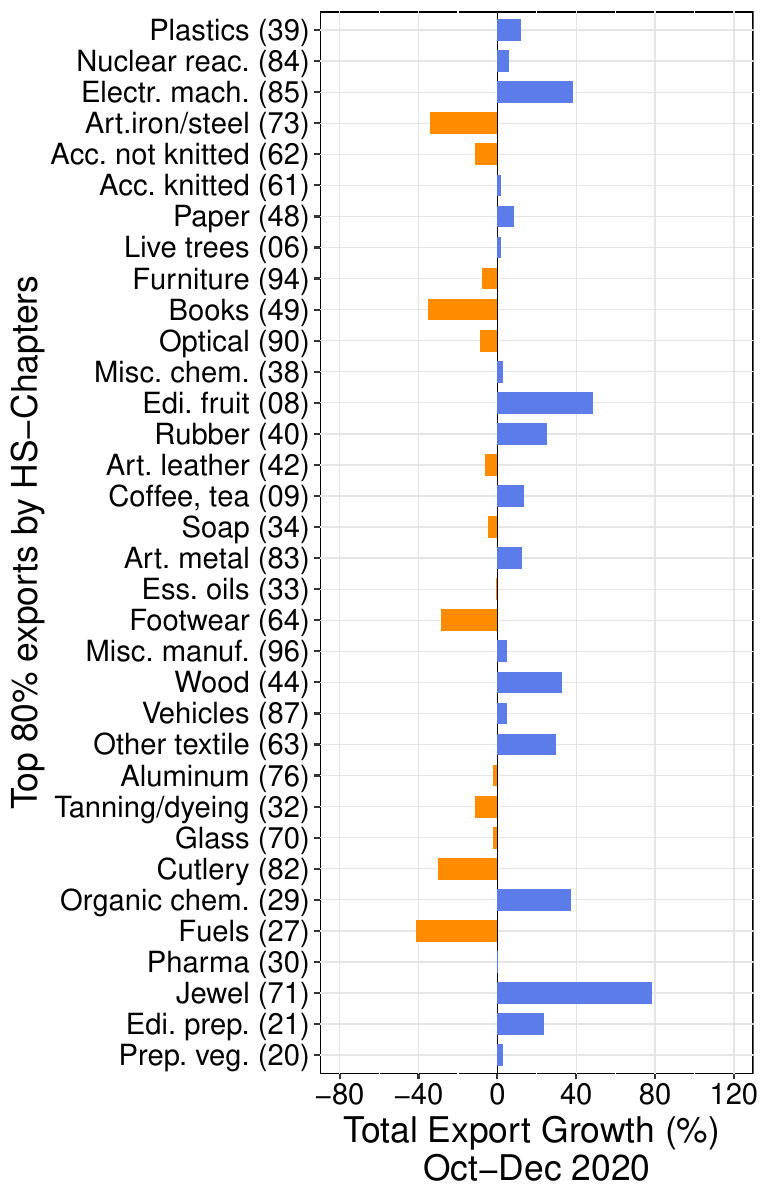}\\
\caption{The growth of the total number of exporters and the total value of exports by sector for the four quarters of 2020. Orange bars represent export reductions and blue bars positive export growth. Product sectors are sorted from top to bottom according to their importance in the share of the number of exporters in 2019. Product sectors correspond to the chapters of the HS code in parenthesis and the full name of the chapters is shortened to improve readability.}
\label{fig:growth_hs}
\end{figure}

Figures~\ref{fig:growth_iso} and~\ref{fig:growth_hs} show the growth of the total number of exporters and the growth of the total volume of exports between 2019 and 2020, by country of destination and product sector. We consider the first and the second quarter separately, and we select destinations and product sectors that account for 80\% of the total exporters in 2019. In both figures, the product sectors and the destinations are arranged by importance from top to bottom.

\begin{figure}[h!]
\centering
\includegraphics[width=0.24\linewidth]{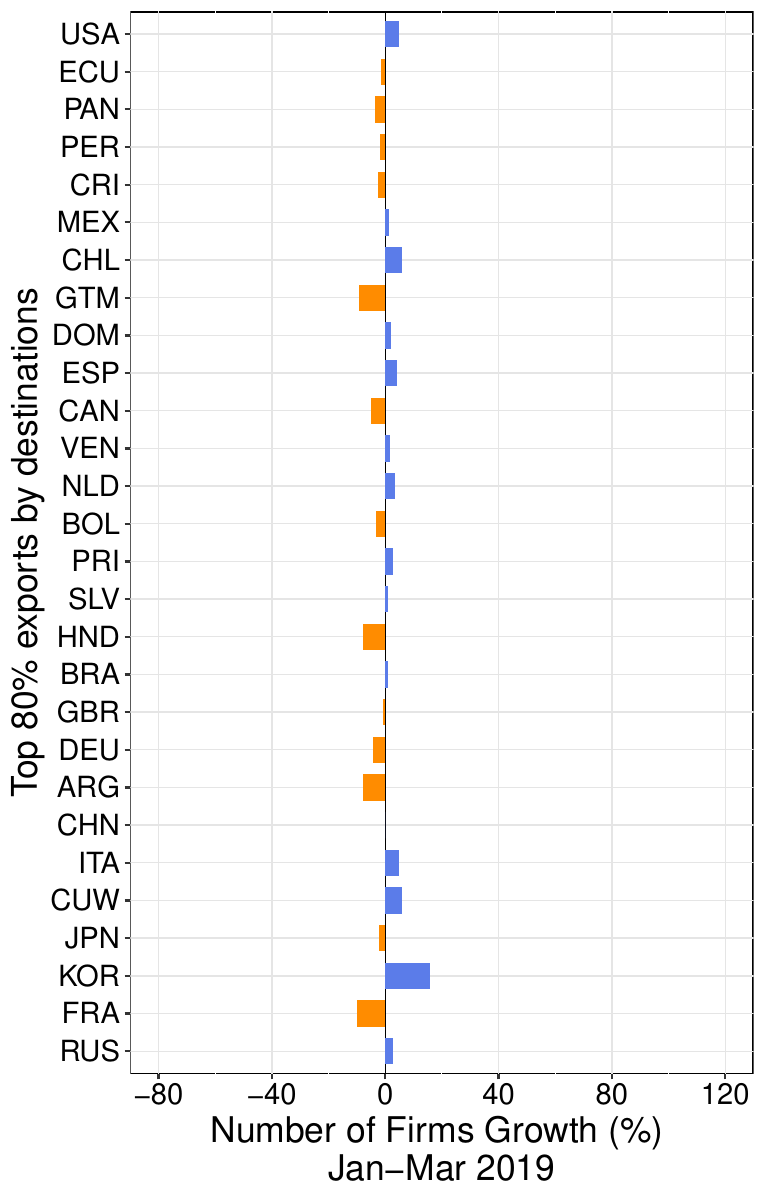}
\includegraphics[width=0.24\linewidth]{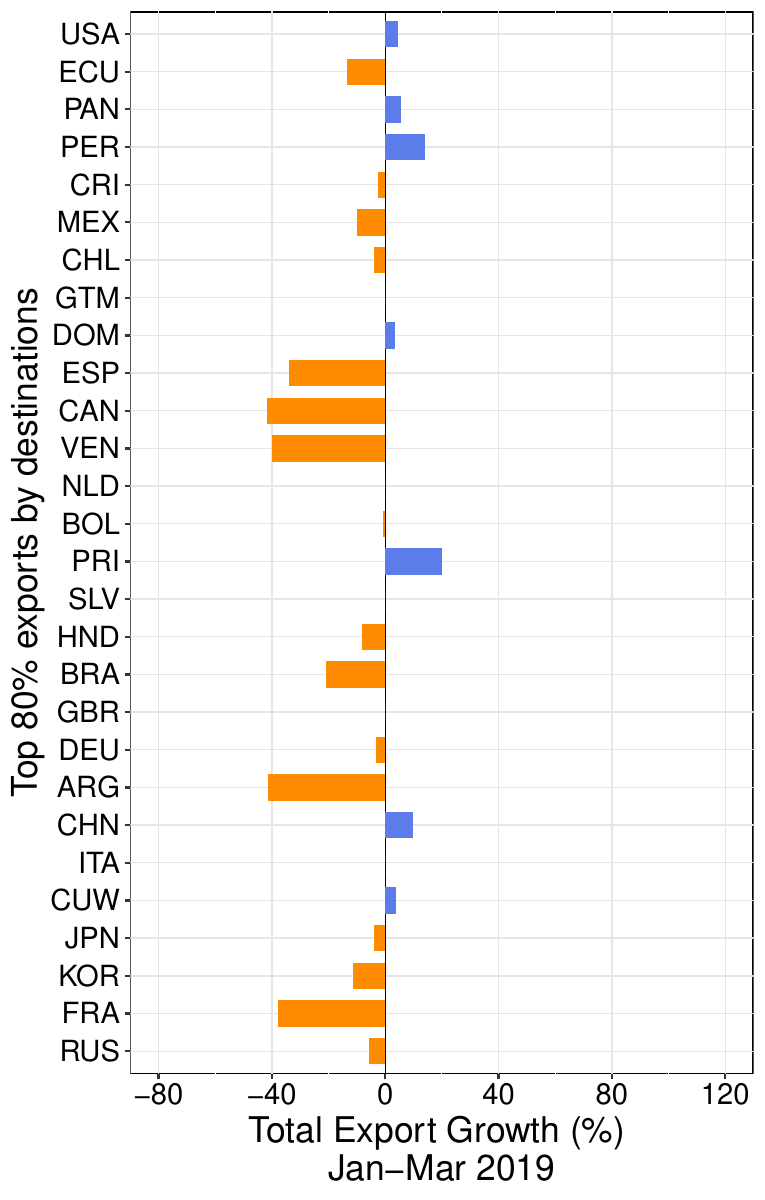}
\includegraphics[width=0.24\linewidth]{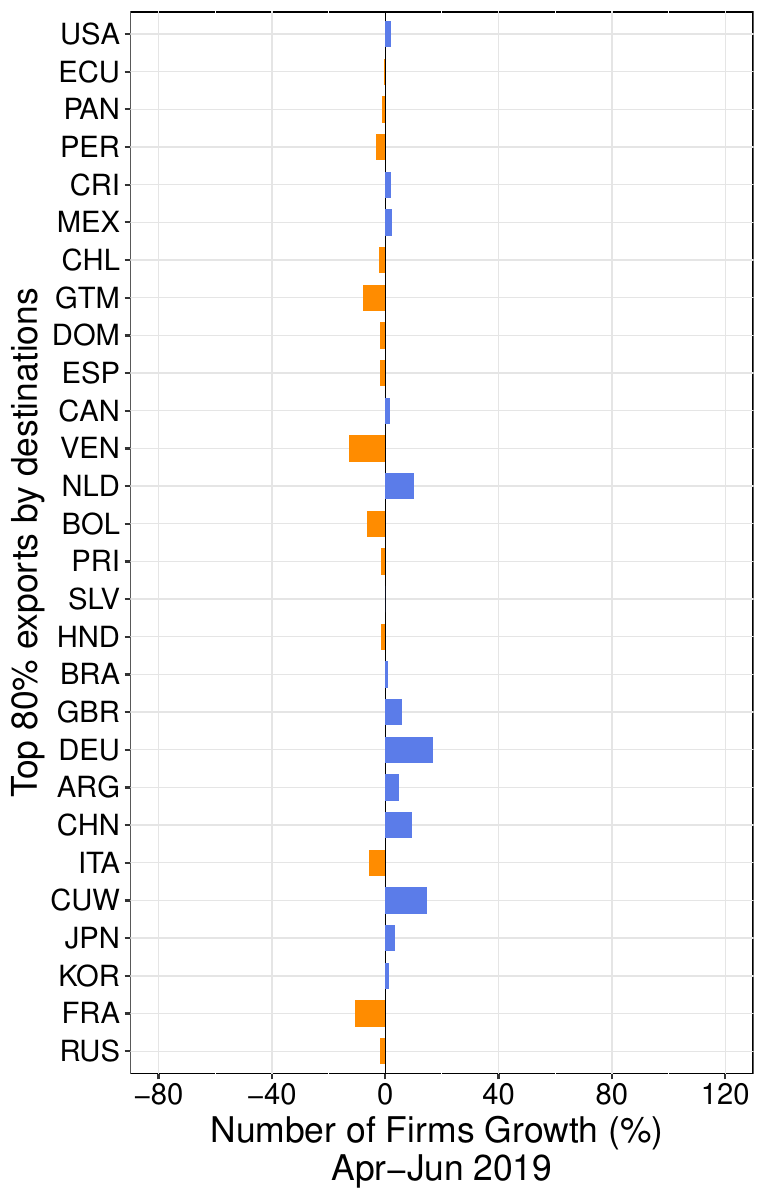}
\includegraphics[width=0.24\linewidth]{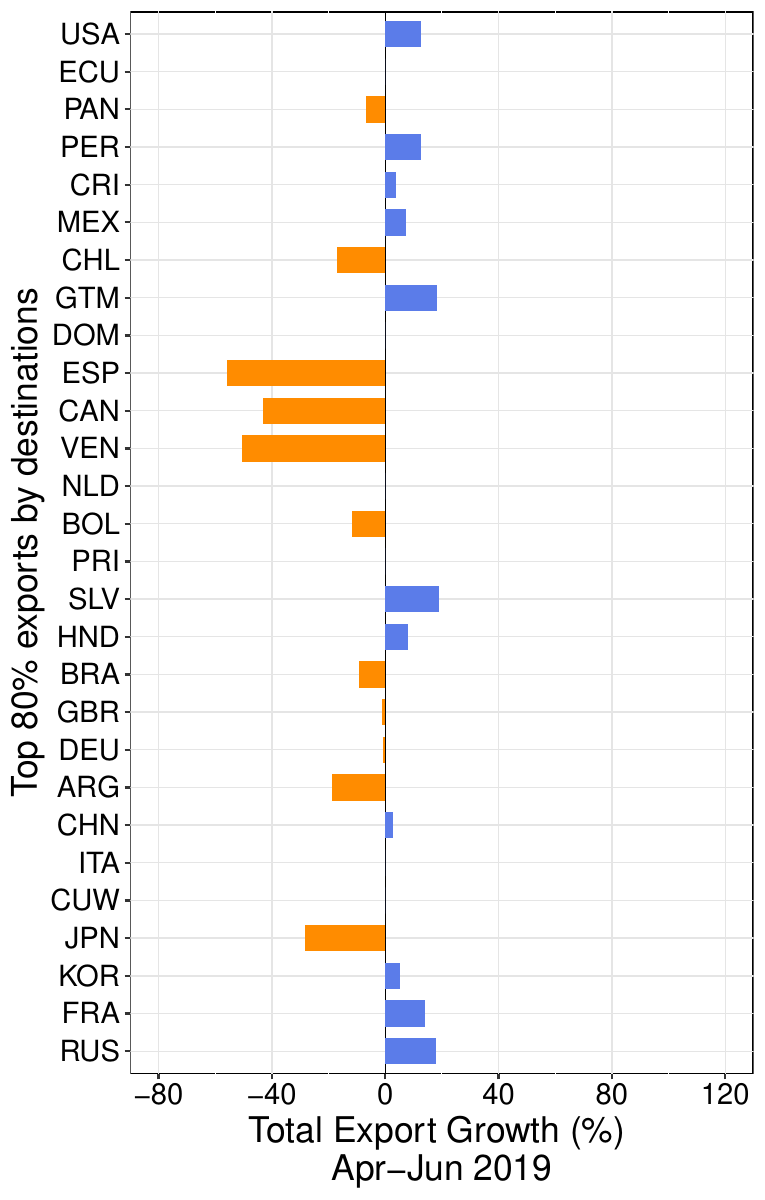}
\caption{The growth of the total number of exporters and the total value of exports by destination country for the first and the second quarters of 2019. Orange bars represent negative growth and blue bars positive growth. Destination countries are sorted from top to bottom accordingly with their importance in the share of number of exporters in 2019.}
\label{fig:app_growth_iso}
\end{figure}

Figure~\ref{fig:growth_iso} shows that, compared to the first quarter of 2020, the second quarter of the year is characterized by a severe and pervasive drop in the number of exporting firms and the volume of exports in almost all the destinations reported. Figure~\ref{fig:app_growth_iso} shows that the same drop is not observed during the second quarter of 2019.
During the third and fourth quarters of 2020, the value exported experienced more volatility than the number of firms. Nevertheless, the latter did not recover to the growth rates of the first quarter of the year. 

\begin{figure}[h!]
\centering
\includegraphics[width=0.24\linewidth]{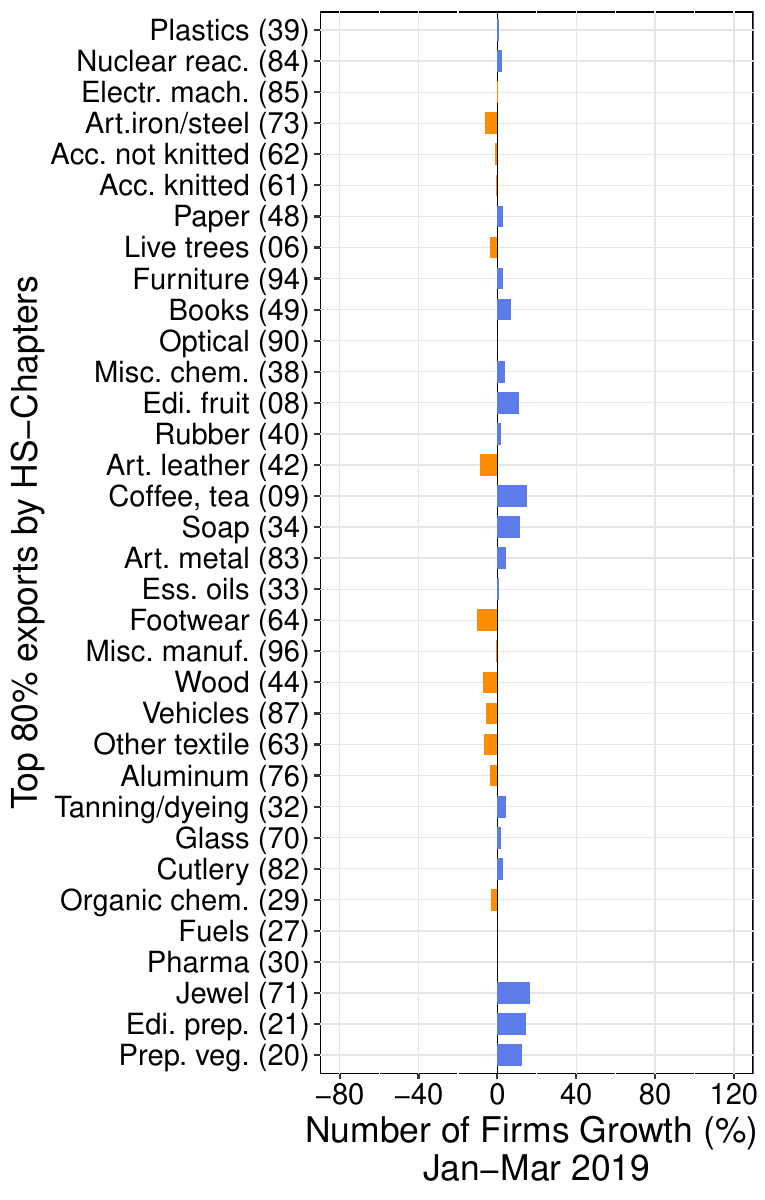}
\includegraphics[width=0.24\linewidth]{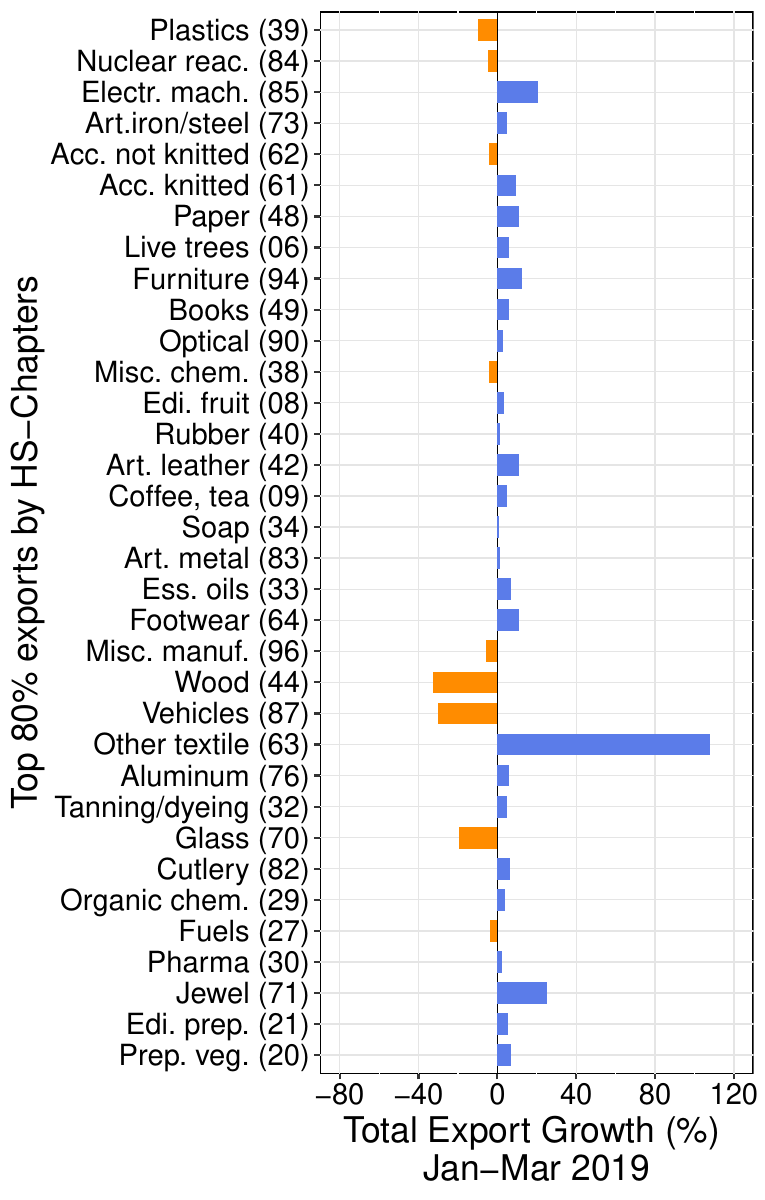}
\includegraphics[width=0.24\linewidth]{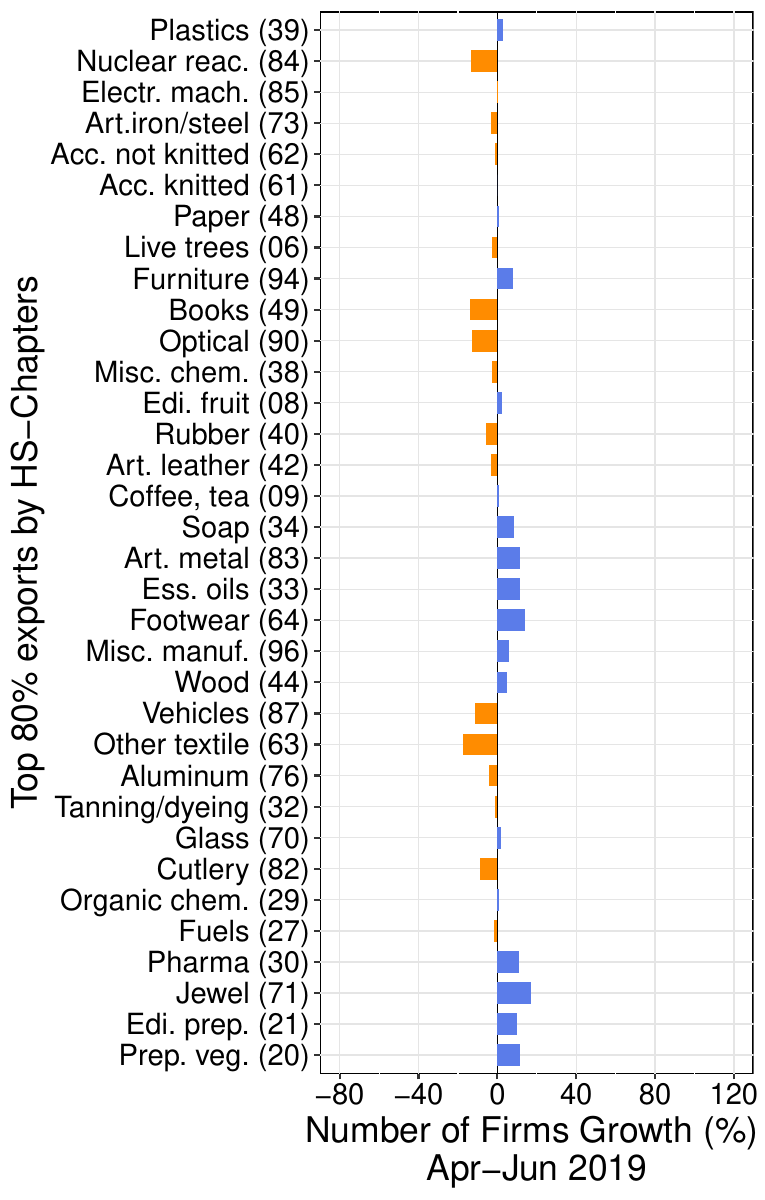}
\includegraphics[width=0.24\linewidth]{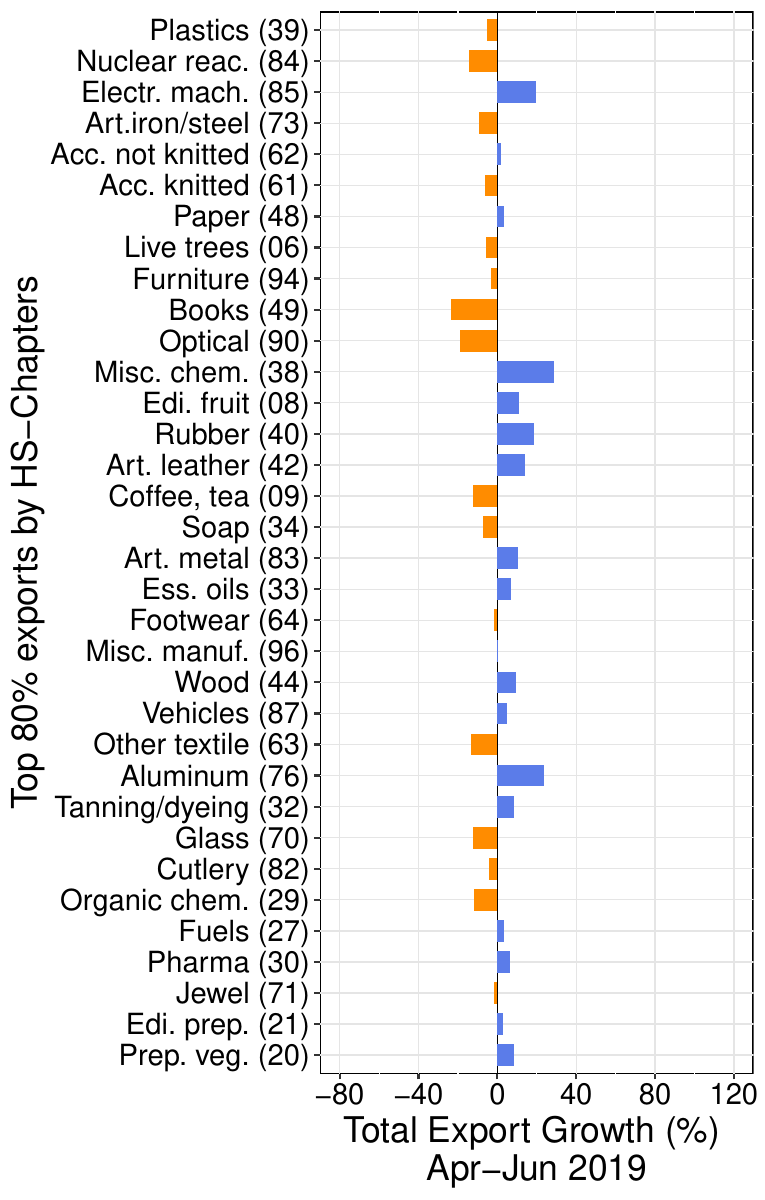}
\caption{The growth of the total number of exporters and the total value of exports by sector for the first and the second quarters of 2019. Orange bars represent export reductions and blue bars positive export growth. Product sectors are sorted from top to bottom accordingly with their importance in the share of number of exporters in 2019. Product sectors correspond to the chapters of the HS-code in parenthesis, the full name of the chapters is shortened to improve readability.}
\label{fig:app_growth_hs}
\end{figure}

Exports by product sectors in the second quarter of 2020 (see Figure~\ref{fig:growth_hs}) reveal a generalized decrease in the number of exporting firms and trade values, while the first quarter exhibits very heterogeneous patterns. The sectors that appear to be more severely affected in the second quarter are Footwear (HS64), Leather Articles (HS42), Furniture (HS94), Books (HS49), Articles of Metal (HS83), Knitted and Not-Knitted Accessories (HS61-62), Vehicles (HS87) and Articles of Iron or Steel (HS73). Interestingly, these sectors are relatively more labor-intensive in Colombia, and therefore they could be susceptible to disruptions connected to social distancing. 
Finally, only for Coffee and Tea (HS08), Other textiles (HS63), and Jewelries (HS71) exports in value significantly grew in the second quarter. Instead, in terms of the number of exporting firms, no product sectors exhibit notable positive dynamics. During the third and fourth quarters of 2020, there is a rapid back to normality in both the growth of value exported and in the number of exporters' growth rate by sector. Figure~\ref{fig:app_growth_hs} shows the same figures for 2019, suggesting that in periods without relevant shocks -- such as the ones of the first quarter of 2020 -- the changes in exports are also very heterogeneous, but there are not such extreme changes.


\newpage

\section{Appendix - Statistical Significance of the $CADiff$} 
\label{app:jpvals}

Starting from $B$ bootstrap replications of all the estimation steps (including the prediction stage), we calculate the $CADiff$ $B$ times. To determine the significance of the $CADiff$ we perform a two-tailed test. The p-values are constructed as follows: $$2\cdot min\{Pr(S\geq t|H_0), Pr(S\leq t|H_0)\}$$ being $t$ the observed $t$ test statistic, $t=\frac{\widetilde{CADiff_{original}}}{\Tilde{\sigma}}$, drawn from the unknown distribution $S$. $\Tilde{\sigma}$ represents the standard deviation of the bootstrapped $CADiff$. To adjust the p-values and obtain the joint p-values taking into account that we are testing hypotheses jointly on many covariates, we reproduce the ``single-step'' method employed by \cite{chernozhukov2018sorted} to control for the family-wise error rate.

In the following is described the single-step algorithm. We will indicate the bootstrap version of a variable, $v$, as $\Tilde{v}$ and its estimated version (on the original data) as $\hat{v}$. Moreover, $\Lambda(x)^{-u}$ will denote the first moment for the feature $x$ of interest in the least affected group including the observational units $i$ such that $\alpha_i<\alpha^{*}(u)$. Similarly, $\Lambda(x)^{+u}$ defines the first moment for the variable $x$ of interest in the most affected group including the observational units $i$ such that $\alpha_i>\alpha^{*}(1-u)$. Since we do not observe $\alpha$ directly, the mentioned quantities are estimated. According to the above convention, the estimated value of $\Lambda(x)^{-u}$ ($\Lambda(x)^{+u}$) will be $\hat{\Lambda}(x)^{-u}$ ($\hat{\Lambda}(x)^{+u}$) indicating the first moment for the variable $x$ of interest for firms $i$ such that $\hat{\alpha}_i>\hat{\alpha}^{*}(u)$ ($\hat{\alpha}_i<\hat{\alpha}^{*}(1-u)$). In the present paper $u=25$, however, we will maintain the more general $u$ notation for the sake of consistency with Section 4. \\
The single-step algorithm proceeds as follows: 1) for each variable $x \in X_t$, compute $\Tilde{\Lambda}(x)^{+u}$ and $\Tilde{\Lambda}(x)^{-u}$, bootstrap draws of $\hat{\Lambda}(x)^{+u}$ and $\hat{\Lambda}(x)^{-u}$ respectively. We want to test the null hypothesis, $H_0$, that $\Lambda^u(x)=0$, for $\Lambda^u(x)=[\Lambda(x)^{-u},\Lambda(x)^{+u}]$. 2) Construct a bootstrap draw of the distribution of ($\hat{\Lambda}^{+u}(x)-\hat{\Lambda}^{-u}(x)$), $Z^u_{\infty}(x)$. The latter is obtained by exploiting the bootstrap version of $\Lambda^{+u}(x)$ and $\Lambda^{-u}(x)$, namely: $\Tilde{Z}_{\infty}(x)= \sqrt{n}(\Tilde{\Lambda}^{u}(x) -\hat{\Lambda}^{u}(x))$ where $\Tilde{\Lambda}^u(x)=[\Tilde{\Lambda}(x)^{-u},\Tilde{\Lambda}(x)^{+u}]$\footnote{Similarly, $\hat{\Lambda}^u(x)=[\hat{\Lambda}(x)^{-u},\hat{\Lambda}(x)^{+u}]$}. 3) Repeat steps 1) and 2) $B$ times; 4) compute a bootstrap estimator of the variance of $Z_{\infty}$ as  $\hat{\Sigma}^u(x)=\frac{q^u_{0.75}(x)-q^u_{0.25}(x)}{z_{0.75}-z_{0.25}}$ being $q^u_p(x)$ the $p^{th}$ sample quantile of $\Tilde{Z}_{\infty}(x)$ and $z_p$ the $p^{th}$ quantile of a standard normal distribution. 5) Use the latter to construct the test statistic $\Tilde{\tau}(X_t)=sup_{x \in X_t}|\Tilde{Z}_{\infty}(x)|\cdot|\hat{\Sigma}^u(x)|^{-1/2}$. A p-value for the null $H_0$ that $\Lambda^u(x)=0$ for all $x\in X_t$ of the realization of the estimated statistic, $sup_{x \in X_t}|\hat{\Lambda}^u(x)| \cdot |\hat{\Sigma}(x)|^{-1/2} =s$, is given by the average number of times that $\Tilde{\tau}(X_t)$ is greater than $s$, where $s = \frac{\Tilde{\beta_{1,f}^{m}}}{\hat{\Sigma}^u(x)}$. The $\Tilde{.}$ indicates simply that the $\beta_{1,f}^{m}$ has been projected to the bootstrap dimension. \\

\newpage
\section{Appendix - Comparison with \cite{chernozhukov2020generic}} 
\label{app:comparison}

We show that our empirical strategy is built on the same pillars as \cite{chernozhukov2020generic}, but applies them to a different setting. To simplify the exposition, we refer to Table \ref{tab:toy_ex} which provides a simplified representation of our empirical setting.

 \cite{chernozhukov2020generic} deal with an experimental empirical setting in which one can easily separate a treated group from a control group. In order to study the heterogeneity of the average treatment effect, the first step of \cite{chernozhukov2020generic} is to split randomly the sample under analysis in an auxiliary ($A$) and a main sample ($M$) of approximately the same size. Then, they employ ML techniques to learn in $A$ the function approximating the potential outcomes in the treatment and non-treatment scenarios, while $M$ is used to make inferences on the key features of treatment effect heterogeneity. In other words, they estimate the function describing the outcome in case of treatment (no treatment) on the subset of treated (non-treated) firms contained in $A$. These two estimated functions are used to impute the two potential outcomes for each firm contained in the $M$ sample (the difference represents the estimated individual treatment effects) and study the treatment effect heterogeneity estimated for these firms by using, inter alia, the Sorted Effects method \citep{chernozhukov2018sorted}. This procedure is designed in this way to avoid overfitting (i.e., doing learning and prediction using the same sample), and, starting from the random splitting, it is repeated many times in order to obtain many distributions of estimated treatment effects to which the Sorted Effects method is applied.
\begin{table}[h]
\begin{center}
\begin{tabular}{cccc}
\toprule
  \multicolumn{3}{c}{\thead{Our Setting}} &
  \multicolumn{1}{c}{\thead{Chernozhukov \\ (2020) Setting}} \\
\cmidrule(lr){1-3} 
 \multicolumn{2}{c}{SUM}   &SAM &  $A-M$ splitting \\
\midrule
\tikzmark{startup}$(X_{2018},Y_{2019})_1$\tikzmark{endup} &  \tikzmark{st}$(X_{2019},Y_{2020})_{11}$ \tikzmark{en}& \tikzmark{st2}$(X_{2019},Y_{2020})_{11}$\tikzmark{nd2} &\tikzmark{stt}$(X_{2019},Y_{2020})_{11}$\tikzmark{ndd} \tikzmark{A} \\ 
$(X_{2018},Y_{2019})_2$ &  $(X_{2019},Y_{2020})_{12}$ & \tikzmark{st3}$(X_{2019},Y_{2020})_{12}$\tikzmark{nd3} &$(X_{2019},Y_{2020})_{12}$\\  
$(X_{2018},Y_{2019})_3$ &  $(X_{2019},Y_{2020})_{13}$ & \tikzmark{st4}$(X_{2019},Y_{2020})_{13}$\tikzmark{nd4} &$(X_{2019},Y_{2020})_{13}$ \\ 
$(X_{2018},Y_{2019})_4$ &  $(X_{2019},Y_{2020})_{14}$ & $(X_{2019},Y_{2020})_{14}$ &$(X_{2019},Y_{2020})_{14}$ \\ 
$(X_{2018},Y_{2019})_5$ &  $(X_{2019},Y_{2020})_{15}$ & $(X_{2019},Y_{2020})_{15}$ &$(X_{2019},Y_{2020})_{15}$\tikzmark{B}\\ 
$(X_{2018},Y_{2019})_6$ &  $(X_{2019},Y_{2020})_{16}$ & $(X_{2019},Y_{2020})_{16}$ &\tikzmark{stt1}$(X_{2019},Y_{2020})_{16}$\tikzmark{ndd1} \tikzmark{C}\\ 
$(X_{2018},Y_{2019})_7$ &  $(X_{2019},Y_{2020})_{17}$ & $(X_{2019},Y_{2020})_{17}$ &$(X_{2019},Y_{2020})_{17}$ \tikzmark{D}\\
$(X_{2018},Y_{2019})_8$ &  $(X_{2019},Y_{2020})_{18}$ & $(X_{2019},Y_{2020})_{18}$ &$(X_{2019},Y_{2020})_{18}$ \tikzmark{D}\\
$(X_{2018},Y_{2019})_9$ &  $(X_{2019},Y_{2020})_{19}$ & $(X_{2019},Y_{2020})_{19}$ &$(X_{2019},Y_{2020})_{19}$ \tikzmark{D}\\
$(X_{2018},Y_{2019})_{10}$ &  $(X_{2019},Y_{2020})_{20}$ & $(X_{2019},Y_{2020})_{20}$ &$(X_{2019},Y_{2020})_{20}$ \tikzmark{D}\\
\bottomrule
\end{tabular}
\end{center}
    \caption{A simplified representation of our empirical setting in which we compare the methods used in the present paper to those described in \cite{chernozhukov2020generic}.}
    \label{tab:toy_ex}
\end{table}

\begin{tikzpicture}[remember picture,overlay]
\foreach \Val in {up}
{
\draw[rounded corners,red,thick]
  ([shift={(-0.5\tabcolsep,2.5ex)}]pic cs:start\Val) 
    rectangle 
  ([shift={(0.5\tabcolsep,-28ex)}]pic cs:end\Val);
}

\draw[rounded corners,green,thick]
  ([shift={(-0.5\tabcolsep,2.5ex)}]pic cs:st) 
    rectangle 
  ([shift={(0.5\tabcolsep,-28ex)}]pic cs:en);

\draw[rounded corners,gray,dashed,thick]
  ([shift={(-0.5\tabcolsep,1.8ex)}]pic cs:st2) 
    rectangle 
  ([shift={(0.5\tabcolsep,-28ex)}]pic cs:nd2);
  
\draw[rounded corners,orange,dashed,thick]
  ([shift={(-0.5\tabcolsep,1.8ex)}]pic cs:st3) 
    rectangle 
  ([shift={(0.5\tabcolsep,-28ex)}]pic cs:nd3);
  
  \draw[rounded corners,green,dashed,thick]
  ([shift={(-0.5\tabcolsep,1.8ex)}]pic cs:st4) 
    rectangle 
  ([shift={(0.5\tabcolsep,-28ex)}]pic cs:nd4);

\draw[rounded corners,red,thick]
  ([shift={(-0.5\tabcolsep,1.8ex)}]pic cs:stt) 
    rectangle 
  ([shift={(0.5\tabcolsep,-13ex)}]pic cs:ndd);
  
  
  \draw[rounded corners,blue,thick]
  ([shift={(-0.5\tabcolsep,1.8ex)}]pic cs:stt1) 
    rectangle 
  ([shift={(0.5\tabcolsep,-14ex)}]pic cs:ndd1);
  
\end{tikzpicture}

\begin{tikzpicture}[remember picture, red, overlay]
\draw [thick,decorate,decoration={brace,amplitude=10pt,raise=4pt}]
 ($(pic cs:A) + (0.3, 0.2)$)   --  ($(pic cs:B) + (0.3,-0.1)$) node [midway,xshift=0.9cm] {$A$};
 \end{tikzpicture}
\begin{tikzpicture}[remember picture, blue, overlay]
 \draw [thick,decorate,decoration={brace,amplitude=10pt,raise=4pt}]
 ($(pic cs:C) + (0.3, 0.2)$)   --  ($(pic cs:D) + (0.3,-0.1)$) node [midway,xshift=0.9cm] {$M$};
 \end{tikzpicture}
 

As an example, in Table \ref{tab:toy_ex}, we represent 20 exporting firms observed in 2018 or in 2019.
 In the context of our setting, the strategy of \citep{chernozhukov2018sorted} would imply that the 2019 sample, for which we are interested in estimating the average treatment effect, should be divided in two, as shown in the last column of Table \ref{tab:toy_ex}. However, in the COVID-19 scenario, one cannot easily separate treated and control units because COVID-19 imposes a (at least indirect) treatment over all units, hence preventing the possibility of discerning between treated and controls.\footnote{Furthermore, even if we assume that during the first three months of the year there was no COVID-19 effect going on, and therefore we categorize as non-treated (treated) firms operating in those months (in the other remaining months), and we use the non treated firms in the auxiliary sample to learn, it would be problematic to use the learning outcome in case of no treatment during the first three months to predict the outcome in case of no treatment for the treated firms that are those in the last 8 months because of the strong seasonality effects we have. So the outcome during the first three months in case of no treatment would be very different from the outcome of the last months in case of no treatment just because of seasonality effects. 
 }
  Moreover, with respect to \cite{chernozhukov2020generic}, in our empirical setting, we do not have the necessity to predict the outcome of controls in the case of ``no treatment'' because we are not interested in estimating the COVID-19 effect on 2018 exporters. Therefore, we do not have to split the controls observed in 2018 in two halves to avoid overfitting and this enables us to reconstruct a counterfactual outcome of no treatment for each 2019 exporter without incurring in overfitting problems. Therefore, in this paper for the SUM we use as an auxiliary sample all the Colombian exporters observed in 2018 ($A$) and as the main sample ($M$) all the Colombian exporters observed in 2019.  For the SAM, we perform instead a K-Fold splitting in which, iteratively we select 80\% of the firms in 2019-2020 as being part of $A$ and the remaining 20\% as being part of $M$. This is shown in the column ``Our Setting (SAM)'' of Table \ref{tab:toy_ex}, where different $A$ (and, accordingly different $M$) groups are selected according to the different colors of the dashed circles.
 In this way we avoid overfitting problems and, at the same time, we exploit all the available data by being able to compare the predicted probabilities to export in the COVID-19 with those in the non-COVID-19 scenario for all the observed 2019 exporters.

\end{document}